\documentclass[11pt,prd,superscriptaddress,nofootinbib]{revtex4}
\usepackage[english]{babel}
\usepackage{graphicx}
\usepackage{mathtext}
\usepackage{indentfirst}
\usepackage{epsfig,amsmath,amsfonts}
\newcommand{\beq}[1]{\begin{eqnarray}\label{#1}}
\newcommand\eeq {\end{eqnarray}}
\newcommand\bqa {\begin{eqnarray}}
\newcommand\eqa {\end{eqnarray}}

\newcommand{\bear}{\begin{array}}
\newcommand{\enar}{\end{array}}

\begin{document}

\hfill {\it To my mother}

\vspace{20mm}

\centerline{\bf \Large Lectures on General Theory of Relativity}

\vspace{10mm}

\centerline{Emil T. Akhmedov}

\vspace{10mm}

These lectures are based on the following books:

\begin{itemize}

\item 	{\it Textbook On Theoretical Physics. Vol. 2: Classical Field Theory},
by {\bf L.D. Landau} and {\bf E.M. Lifshitz}

\item {\it Relativist's Toolkit: The mathematics of black-hole mechanics}, by {\bf E.Poisson}, Cambridge University Press, 2004

\item {\it General Relativity}, by {\bf R. Wald}, The University of Chicago Press, 2010

\item {\it General Relativity}, by {\bf I.Khriplovich}, Springer, 2005

\item {\it An Introduction to General Relativity}, by {\bf L.P. Hughston} and {\bf K.P. Tod}, Cambridge University Press, 1994

\item {\it Black Holes (An Introduction)}, by {\bf D. Raine} and {\bf E.Thomas}, Imperial College Press, 2010

\end{itemize}

They were given to students of the Mathematical Faculty of the Higher School of Economics in Moscow.
At the end of each lecture I list some of those subjects which are not covered in the lectures.
If not otherwise stated, these subjects can be found in the above listed books.
I have assumed that students that have been attending these lectures were familiar with the classical electrodynamics and Special Theory of Relativity,
e.g. with the first nine chapters of the second volume of Landau and Lifshitz course. I would like to thank Mahdi Godazgar and Fedor Popov for useful comments, careful reading and corrections to the text. The work was done under the support of the RFBR grant 15-01-99504.

\newpage

\tableofcontents

\newpage

\section*{LECTURE I \\ {\it General covariance. Transition to non--inertial reference frames in Minkowski space--time. Geodesic equation.
Christoffel symbols.}}

\vspace{10mm}

{\bf 1.} Minkowski space-time metric is as follows:

\bqa
ds^2 = \eta_{\mu\nu} dx^\mu dx^\nu = dt^2 - d\vec{x}^2.
\eqa
Throughout these lectures we set the speed of light to one $c=1$, unless otherwise stated.
Here $\mu,\nu = 0, \dots, 3$ and Minkowskian metric tensor is

\bqa
||\eta_{\mu\nu}|| = Diag\left(1,-1,-1,-1\right).
\eqa
The bilinear form defining the metric tensor is invariant under the hyperbolic rotations:

\bqa
t' = t \cosh\alpha + x \sinh\alpha, \nonumber \\
x' = t \sinh\alpha + x \cosh\alpha, \nonumber \\ \alpha=const, \quad
y' = y, \quad z'=z,
\eqa
i.e. $\eta_{\mu\nu} \, dx^\mu \, dx^\nu = dt^2 - d\vec{x}^2 = \left(dt'\right)^2 - \left(d\vec{x}'\right)^2 = \eta_{\mu\nu} \, dx^{'\mu} \, dx^{'\nu}$.

This is the so called Lorentz boost, where $\cosh \alpha = \gamma = 1/\sqrt{1-v^2}$, $\sinh\alpha = v\,\gamma$. Its physical meaning is the transformation from an {\it inertial} reference system to another {\it inertial} reference system. The latter one moves along the $x$ axis with the constant velocity $v$ with respect to the initial reference system.

Under an arbitrary coordinate transformation (not necessarily linear),
$x^\mu = x^\mu\left(\bar{x}^\nu\right)$, the metric can change in an unrecognizable way, if it is transformed as the second rank tensor (see the next lecture):

\bqa
g_{\alpha\beta}\left(\bar{x}\right) = \eta_{\mu\nu} \, \frac{\partial x^\mu}{\partial \bar{x}^\alpha} \, \frac{\partial x^\nu}{\partial \bar{x}^\beta}.
\eqa
But it is important to note that, as the consequence of this transformation of the metric, the interval does not change under such a coordinate transformation:

\bqa
ds^2 = \eta_{\mu\nu} dx^\mu dx^\nu = g_{\alpha\beta} \left(\bar{x}\right) d\bar{x}^\alpha d\bar{x}^\beta.
\eqa
In fact, it is natural to expect that if one has a space--time, then the distance between any its
two--points does not depend on the way one draws the coordinate lattice on it. (The lattice is obtained by drawing three--dimensional hypersurfaces of constant coordinates $\bar{x}^\mu$ for each $\mu=0,\dots, 3$ with fixed lattice spacing in every direction.) Also it is natural to expect that the laws of physics should not depend on the choice of the coordinates in the space--time. This axiom is referred to as \underline{general covariance} and is the basis of the General Theory of Relativity.

{\bf 2.} Lorentz transformations in Minkowski space--time have the meaning of transitions between inertial reference systems. Then what is a meaning of an arbitrary coordinate transformation? To answer this question let us start with the transition into a non--inertial reference system in Minkowski space--time.

The simplest non--inertial motion is the one with the constant linear acceleration. Three--acceleration cannot be constant in a relativistic situation. Hence, we have to consider a motion of a particle with a constant four--acceleration, $w^\mu \, w_\mu = - a^2 = const$, where $w^\mu = d^2z^\mu(s)/ds^2$ and $z^\mu(s) = \left[z^0(s), \vec{z}(s)\right]$ is the world--line of the particle parametrized by the proper time\footnote{Note that four--velocity, $u_\mu = dz_\mu(s)/ds$, obeys the relation $u_\mu \, u^\mu = 1$, i.e. it is time--like vector. Differentiating both sides of this equality we obtain that $w^\mu \, u_\mu = 0$. Hence, $w_\mu$ should be space--like. As the result $w^\mu \, w_\mu = - a^2 < 0$.} $s$. Let us choose the spatial reference system such that the acceleration will be directed along the first axis. Then we have that:

\bqa
\left(\frac{d^2z^0}{ds^2}\right)^2 - \left(\frac{d^2z^1}{ds^2}\right)^2 = - a^2.
\eqa
Thus, the components of the four--acceleration compose a hyperbola. Hence, the standard solution of this equation is as follows:

\bqa\label{accel}
z^0(s) = \frac{1}{a}\, \sinh(a\,s), \quad z^1(s) = \frac{1}{a}\left[\cosh(a\, s) - 1\right].
\eqa
The integration constant in $z^1(s)$ is chosen for the future convenience.

\begin{figure}
\begin{center}
\includegraphics[scale=0.5]{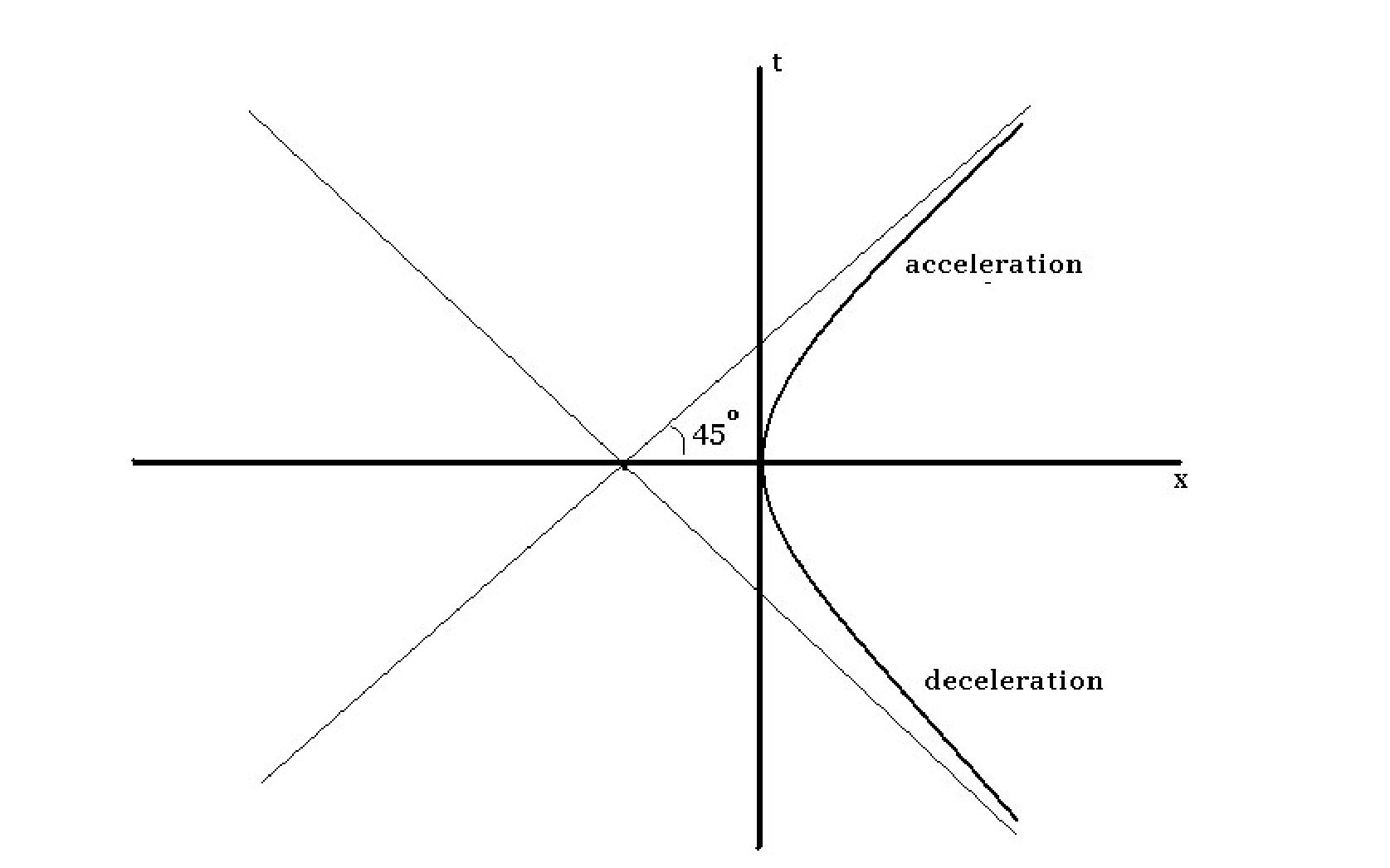}\caption{In this picture and also in the other pictures of this lecture we show only slices of fixed $y$ and $z$.}\label{fig1}
\end{center}
\end{figure}

Thus, one has the following relation between $z^1$ and $z^0$ themselves:

\bqa\label{hyperb}
\left(z^1 + \frac{1}{a}\right)^2 - \left(z^0\right)^2 = \frac{1}{a^2}.
\eqa
I.e. the world--line of a particle which moves with constant eternal acceleration is just a hyperbola (see fig. (\ref{fig1})). Note that the three--dimensional part of the acceleration is always along the positive direction of the $x$ axis: $d^2z^1/(dz^0)^2 = \frac{a}{\cosh(a \, s)} \, \left[1 - \tanh^2(a\,s)\right] > 0$. Hence, for the negative $s$ the particle is actually decelerating, while for the positive $s$ it accelerates. (Note that $s=0$ corresponds to $t=0$, as is shown on the fig. (\ref{fig1}).) The asymptotes of the hyperbola are the light--like lines, $z^1 = \pm z^0 - 1/a$. Hence, even if one moves with eternal constant acceleration, he cannot exceed the speed of light, because the motion with the speed of light would be along one of the above asymptotes of the hyperbola.

Moreover, for small $a\,z^0$ we find from (\ref{hyperb}) that: $z^1 \approx a \left(z^0\right)^2/2.$ In fact, for small proper times, $a \, s \ll 1$, we have that $z^0 \approx s$, $v \approx dz^1/ds \approx a \, z^0 \ll 1$ and obtain the standard nonrelativistic acceleration, which, however, gets modified according to (\ref{hyperb}) once the particle reaches high enough velocities. It is important to stress at this point that eternal constant acceleration is physically impossible due to the infinite energy consumption. I.e. here we are just discussing some mathematical abstraction, which, however, is helpful to clarify some important issues.

These observations will allow us to find the appropriate coordinate system for accelerated observers.
The motion with a constant eternal acceleration is homogeneous, i.e. accelerated observer cannot distinguish any moment of its proper time from any other. Hence, it is natural to expect that there should be static (invariant under both time--translations and time--reversal transformations) reference frame seen by accelerated observers. Inspired by (\ref{accel}), we propose the following coordinate
change:

\bqa\label{coordchange}
t = \rho \, \sinh \tau, \quad x = \rho \, \cosh \tau, \quad \rho \geq 0, \nonumber \\
y' = y, \quad {\rm and} \quad z' = z.
\eqa
Please note that these coordinates cover only quarter of the entire Minkowski space--time. Namely --- the right quadrant. In fact, since $\cosh\tau \geq |\sinh\tau|$, we have that $x \geq |t|$. It is not hard to guess the coordinates which will cover the left quadrant. For that one has to choose $\rho \leq 0$ in (\ref{coordchange}).

\begin{figure}
\centering
\begin{minipage}{0.35\linewidth}
\includegraphics[scale=1]{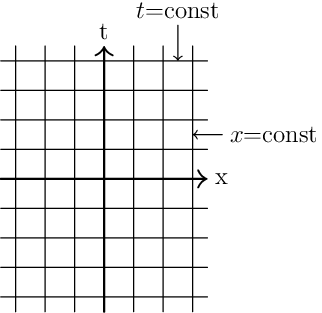}
\end{minipage}
% \hfill
\begin{minipage}{0.35\linewidth}
\includegraphics[scale=1]{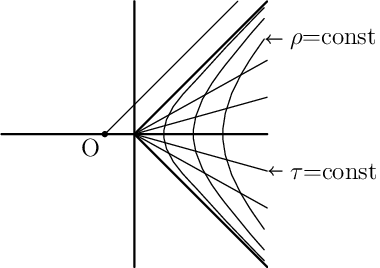}
\end{minipage}
\caption{}\label{fig2}
\end{figure}

Under such a coordinate change we have:

\bqa
dt = d\rho \, \sinh \tau + \rho \, d\tau \, \cosh \tau, \quad dx = d\rho \, \cosh \tau + \rho \, d\tau \, \sinh\tau.
\eqa
Then $dt^2 - dx^2 = \rho^2 \, d\tau^2 - d\rho^2$ and:

\bqa\label{Rindler}
ds^2 = dt^2 - dx^2 - dy^2 - dz^2 = \rho^2 \, d\tau^2 - d\rho^2 - dy^2 - dz^2
\eqa
is the so called \underline{Rindler metric}. It is not constant, $||g_{\mu\nu}|| = Diag(\rho^2, -1,-1,-1)$, but is time--independent and diagonal (i.e. static), as we have expected.

In this metric the levels of the constant coordinate time $\tau$ are straight lines
$t/x = \tanh \tau$ in the $x-t$ plane (or three--dimensional flat planes in the entire Minkowski space). The levels of the constant $\rho$ are the hyperbolas $x^2 - t^2 = \rho^2$ in the $x-t$ plane. The latter ones correspond to world--lines of observers which are moving with constant four--accelerations equal to $1/\rho$ on a slice of fixed $y$ and $z$.  The hyperbolas degenerate to light-like lines $x = \pm t$ as $\rho \to 0$. These are asymptotes of the hyperbolas for all $\rho$. As one takes $\rho$ closer and closer to zero the corresponding hyperbolas are closer and closer to their asymptotes. Note also that $\tau = -\infty$ corresponds to $x = -t$ and $\tau = + \infty$ --- to $x=t$. As a result we get a change of the coordinate lattice, which is depicted on the fig. (\ref{fig2}).

{\bf 3.} The important feature of the Rindler's metric (\ref{Rindler}) is that it degenerates at $\rho = 0$. This is the so called \underline{coordinate singularity}. It is similar to the singularity of the polar coordinates $dr^2 + r^2 d\varphi^2$ at $r = 0$. The space--time itself is regular at $\rho=0$. It is just flat Minkowski space--time  at the light--like lines $x = \pm t$. Another important feature of the Rindler's metric is that the speed of light is coordinate dependent:

\bqa
{\rm If} \quad ds^2 = 0, \quad {\rm then} \quad \left|\frac{d\rho}{d\tau}\right| = \rho, \quad {\rm when} \quad dy = dz = 0.
\eqa
At the same time, in the proper coordinates the speed of light is just equal to one $d\rho/ds = d\rho/\rho d\tau = 1$. Furthermore, as $\rho \to 0$ the speed of light, $d\rho/d\tau$, becomes zero. This phenomenon is related to the fact that if an observer starts an eternal acceleration with $a = 1/\rho$, say at the moment of time $t = 0 = \tau$, then there is a region in Minkowski space--time from which light rays cannot reach him. In fact, as shown on the fig. (\ref{fig2}) if a light ray was emitted from a point like O it is parallel to the asymptote $x=t$ of the world--line of the observer in question. As the result, the light ray never intersects with hyperbolas, i.e. never catches up with eternally accelerating observer. These are the reasons why one cannot extend the Rindler metric beyond the light--like lines $x = \pm t$. The three--dimensional surface $x=t$ of the entire Minkowski space--time is referred to as the \underline{future event horizon} of the Rindler's observers (those which are staying at the constant $\rho$ positions throughout their entire life time). At the same time $x=-t$ is the \underline{past event horizon} of the Rindler's observers.

Note that if an observer accelerates during a finite period of time, then, after the moment when the acceleration is switched off his world--line will be a straight line, which is tangential to the corresponding hyperbola. (I.e. the observer will continue moving with the gained velocity.) The angle this tangential line
has with the Minkowskian time axis is less that $45^o$. Hence, sooner or later the light ray emitted from a point like $O$ will actually reach such an observer. I.e. this observer does not have an event horizon.

\begin{figure}
\begin{center}
\includegraphics[scale=0.5]{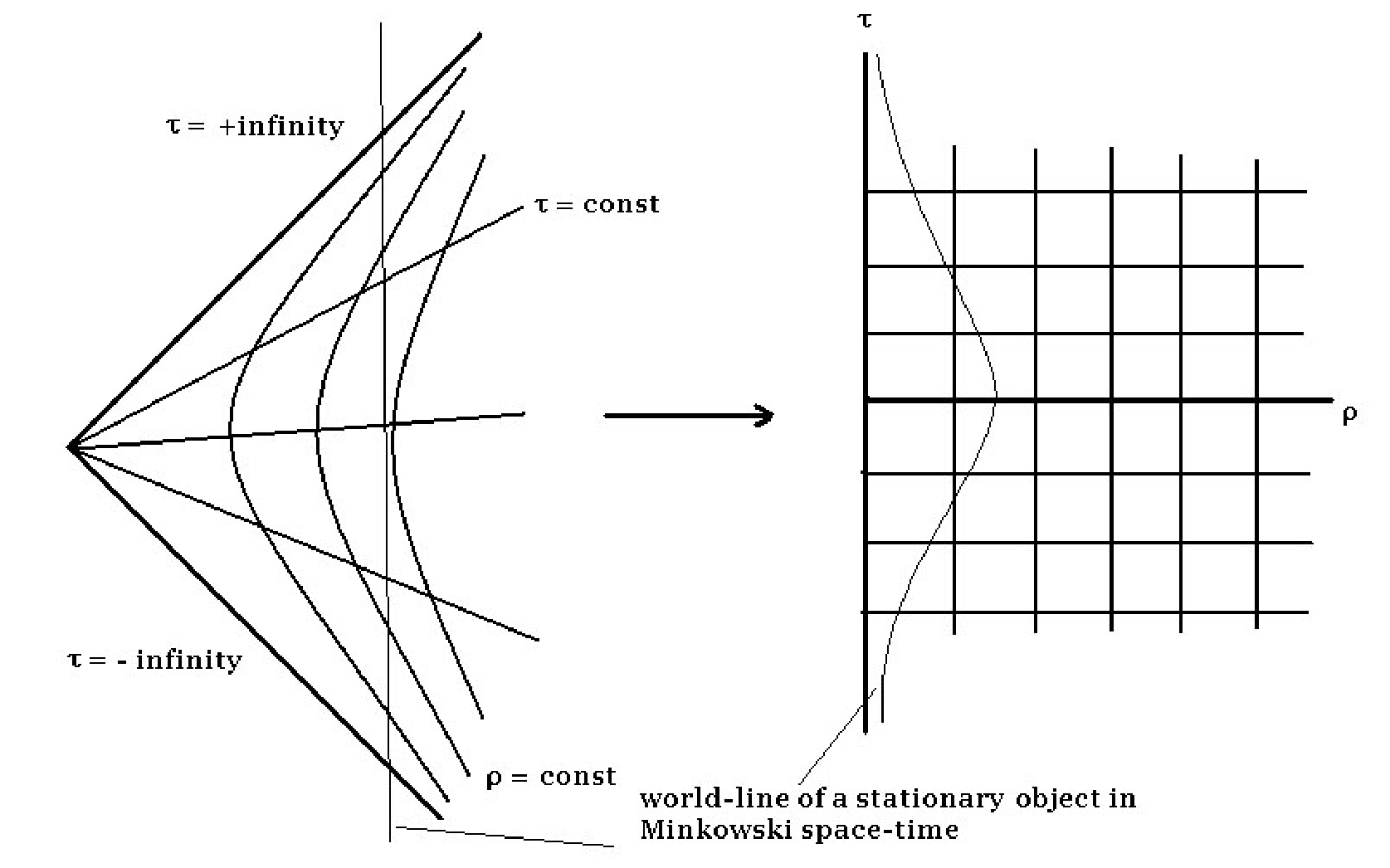}\caption{}\label{fig3}
\end{center}
\end{figure}

Another interesting phenomenon which is seen by the Rindler's observers is shown on the fig. (\ref{fig3}). A stationary object, $x=const$, in Minkowski space--time cannot cross the event horizon of the Rindler's observers during any finite period of the coordinate time $\tau$, which according to (\ref{accel}) is linearly related to the proper times of the eternally accelerating observers. This object just slows down and only asymptotically approaches the horizon. Note that, as $\rho\to 0$ a fixed finite portion of the proper time, $ds = \rho d\tau$, corresponds to increasing portions of the coordinate time, $d\tau$. Recall also that $\tau = -\infty$ corresponds to $x = -t$ and $\tau = + \infty$ --- to $x=t$.

All these peculiarities of the Rindler metric is the price one has to pay for the consideration of the physically impossible eternal acceleration. However, if one were transferring to a reference system of observers which are moving with accelerations only during finite proper times, then he would obtain a non--stationary metric due to the inhomogeneity of such a motion.

The main lesson to draw from these observations is as follows. The physical meaning of a general coordinate transformation that mixes spatial and time coordinates is a transition to another, not necessary inertial, reference system. In this case curves corresponding to fixed spatial coordinates (e.g. $d\rho = dy = dz = 0$) are world--lines of (non--)inertial observers. As the result, the essence of the general covariance is that physical laws, in a suitable form/formualtion, should not depend on the choice of observers.

{\bf 4.} If even in flat space--time one can choose curvilinear coordinates and obtain a non--trivial metric tensor $g_{\mu\nu}(x)$, then how can one distinguish flat space--time from the curved one? Furthermore, since we understood the physics behind the curvilinear coordinates in flat space--time, then it is also natural to ask: What is the physics behind curved space--times? To start answering these questions in the following lectures let us solve here a simple problem.

Namely, let us consider a free particle moving in a space--time with a metric $ds^2 = g_{\mu\nu}(x) \, dx^\mu \, dx^\nu$. Let us find its world--line via the minimal action principle. If one considers a world--line $z^\mu(\tau)$ parametrized by a parameter $\tau$ (that, e.g., could be either a coordinate time or the proper one), then the simplest invariant characteristic that one can associate to the world--line is its length. Hence, the natural action for the free particle should be proportional to the length of its world--line. The reason why we are looking for an invariant action is that we expect the corresponding equations of motion to be covariant (i.e. to have the same form in all coordinate systems), according to the above formulated principle of general covariance.

If one approximates the world--line by a broken line consisting of a chain of small intervals, then its length can be approximated by the expression as follows:

\bqa
L = \sum_{i=1}^N \sqrt{g_{\mu\nu}\left[z_i\right] \, \left[z_{i+1} - z_i\right]^\mu \, \left[z_{i+1} - z_i\right]^\nu},
\eqa
which follows from the definition of the metric. In the limit $N\to \infty$ and $\left|z_{i+1} - z_i\right| \to 0$
we obtain an integral instead of the sum. As a result, the action should be as follows:

\bqa\label{action}
S = - m\, \int_1^2 ds = - m \, \int_{\tau_1}^{\tau_2} d\tau \sqrt{g_{\mu\nu}\left[z(\tau)\right] \, \dot{z}^\mu\, \dot{z}^\nu}.
\eqa
Here $\dot{z} = dz/d\tau$. The coefficient of the proportionality between the action, $S$, and the length, $L$, is minus the mass, $m$, of the particle. This coefficient follows from the complementarity --- from the fact that when $g_{\mu\nu}(x) = \eta_{\mu\nu}$ we have to obtain the standard action for the relativistic particle in the Special Theory of Relativity.

Note that the action (\ref{action}) is invariant under the coordinate transformations, $z^\mu \to \bar{z}^\mu(z)$, and also under the reparametrizations, $\tau \to f(\tau)$, if one respects the ordering of points along the world--line, $df/d\tau \geq 0$. In fact, then:

$$
d\tau \sqrt{g_{\mu\nu} \frac{dz^\mu}{d\tau} \, \frac{dz^\nu}{d\tau}} = df \sqrt{g_{\mu\nu} \frac{dz^\mu}{df} \, \frac{dz^\nu}{df}}.
$$
Let us find equations of motion that follow from the minimal action principle applied to (\ref{action}). The first variation of the action is:

\bqa\label{ac1}
\delta S = - m \int_{\tau_1}^{\tau_2} d\tau \, \frac{\delta\left[g_{\mu\nu}(z) \, \dot{z}^\mu \, \dot{z}^\nu\right]}{2\, \sqrt{\dot{z}^2}} = \nonumber \\ = - m \int_{\tau_1}^{\tau_2} \frac{d\tau \, \sqrt{\dot{z}^2}}{2\, \sqrt{\dot{z}^2}\, \sqrt{\dot{z}^2}} \,\left[\delta g_{\mu\nu}(z)^{\phantom{\frac12}} \dot{z}^\mu \, \dot{z}^\nu + g_{\mu\nu}(z) \, \delta \dot{z}^\mu \, \dot{z}^\nu + g_{\mu\nu}(z)\, \dot{z}^\mu \, \delta\dot{z}^\nu\right].
\eqa
Here we denote $\dot{z}^2 \equiv g_{\alpha\beta}(z) \, \dot{z}^\alpha \dot{z}^\beta$. Using the fact that $\sqrt{\dot{z}^2} \, d\tau = \sqrt{g_{\mu\nu} \, dz^\mu \, dz^\nu} = ds$ we can change in (\ref{ac1}) the parametrization from $\tau$ to the proper time $s$. After that we integrate by parts in the last two terms in the last line of (\ref{ac1}). This way we get rid from the differential operator acting on $\delta z$: $\delta\dot{z} = \frac{d}{ds}\delta z$. Then, using the Dirichlet boundary conditions, i.e. assuming that $\delta z(s_1) = \delta z(s_2) = 0$, we arrive at the following expression for the first variation of $S$:

\bqa
\delta S = - m \int_{s_1}^{s_2} \frac{ds}{2} \, \left\{\partial_\alpha g_{\mu\nu}(z) \,\delta z^\alpha \,  \dot{z}^\mu \, \dot{z}^\nu - \frac{d}{ds}\left[g_{\mu\nu}(z)^{\phantom{\frac12}} \dot{z}^\nu \right]\, \delta z^\mu  - \frac{d}{ds} \left[g_{\mu\nu}(z)^{\phantom{\frac12}} \dot{z}^\mu\right] \, \delta z^\nu\right\} = \nonumber \\ =  - m \int_{s_1}^{s_2} \frac{ds}{2} \, \left\{\partial_\alpha g_{\mu\nu}^{\phantom{\frac12}}\delta z^\alpha \,  \dot{z}^\mu \, \dot{z}^\nu - \partial_\alpha g_{\mu\nu} \, \dot{z}^\alpha \, \dot{z}^\nu \, \delta z^\mu  -  \partial_\alpha g_{\mu\nu}\, \dot{z}^\alpha \, \dot{z}^\mu \, \delta z^\nu - 2 \, g_{\mu\nu} \ddot{z}^\mu \, \delta z^\nu\right\} = \nonumber \\ = - m \int_{s_1}^{s_2} ds \left[\frac12 \, \left(\partial_\alpha \, g_{\mu\nu}^{\phantom{\frac12}} - \partial_\mu g_{\alpha\nu} - \partial_\nu g_{\mu\alpha}\right) \, \dot{z}^\mu \, \dot{z}^\nu - g_{\mu\alpha} \, \ddot{z}^\mu\right] \, \delta z^\alpha.
\eqa
In these expressions $\dot{z} = dz/ds$ and also we have used that $g_{\mu\nu} \, \ddot{z}^\mu \, \delta z^\nu = g_{\mu\nu} \, \delta z^\mu \, \ddot{z}^\nu$ because $g_{\mu\nu} = g_{\nu\mu}$. Taking into account that according to the minimal action principle $\delta S$ should be equal to zero for any $\delta z^\alpha$, we arrive at the following relation:

\bqa
\ddot{z}^\mu + \Gamma^\mu_{\nu\alpha}(z) \, \dot{z}^\nu \, \dot{z}^\alpha = 0,
\eqa
which is referred to as the \underline{geodesic equation}. Here

\bqa
\Gamma^\mu_{\nu\alpha} = \frac12 \, g^{\mu\beta} \left(\partial_\nu \, g_{\alpha\beta}^{\phantom{\frac12}} + \partial_\alpha \, g_{\beta\nu} - \partial_\beta g_{\nu\alpha}\right)
\eqa
are the so called \underline{Christoffel symbols} and $g^{\mu\beta}$ is the inverse metric tensor, $g^{\mu\beta} \, g_{\beta \nu} = \delta^\mu_\nu$.

\vspace{10mm}

\centerline{\bf Problems:}

\vspace{5mm}

\begin{itemize}

\item Show that the metric $ds^2 = \left(1 + a \, h\right)^2 \, d\tau^2 - dh^2 - dy^2 - dz^2$ (homogeneous gravitational field) also covers the Rindler space--time. Find the coordinate change from this metric to the one used in the lecture.

\item Find the coordinates which cover the lower and upper quadrants (complementary to those which are covered by Rindler's coordinates) of the Minkowski space--time.

\item Find the coordinate transformation and the stationary (invariant only under time--translations, but not under time--reversal transformation) metric in the rotating reference system with the angular velocity $\omega$. (See the corresponding paragraph in Landau--Lifshitz.)

\item ({\bf *}) Find the coordinate transformation and the stationary metric in the orbiting reference system, which moves on the radius $R$ with the angular velocity $\omega$.

\item ({\bf *}) Consider a particle which was stationary in an inertial reference system. Then its acceleration was adiabatically turned on and kept finite for long period of time. And finally its acceleration was adiabatically switched off. I.e. this particle for the beginning is stationary then accelerates for a while,
    and finally proceeds its motion with a constant gained velocity. Find the world--line for such a motion. Find a metric which is seen by such observers.

\item ({\bf *}) Find the equation for geodesics in the non--Riemanian metric:

$$ds^n = g_{\mu_1 \dots \mu_n}(x) \, dx^{\mu_1} \dots dx^{\mu_n}.$$
({\bf **}) What kind of geometries (instead of the Minkowskian one) are there, if
$g_{\mu_1\dots \mu_n}$ has only constant (coordinate independent) components? (We know that in the case
of constant metrics with two indexes we can reduce them by coordinate transformations to one of the
standard forms --- $(1,1,1,1)$, $(-1, 1, 1, 1)$, $(-1,-1, 1, 1)$ and etc.. What are the standard types of constant metrics with more indexes? Furthermore, in the case of Minkowsian signature there is a light--cone, which allows one to specify which events are causally connected. What does one have instead of that in the case of constant metrics with more indexes?)

\item ({\bf **}) What kind of geometry (instead of the Minkowskian one) is there, if $g_{\mu\nu} = Diag(1,1,-1,-1)$ instead of Minkowskian metric? What is there instead of the light--cone and causality?

\end{itemize}

\vspace{10mm}

\centerline{\bf Subjects for further study:}

\vspace{5mm}

\begin{itemize}

\item Radiation of the homogeneously accelerating charges: What is the intensity seen by a distant inertial observer? What is the intensity seen by a distant co--moving non--inertial observer? What is the invariant energy loss of the homogeneously accelerating charge? Does a free falling charge in a homogeneous gravitational field create a radiation? Does a charge, which is fixed in a homogeneous gravitational field, create a radiation? (``Radiation from a Uniformly Accelerated Charges'', D.G.Bouleware, Annals of Physics 124 (1980) 169. \\ ``On radiation due to homogeneously accelerating sources'', D.Kalinov, e-Print: arXiv:1508.04281)

\item Action and minimal action principle for strings and membranes in arbitrary dimensions. (Gauge fields and strings, A.Polyakov, Harwood Academic Publishers, 1987.)

\item Unruh effect ( On the physical meaning of the Unruh effect,
Emil T. Akhmedov, Douglas Singleton,
Published in Pisma Zh.Eksp.Teor.Fiz. 86 (2007) 702-706, JETP Lett. 86 (2007) 615-619;
e-Print: arXiv:0705.2525;\\
On the relation between Unruh and Sokolov-Ternov effects
Emil T. Akhmedov, Douglas Singleton,
Published in Int.J.Mod.Phys. A22 (2007) 4797-4823;
e-Print: hep-ph/0610391.)

\end{itemize}

\newpage

\section*{LECTURE II \\{\it Tensors. Covariant differentiation. Parallel transport. Locally Minkowskian reference system. Curvature or Riemann tensor and its properties.}}

\vspace{10mm}

{\bf 1.} This lecture is rather formal. Here we answer some of the questions posed
in the first lecture and also clarify the geometric meaning of the Christoffel symbols.

For the beginning let us recall what is tensor. Under a transformation $x^\mu = x^\mu\left(\bar{x}^\nu\right)$ the space--time coordinates tautologically transform as:

\bqa
dx^\mu = \frac{\partial x^\mu}{\partial \bar{x}^\nu} \, d\bar{x}^\nu.
\eqa
A vector $A^\mu$ is referred to as \underline{contravariant} if it transforms, under the coordinate transformation, in the same way as coordinates do:

\bqa
A^\mu(x) = \frac{\partial x^\mu}{\partial \bar{x}^\nu} \, \bar{A}^\nu\left(\bar{x}\right).
\eqa
At the same time a vector $A_\mu$ is referred to as \underline{covariant} if it transforms as a one--form:

\bqa
A_\mu(x) \, dx^\mu = \bar{A}_\nu\left(\bar{x}\right) \, d \bar{x}^\nu, \quad {\rm then} \quad A_\mu(x) = \frac{\partial \bar{x}^\nu}{\partial x^\mu} \, \bar{A}_\nu\left(\bar{x}\right).
\eqa
With the use of the metric tensor $g_{\mu\nu}$ and its inverse, $g^{\mu\nu} \, g_{\nu\alpha} = \delta^\mu_\alpha$, one can map covariant indexes onto contravariant ones and back:

\bqa
A_\mu = g_{\mu\nu} \, A^\nu, \quad {\rm and} \quad A^\mu = g^{\mu\nu} \, A_\nu.
\eqa
In particular $x_\mu = g_{\mu\nu}\, x^\nu$.

Then, $n$--tensor with the corresponding number of covariant and contravariant indexes is the quantity, which changes under the coordinate transformations, as follows ($l+k = n$):

\bqa
T_{\mu_1 \dots \mu_k}^{\nu_1 \dots \nu_l}(x) = \frac{\partial \bar{x}^{\alpha_1}}{\partial x^{\mu_1}} \dots \frac{\partial \bar{x}^{\alpha_k}}{\partial x^{\mu_k}} \, \frac{\partial x^{\nu_1}}{\partial \bar{x}^{\beta_1}} \dots \frac{\partial x^{\nu_l}}{\partial \bar{x}^{\beta_l}} \bar{T}_{\alpha_1 \dots \alpha_k}^{\beta_1 \dots \beta_l} \left(\bar{x}\right).
\eqa
In principle the order of the upper and lower indexes is important, but to simplify this formula we ignore this detail here. For example, for the metric we have that

\bqa
ds^2 = g_{\mu\nu}(x) \, dx^\mu dx^\nu = \bar{g}_{\alpha\beta}\left(\bar{x}\right) \, d\bar{x}^\alpha d\bar{x}^\beta,
\quad {\rm then} \quad g_{\mu\nu}(x) = \frac{\partial \bar{x}^\alpha}{\partial x^\mu} \, \frac{\partial \bar{x}^\beta}{\partial x^\nu}\, \bar{g}_{\alpha \beta} \left(\bar{x}\right).
\eqa
With the use of the metric tensor and its inverse tensor one also can rise and lower indexes of higher rank tensors: e.g., ${T_{\mu\nu}}^\alpha \, g^{\nu\beta} = {T_\mu}^{\beta\alpha}$. In the last equation we show that the order of indexes is important.

All these definitions are necessary to make contractions of tensors to transform also as tensors. For example,
${{T^{\mu}}_\nu}^{\alpha\beta} \, {M_\beta}^\nu$ should transform as two--tensor and it does, if one uses the above definitions. In particular, the scalar product of two vectors $A_\mu\, B^\mu = A^\mu \, B_\mu = g_{\mu\nu} \, A^\mu \, B^\nu = g^{\mu\nu} \, A_\mu \, B_\nu$ should be (and is) invariant. That is all essence and convenience of the tensor notations, because then every expression has obvious properties under the general coordinate transformations.

{\bf 2.} Now we are ready to define the \underline{covariant differential}. The ordinary differential is defined as

\bqa
\partial_\nu A^\mu \, dx^\nu \equiv A^\mu\left(x + dx\right) - A^\mu(x).
\eqa
We will frequently use several different notations for the ordinary differential: $\frac{\partial A^\mu}{\partial x^\nu} \equiv \partial_\nu A^\mu \equiv A^\mu_{,\nu}.$
The problem with the ordinary differential, $\partial_\nu A^\mu$, is that, despite the fact that it has two indexes, it does not transform as two--tensor. In fact,

\bqa\label{Atrans}
\bar{A}^{\alpha}_{,\beta}\left(\bar{x}\right) = \frac{\partial}{\partial \bar{x}^\beta}\, \frac{\partial \bar{x}^\alpha}{\partial x^\mu} \, A^\mu(x) = \frac{\partial \bar{x}^\alpha}{\partial x^\mu} \, \frac{\partial x^\nu}{\partial \bar{x}^\beta} \, A^\mu_{,\nu}(x) + \frac{\partial^2 \bar{x}^\alpha}{\partial x^\mu \partial x^\nu} \, \frac{\partial x^\nu}{\partial \bar{x}^\beta} \, A^\mu(x).
\eqa
It is the covariant differential of a vector $A^\mu$ which transforms as two--tensor. To define it let us subtract a quantity $\delta A^\mu$ from the ordinary differential:

\bqa
D_\alpha A^\mu \, dx^\alpha = \partial_\alpha A^\mu \, dx^\alpha - \delta A^\mu.
\eqa
We will frequently use different notations for the covariant differential: $D_\alpha A^\mu \equiv A^\mu_{;\alpha}.$
The geometric interpretation of $\delta A^\mu$ is as follows.
The above problems with the ordinary differential are due to the fact that to find it we subtract two vectors $A^\mu\left(x+dx\right)$ and $A^\mu(x)$, which are defined at two different points --- $x+dx$ and $x$. To overcome these problems, one has to parallel transport $A^\mu(x)$ to the point $x+dx$. That is exactly what the addition of $\delta A^\mu$ does:

\bqa\label{covar}
D_\alpha A^\mu \, dx^\alpha \equiv A^\mu\left(x+dx\right) - \left[A^\mu(x) + \delta A^\mu(x)\right].
\eqa
For small $dx$ the quantity $\delta A^\mu$ should be linear in $dx$ and also in $A^\mu$. Hence, we define it to have the following form:

\bqa\label{Gamma}
\delta A^\mu(x) \equiv - \Gamma^\mu_{\nu\alpha}(x) \, A^\nu(x) \, dx^\alpha,
\eqa
where $\Gamma^\mu_{\nu\alpha}(x)$ is referred to as the \underline{connection}. Clearly $\Gamma^\mu_{\nu\alpha}(x) \, dx^\alpha = M^\mu_\nu$ is a matrix that transforms the vector $A_\mu$ during the parallel transport.

\begin{figure}
\begin{center}
\includegraphics[scale=0.5]{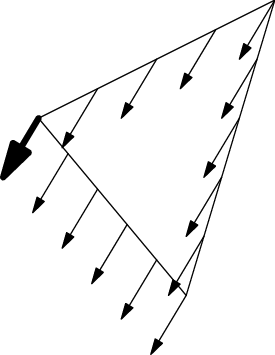}\caption{}\label{fig4a}
\end{center}
\end{figure}

\begin{figure}
\begin{center}
\includegraphics[scale=0.5]{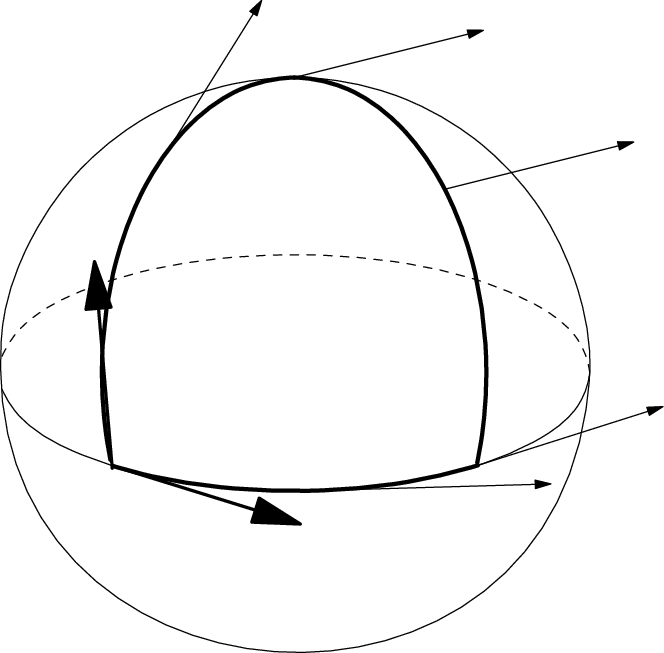}\caption{}\label{fig4b}
\end{center}
\end{figure}

\begin{figure}
\begin{center}
\includegraphics[scale=0.5]{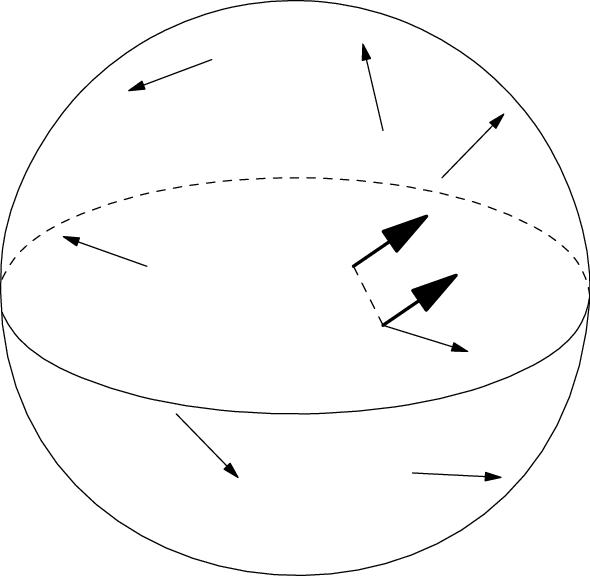}\caption{}\label{fig4c}
\end{center}
\end{figure}

To clarify what means connection let us illustrate it on the simplest textbook example. Consider flat two--dimensional space and a closed triangular path in it (see fig. (\ref{fig4a})). Let us parallel transport a vector along this path. The rule for the parallel transport is that the angle between the vector and the path is always the same along the path. (This just means that we have specified connection of matrix $M^\mu_\nu$ defined above.) Then, as can be seen from the fig. (\ref{fig4a}) in flat space the vector returns back to the same position after the parallel transport along the closed path. Let us see now how the picture is changed in the simplest curved space --- sphere (see fig. \ref{fig4b}). We choose segments of three different equators as parts of the closed triangular path on the sphere. One can see from the fig. \ref{fig4b} that the vector does not return to the same position after the parallel transport (pay attention to the bold face vectors/arrows). Finally, to clarify the meaning of the covariant differentiation consider a vector field on the sphere, as is shown on the fig. (\ref{fig4c}). To subtract from a value of the vector field at one position its value at a nearby position we parallel transport the vector from the last position to the first one, as is shown on the fig. (\ref{fig4c}) by bold face vectors. That is how we obtain the covariant differential.

As is seen from (\ref{Atrans}) and (\ref{covar}) for $D_\alpha A^\mu$ to be a two--tensor the connection $\Gamma^\mu_{\nu\alpha}$ should transform as:

\bqa\label{Gtrans}
\Gamma^\mu_{\nu\alpha}(x) = \bar{\Gamma}^\beta_{\gamma\sigma}\left(\bar{x}\right)\, \frac{\partial x^\mu}{\partial \bar{x}^\beta} \, \frac{\partial \bar{x}^\gamma}{\partial x^\nu} \, \frac{\partial \bar{x}^\sigma}{\partial x^\alpha} + \frac{\partial^2 \bar{x}^\gamma}{\partial x^\nu \partial x^\alpha} \, \frac{\partial x^\mu}{\partial \bar{x}^\gamma}.
\eqa
At the same time scalar product should not change under the parallel transport. Hence, from $\delta \left(A_\mu \, B^\mu\right) = 0$ we have that:

\bqa\label{BA}
B^\mu \, \delta A_\mu = - A_\mu \delta B^\mu = A_\mu \, \Gamma^\mu_{\nu\alpha} \, B^\nu \, dx^\alpha,
\eqa
where to obtain the last equality we used eq. (\ref{Gamma}) for $\delta B^\mu$. Because (\ref{BA}) should be valid for any $B^\mu$, we have that:

\bqa
\delta A_\mu = \Gamma^\nu_{\mu\alpha}(x) \, A_\nu(x) \, dx^\alpha,
\eqa
in addition to (\ref{Gamma}).

As the result we have the following definition of the covariant derivative:

\bqa
A^\mu_{;\alpha} \equiv D_\alpha A^\mu = \partial_\alpha A^\mu + \Gamma^\mu_{\nu\alpha} \, A^\nu = \left(\partial_\alpha \, \delta^\mu_\nu + \Gamma^\mu_{\nu\alpha}\right) \, A^\nu, \nonumber \\
A_{\mu;\, \alpha} \equiv D_\alpha A_\mu = \partial_\alpha A_\mu - \Gamma^\nu_{\mu\alpha} \, A_\nu = \left(\partial_\alpha \, \delta^\nu_\mu - \Gamma^\nu_{\mu\alpha}\right)\, A_\nu.
\eqa
Similarly, the covariant differential of higher rank tensors is as follows:

\bqa\label{AAA}
A^{\mu\nu}_{;\alpha} = D_{\alpha} A^{\mu\nu} & = & \partial_\alpha A^{\mu\nu} + \Gamma^\mu_{\beta\alpha} \, A^{\beta\nu} + \Gamma^\nu_{\beta\alpha} \, A^{\mu\beta}, \nonumber \\
A^\mu_{\nu; \, \alpha} = D_\alpha A^\mu_\nu & = & \partial_\alpha A^\mu_\nu + \Gamma^\mu_{\beta\alpha} \, A^\beta_\nu - \Gamma^\beta_{\nu\alpha} \, A^\mu_\beta, \nonumber \\
A_{\mu\nu; \, \alpha} = D_\alpha A_{\mu\nu} & = & \partial_\alpha A_{\mu\nu} - \Gamma^\beta_{\mu\alpha} \, A_{\beta\nu} - \Gamma^\beta_{\nu\alpha} \, A_{\mu\beta}, \quad {\rm etc.}.
\eqa
Along with $\Gamma^\mu_{\nu\alpha}$ we will use:

\bqa\label{GG}
\Gamma_{\beta | \, \nu\alpha} = g_{\beta\mu} \, \Gamma^\mu_{\nu\alpha}.
\eqa
It is instructive to have in mind that for Minkowski metric, $\eta_{\mu\nu}$, one has that $\Gamma^\mu_{\nu\alpha} = 0$.

{\bf 3.} For the future convenience here we define the \underline{Locally Minkowskian Reference System} (LMRS).
It is such a reference frame in a vicinity of an arbitrary point $x_0$ in which

\bqa\label{LMRS}
g_{\mu\nu}(x_0) = \eta_{\mu\nu}, \quad {\rm and} \quad \Gamma^\alpha_{\beta\gamma} (x_0) = 0,
\eqa
but it does not mean that the derivatives of $g_{\mu\nu}$ and $\Gamma^\mu_{\nu\alpha}$ are vanishing. Below we will see the condition when it is impossible to put the derivatives to zero.

Let us discuss under what conditions one can fix such a gauge as (\ref{LMRS}). We put $x_0$ to the origin of two reference systems --- of an original one, $K$, and of a new, $\bar{K}$, reference system. Then, if $\xi = x - x_0$ and $\bar{\xi} = \bar{x} - x_0$, we can expand:

\bqa
\bar{\xi}^\alpha = A^\alpha_\beta \, \xi^\beta + \frac12\, B^\alpha_{\beta\gamma} \, \xi^\beta \, \xi^\gamma + {\cal O}\left(\xi^3\right).
\eqa
where $A^\mu_\nu$ and $B^\mu_{\nu\alpha}$ are some constant tensor parameters. Note that $B^\alpha_{\beta\gamma} = B^\alpha_{\gamma\beta}$.

Under such a transformation we have that:

\bqa
\bar{g}_{\alpha\beta}(x_0) & = & A^\mu_\alpha \, A^\nu_\beta \, g_{\mu\nu} (x_0) + {\cal O}(\xi), \nonumber \\
\bar{\Gamma}^\alpha_{\beta\gamma}(x_0) & = & A^\alpha_\mu \, A^\nu_\beta \, A^\sigma_\gamma \, \Gamma^\mu_{\nu\sigma}(x_0) - B^\alpha_{\mu\nu} \, A^\mu_\beta \, A^\nu_\gamma + {\cal O}(\xi).
\eqa
Using 16 components of $A^\mu_\alpha$ we can always solve 10 equations $\bar{g}_{\mu\nu}(x_0) = \eta_{\mu\nu}$. The remaining 6 parameters of $A^\mu_\nu$ correspond to the 3 rotations and 3 Lorentz boosts, under which the Minkowskian metric tensor, $\eta_{\mu\nu}$, does not change. Furthermore, one can put $\bar{\Gamma}^\mu_{\nu\sigma}(x_0) = 0$ by choosing $B^\alpha_{\mu\nu} = A^\alpha_\gamma \, \Gamma^\gamma_{\mu\nu}(x_0)$.

The physical meaning of the reference system under consideration is very simple. Any space--time in a sufficiently small vicinity of any its point looks as almost flat. Of course in this almost flat vicinity of any point one can fix Minkowskian coordinates. As it follows from the above calculations, in a vicinity of a point $x_0$:

\bqa
g_{\mu\nu}(x) = \eta_{\mu\nu} + {\cal O}\left(\left|x-x_0\right|^2\right), \quad {\rm and} \quad \Gamma^\alpha_{\beta\gamma} (x) = 0 + {\cal O}(x-x_0),
\eqa
A choice of a reference system/frame we will frequently call as a choice of a gauge.

{\bf 4.} Let us define now the so called \underline{torsion}:

\bqa
S^\mu_{\nu\alpha} \equiv \Gamma^\mu_{\nu\alpha} - \Gamma^\mu_{\alpha\nu}.
\eqa
According to (\ref{Gtrans}) it transforms under the coordinate transformations as a three--tensor. If one can
choose LMRS at any point $x$, he obtains that $S^\mu_{\nu\alpha} = 0$, because $\Gamma^\mu_{\nu\alpha} = 0$. But, $S^\mu_{\nu\alpha}$ transforms as a tensor, i.e. multiplicatively. Hence, if its components are smooth functions, then this tensor is also zero in any other reference system. Due to the arbitrariness of the point $x$, we conclude that if the metric is smooth enough and the gauge LMRS is possible, then

\bqa\label{symm}
\Gamma^\mu_{\nu\alpha} = \Gamma^\mu_{\alpha\nu},
\eqa
i.e. the connection is symmetric under the exchange of its lower indexes. Manifolds with vanishing torsion are referred to as \underline{Riemanian}.

{\bf 5.} Here we express the connection via the metric tensor. For the beginning we show that any metric tensor should be covariantly constant. In fact,

\bqa\label{covar1}
D_\alpha A_\mu = D_\alpha \left(g_{\mu\nu}\, A^\nu\right) = \left(D_\alpha g_{\mu\nu}\right)\, A^\nu + g_{\mu\nu}\, D_\alpha A^\nu.
\eqa
But $D_\alpha A_\mu$ is two--tensor. Hence, by the definition of the relation between covariant and contravariant indexes, we should have that $D_\alpha A_\mu = g_{\mu\nu}\, D_\alpha A^\nu$. Hence, from (\ref{covar1}) it follows that the metric tensor should be covariantly constant: $D_\alpha g_{\mu\nu} = 0$. Using (\ref{AAA}) and (\ref{GG}), we can write this condition as:

\bqa
g_{\mu\nu; \, \alpha} \equiv D_\alpha g_{\mu\nu} = \partial_\alpha g_{\mu\nu} - \Gamma_{\mu | \, \nu\alpha} - \Gamma_{\nu | \, \mu\alpha} = 0.
\eqa
Reshuffling the indexes in this equation we also find that:

\bqa
\partial_\nu g_{\alpha\mu} - \Gamma_{\alpha | \, \mu\nu} - \Gamma_{\mu | \, \alpha \nu} = 0, \nonumber \\
\partial_\mu g_{\nu\alpha} - \Gamma_{\alpha | \, \nu\mu} - \Gamma_{\nu | \, \alpha \mu} = 0.
\eqa
Then, using the obtained system of three linear algebraic equations on $\Gamma^\mu_{\nu\alpha}$ and the identity (\ref{symm}), we find the relation between the connection and the metric tensor:

\bqa
\Gamma_{\alpha |\, \mu\nu} = \frac12 \, \left(\partial_\nu^{\phantom{\frac12}} g_{\alpha\mu} + \partial_\mu \, g_{\nu\alpha} - \partial_\alpha \, g_{\mu\nu}\right),
\eqa
or

\bqa
\Gamma^\alpha_{\mu\nu} = \frac12 \, g^{\alpha\beta} \, \left(\partial_\nu^{\phantom{\frac12}} g_{\beta\mu} + \partial_\mu \, g_{\nu\beta} - \partial_\beta \, g_{\mu\nu}\right).
\eqa
Thus, for Riemanian manifolds the connection $\Gamma^\mu_{\nu\alpha}$ coincides with the Christoffel symbols defined in the previous lecture.

As the result, the geodesic equation found in the previous lecture acquires a clear geometric meaning,

\bqa
\left[\frac{d}{ds}\delta^\mu_\alpha + \Gamma^\mu_{\nu\alpha}\left[z(s)\right] \, u^\nu(s) \right] u^\alpha(s) = \left[\dot{z}^\nu \frac{\partial}{\partial z^\nu}\delta^\mu_\alpha + \Gamma^\mu_{\nu\alpha}\left[z(s)\right] \, u^\nu(s) \right] u^\alpha(s) = u^\nu(s) \, D_\nu u^\mu(s) = 0,
\eqa
as the condition of the covariant constancy of the four velocity $u^\mu = dz^\mu/ds$ along the geodesic path. In fact, $u^\alpha \, D_\alpha u^\mu$ is just the projection of the covariant derivative $D_\alpha u^\mu$ on to the tangent vector $u^\alpha$ to the geodesic.

\begin{figure}
\begin{center}
\includegraphics[scale=0.5]{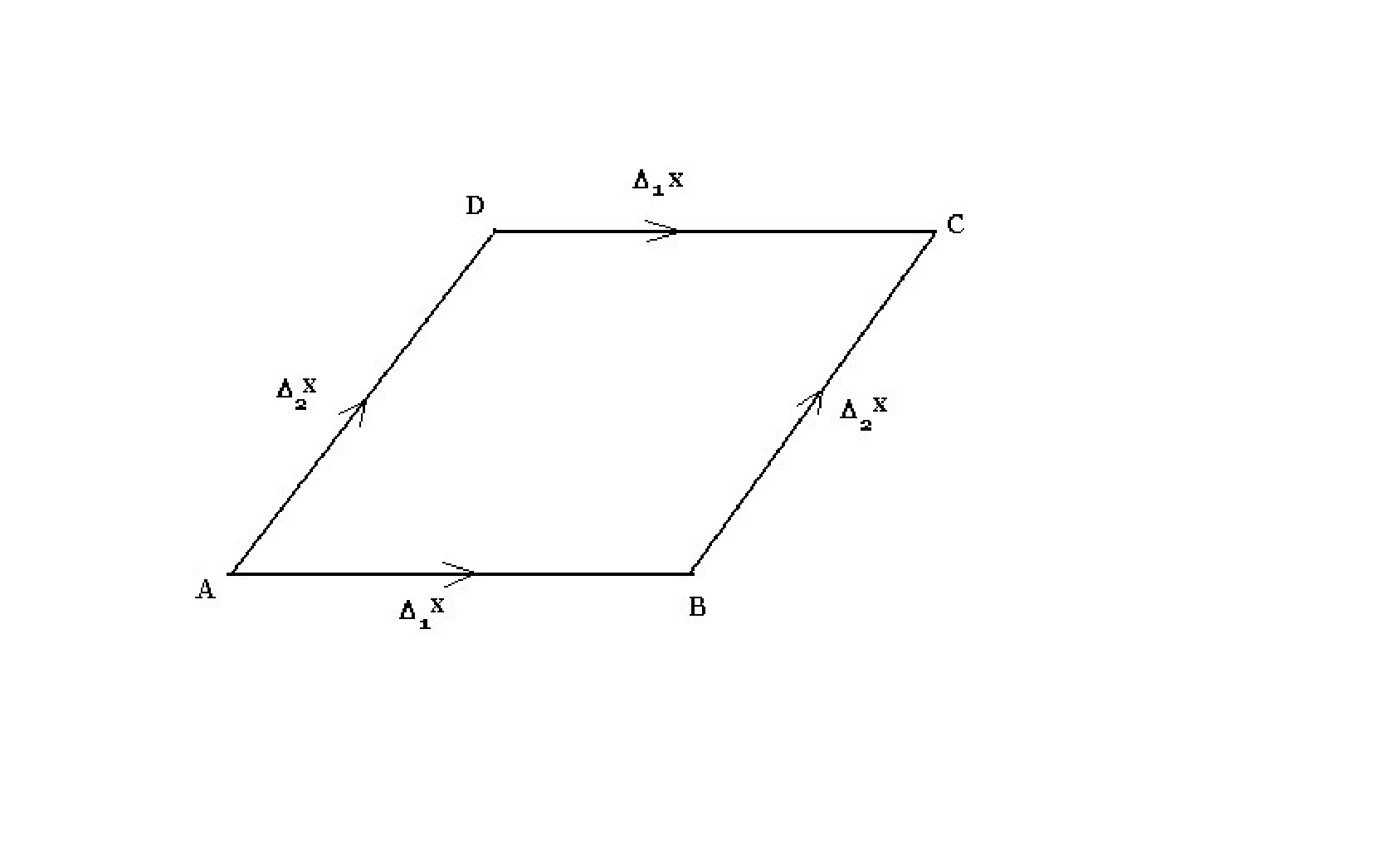}\caption{}\label{fig4}
\end{center}
\end{figure}

{\bf 6.} Now we define the \underline{curvature} or \underline{Riemann tensor}. Let us consider parallel transports of a vector $v^\mu$ from a point $A$ to a nearby point $C$ along two different infinitesimal paths --- $ABC$ and $ADC$, as is shown on the fig. (\ref{fig4}).

If one parallel transports $v^\mu$ from $A$ to $B$, then he finds:

\bqa
v^\mu_{AB} \approx v^\mu - \Gamma^\mu_{\nu\alpha}\left(A\right) \, v^\nu \, \Delta_1 x^\alpha.
\eqa
Here $\Gamma^\mu_{\nu\alpha}\left(A\right)$ is the value of the Christoffel symbol at the point $A$. Then,

\bqa
\Gamma^\mu_{\nu\alpha}\left(B\right) \approx \Gamma^\mu_{\nu\alpha}\left(A\right) + \partial_\beta \, \Gamma^\mu_{\nu\alpha}\left(A\right) \, \Delta_1 x^\beta.
\eqa
Now making further parallel transport from the point $B$ to $C$, we find:

\bqa
v^\mu_{ABC} \approx v^\mu_{AB} - \Gamma^\mu_{\nu\alpha}\left(B\right) \, v^\nu_{AB} \, \Delta_2 x^\alpha \approx \nonumber \\ \approx v^\mu - \Gamma^\mu_{\nu\alpha}\left(A\right) \, v^\nu \, \Delta_1 x^\alpha - \left[\Gamma^\mu_{\nu\alpha}\left(A\right) + \partial_\beta \Gamma^\mu_{\nu\alpha}\left(A\right) \, \Delta_1 x^\beta\right] \, \left[v^\nu - \Gamma^\nu_{\beta\delta}\left(A\right) \, v^\beta \, \Delta_1 x^\delta\right] \, \Delta_2 x^\alpha \approx \nonumber \\ \approx
v^\mu - \Gamma^\mu_{\nu\alpha} \, v^\nu \, \Delta_1 x^\alpha - \Gamma^\mu_{\nu\alpha} \, v^\nu \, \Delta_2 x^\alpha - \partial_\beta \Gamma^\mu_{\nu\alpha} \, v^\nu \, \Delta_1 x^\beta \, \Delta_2 x^\alpha + \Gamma^\mu_{\nu\alpha} \, \Gamma^\nu_{\beta\gamma} \, v^\beta \, \Delta_1 x^\gamma \, \Delta_2 x^\alpha.
\eqa
Similarly doing the parallel transport along the $ADC$ path, one finds:

\bqa
v^\mu_{ADC} \approx v^\mu - \Gamma^\mu_{\nu\alpha} \, v^\nu \, \Delta_2 x^\alpha - \Gamma^\mu_{\nu\alpha} \, v^\nu \, \Delta_1 x^\alpha - \partial_\beta \Gamma^\mu_{\nu\alpha} \, v^\nu \, \Delta_1 x^\alpha \, \Delta_2 x^\beta + \Gamma^\mu_{\nu\alpha} \, \Gamma^\nu_{\beta\gamma} \, v^\beta \, \Delta_1 x^\alpha \, \Delta_2 x^\gamma.
\eqa
Then the difference between the two results of the parallel transport $v^\mu_{ABC}$ and $v^\mu_{ADC}$ is given by:

\bqa
v^\mu_{ABC} - v^\mu_{ADC} \approx - \left[\partial_\alpha \Gamma^\mu_{\nu\beta} - \partial_\beta \Gamma^\mu_{\nu\alpha} + \Gamma^\mu_{\gamma\alpha} \, \Gamma^\gamma_{\nu\beta} - \Gamma^\mu_{\gamma\beta} \, \Gamma^\gamma_{\nu\alpha}\right] \, v^\nu \, \Delta_1 x^\alpha \, \Delta_2 x^\beta \equiv \nonumber \\ \equiv - \frac12 \, {R^{\mu}}_{\nu\alpha\beta} \, v^\nu \, \Delta S^{\alpha\beta},
\eqa
where we have defined $\Delta S^{\alpha\beta} = \Delta_1 x^\alpha \, \Delta_2 x^\beta - \Delta_1 x^\beta \, \Delta_2 x^\alpha$. Modulus of this quantity defines the area of the parallelogram shown on the fig. (\ref{fig4}), and

\bqa
{R^{\mu}}_{\nu\alpha\beta} \equiv \partial_\alpha \Gamma^\mu_{\nu\beta} - \partial_\beta \Gamma^\mu_{\nu\alpha} + \Gamma^\mu_{\gamma\alpha} \, \Gamma^\gamma_{\nu\beta} - \Gamma^\mu_{\gamma\beta} \, \Gamma^\gamma_{\nu\alpha}
\eqa
is the Riemann tensor that we have been looking for. The Riemann tensor is nothing but the curvature for the connection in question:

\bqa\label{curvature}
{R^\mu}_{\nu\alpha\beta} = \left[D^\mu_{\alpha\gamma}, \, D^\gamma_{\beta\nu}\right] \equiv D^\mu_{\alpha\gamma}\, D^\gamma_{\beta\nu} - D^\mu_{\beta\gamma}\, D^\gamma_{\alpha\nu}, \quad {\rm where} \quad D^\mu_{\alpha\nu} = \partial_\alpha \, \delta^\mu_\nu + \Gamma^\mu_{\alpha\nu}.
\eqa
One also uses the following form of this tensor: $R_{\mu\nu\alpha\beta} = g_{\mu\gamma} \, {R^\gamma}_{\nu\alpha\beta}$.

Obviously in flat space--time $v_{ABC} = v_{ADC}$, hence the Riemann tensor is the measure of how the space--time
is curved. Note that the Riemann tensor transforms under the coordinate transformations multiplicatively. Hence, if it is zero in one reference system, then it is also zero in any other system.

{\bf 7.} Let us specify the properties of the Riemann tensor. In a vicinity of any point $x$ in LMRS (\ref{LMRS}) this tensor is equal to:

\bqa\label{RiemLMRS}
R_{\mu\nu\alpha\beta} = \partial_\alpha \Gamma_{\mu |\, \nu\beta} - \partial_\beta \Gamma_{\mu |\, \nu\alpha} = \frac12 \, \left[\partial^2_{\nu\alpha}\, g_{\mu\beta} - \partial^2_{\nu\beta} \, g_{\mu\alpha} - \partial^2_{\mu\alpha} \, g_{\nu\beta} + \partial^2_{\mu\beta} \, g_{\nu\alpha}\right],
\eqa
where $\partial^2_{\alpha\beta} = \partial_\alpha \partial_\beta$ and we also will be using the following notations: $\partial^2_{\alpha\beta} \, g_{\mu\nu} = g_{\mu\nu, \, \alpha\beta}$.
Now one can see that if a space--time is curved, then even in the LMRS one cannot put to zero first derivatives of the Christoffel symbols and/or second derivatives of the metric tensor.

From (\ref{RiemLMRS}) one immediately sees the following identities for the Riemann tensor:

\bqa\label{ident}
R_{\mu\nu\alpha\beta} = - R_{\nu\mu\alpha\beta}, \quad R_{\mu\nu\alpha\beta} = R_{\alpha\beta\mu\nu}, \nonumber \\
R_{\mu\nu\alpha\beta} + R_{\mu\alpha\beta\nu} + R_{\mu\beta\nu\alpha} = 0.
\eqa
Furthermore, differentiating (\ref{RiemLMRS}), we find:

\bqa\label{Bianchi}
{R^\mu}_{\nu\alpha\beta; \, \gamma} + {R^\mu}_{\nu\gamma\alpha; \, \beta} + {R^\mu}_{\nu\beta\gamma; \, \alpha} = 0,
\eqa
which is the so called \underline{Bianchi identity}. Although we have obtained these identities in LMRS, they are valid for any reference system because they relate tensorial quantities, which change multiplicatively under coordinate transformations.

A contraction of the Riemann tensor over any two of its indexes leads either to zero, ${R^\alpha}_{\alpha\mu\nu} = 0$ (due to the anti-symmetry of $R_{\mu\nu\alpha\beta}$ under the exchange of the corresponding indexes), or to the so called \underline{Ricci tensor}, $R_{\mu\nu} \equiv {R^\alpha}_{\mu\alpha\nu}$. The latter one is symmetric $R_{\mu\nu} = R_{\nu\mu}$ tensor as the consequence of (\ref{ident}). Contracting further the remaining two indexes, we obtain the \underline{Ricci scalar}: $R \equiv R_{\mu\nu} \, g^{\mu\nu}$. Furthermore, contracting two indexes in (\ref{Bianchi}), we obtain the useful identity:

\bqa \label{Bianchi1}
R^\nu_{\mu;\, \nu} = \frac12\, \partial_\mu R.
\eqa

{\bf 8.} Let us find the number of independent components of the Riemann tensor in a $D$--dimensional space--time. Riemann tensor $R_{\mu\nu\alpha\beta}$ is anti--symmetric under the exchanges $\mu \longleftrightarrow \nu$ and $\alpha \longleftrightarrow \beta$. Hence, the total number of independent combinations for each pair $\mu\nu$ and $\alpha\beta$ in $D$--dimensions is equal to $D\,(D-1)/2$. On the other hand, $R_{\mu\nu\alpha\beta}$ is symmetric under the exchange of these pairs, $\mu\nu \longleftrightarrow \alpha\beta$. Thus, the total number of independent combinations of the indexes is equal to

\bqa
\frac12 \, \frac{D\,\left(D-1\right)}{2} \, \left[\frac{D\, \left(D-1\right)}{2} + 1\right].
\eqa
However, we have to take into account the cyclic symmetry (the last equation in (\ref{ident})):

\bqa
B_{\mu\nu\alpha\beta} = R_{\mu\nu\alpha\beta} + R_{\mu\alpha\beta\nu} + R_{\mu\beta\nu\alpha} = 0.
\eqa
To find the number of these relations note that $B_{\mu\nu\alpha\beta}$ tensor is totaly anti--symmetric. For example,

\bqa
B_{\mu\nu\alpha\beta} = R_{\mu\nu\alpha\beta} + R_{\mu\alpha\beta\nu} + R_{\mu\beta\nu\alpha} = - R_{\nu\mu\alpha\beta} - R_{\nu\alpha\beta\mu} - R_{\nu\beta\mu\alpha} = - B_{\nu\mu\alpha\beta}.
\eqa
Then, it is not hard to see that the total number of independent conditions, $B_{\mu\nu\alpha\beta} = 0$, is equal to $D\,\left(D-1\right)\,\left(D-2\right)\, \left(D-3\right)/4!$. As the result, the total number of independent components of the Riemann tensor is given by:

\bqa
\frac12 \, \frac{D\,\left(D-1\right)}{2} \, \left[\frac{D\, \left(D-1\right)}{2} + 1\right]
- \frac{D\,\left(D-1\right)\,\left(D-2\right)\, \left(D-3\right)}{4!} = \frac{D^2\, \left(D^2 - 1\right)}{12}.
\eqa
In particular, in four dimensions we have 20 independent components, in $D=3$ --- 6 components, in $D=2$ --- only one.

In principle, around any given point we can further reduce the number of independent components of the Riemann tensor. In fact, LMRS around such a point is defined up to rotations and boosts, as we discuss above. Hence, with the appropriate choice of the parameters of the rotation one can put to zero $D\,\left(D-1\right)/2$ more components of the Riemann tensor.

\vspace{10mm}

\centerline{\bf Problems}

\vspace{5mm}

\begin{itemize}

\item Derive (\ref{Gtrans}).

\item Show that $\partial_\mu \left(a_\nu \, b^\nu\right) = \left(D_\mu a_\nu\right) \, b^\nu + a_\nu \left(D_\mu b^\nu\right)$.

\item Prove that

$$
\int_M d^4 x \, \sqrt{|g|} \, A_{\mu\nu} \, \left(D^\mu B^\nu\right) = \oint_{\partial M} d^3\sigma^\mu \, A_{\mu\nu} \, B^\nu - \int_M d^4 x \, \sqrt{|g|}  \, \left(D^\mu A_{\mu\nu}\right) \, B^\nu,
$$
where $\partial M$ is the three--dimensional boundary of a four--dimensional manifold $M$; $d^3\sigma^\mu$ is a four--dimensional vector perpendicular
to $\partial M$, directing inside $M$, whose modulus is equal to the volume element of $\partial M$.

\item Prove (\ref{curvature}).

\item Prove the Bianchi identity (\ref{Bianchi}).

\item Prove (\ref{Bianchi1}).

\item Express the relative four--acceleration of two particles which move over infinitesimally
      close geodesics via the Riemann tensor of the space--time. (See the corresponding problem in the Landau--Lifshitz.)

\end{itemize}

\vspace{10mm}

\centerline{\bf Subjects for further study:}

\vspace{5mm}

\begin{itemize}

\item Riemann, Ricci and metric tensors in three and two dimensions.

\item Extrinsic curvatures of embedded lines and surfaces.

\item Yang--Mills curvature and theory.

\item Veilbein and spin connection. Riemann tensor via spin connection.

\item Weyl tensor.

\item Fermions and torsion.

\end{itemize}

\newpage

\section*{LECTURE III \\ {\it Einstein--Hilbert action. Einstein equations. Matter energy--momentum tensor.}}

\vspace{10mm}

{\bf 1.} From the first two lectures we have learned about the physical meaning of arbitrary coordinate transformations and how to distinguish the flat space--time in cuverlinear coordinates from curved space--times. Now we are ready to address the question: What is the physics behind curved spaces?

Set of experimental observations tells us that space--time is curved by anything that carries energy (originally this was a working guess, perhaps out of aesthetic considerations). Rephrasing this, the metric tensor is a dynamical variable --- a generalized coordinate --- coupled to energy carried by matter. Our goal here is to see that all we need to formulate the theory of gravity is this guess, general covariance
and the minimal action principle. We would like to find equations of motion for the metric tensor, which relate so to say geometry of space--time to energy carried by matter.

Obviously equations of motion that we are looking for should be covariant under general coordinate transformations, i.e. they should have the same form in all coordinate systems. Hence, the corresponding action for the metric should be invariant under these transformations. If we have a metric tensor, then the simplest invariant that one can write is the volume of space--time, $I = \int \sqrt{|g|}\, d^4x$, where $|g|$ is the modulus of the determinant of the metric tensor, $\left|\det \left(g_{\mu\nu}\right)\right|$, and $d^4x = dx^0 \, dx^1 \, dx^2 \, dx^3$. It is alone is not suitable for the action, because it does not contain derivatives of the metric. In fact, after the application of the minimal action principle to the action proportional to this invariant one will find an algebraic rather than differential (dynamical) equations of motion for the metric.

The simplest invariant that contains derivatives of the metric is the Ricci scalar, $R = R_{\mu\alpha} \, g^{\mu\alpha} = g^{\mu\alpha}\, g^{\nu\beta} \, R_{\mu\nu\alpha\beta}$ (see the previous lecture). Thus, the simplest invariant action for the metric alone is as follows:

\bqa\label{ab}
S = a \, \int d^4x \, \sqrt{|g|} \, R + b\, \int d^4x \, \sqrt{|g|},
\eqa
where $a$ and $b$ are some dimensionful constants, which one can fix only on the basis of experimental data.
What remains to be added to this action is matter. Let $S_M$ be an action describing the coupling of matter
to gravity, i.e. to the metric tensor. We have encountered in the first lecture the simplest example of such an action. That is the action for the point particle, $S_{M} = - m \int ds = - m \int d\tau \, \sqrt{g_{\mu\nu}(z) \, \dot{z}^\mu \, \dot{z}^\nu}$. But below we will also encounter other types of actions for matter. The only thing about $S_M$ that we need to know at this point is that it should be invariant under general coordinate transformations.

All in all, we would like to apply the minimal action principle to the following action:

\bqa\label{EHaction}
S_{EH} = - \frac{1}{16 \pi \kappa} \, \int d^4x \, \sqrt{|g|} \, \left(R + \Lambda\right) + S_M\left(g_{\mu\nu}, \, {\rm matter}\right),
\eqa
which is referred to as the \underline{Einstein--Hilbert action}. Here we have fixed the constants $a$ and $b$ in (\ref{ab}), knowing in advance their correct values. The quantity $\Lambda$ is referred to as \underline{cosmological constant}\footnote{Quantum field theory predicts that $\Lambda$ should be huge due to so called zero point fluctuations of quantum fields. At the same time, observational data show that $\Lambda$ is not zero due to so called dark energy, but is very small. This contradiction is the essence of the so called cosmological constant problem. For us, however, in these lectures, which are directed mostly to mathematicians and mathematical physicists, $\Lambda$ is just an arbitrary parameter in the theory, whose choice is at our disposal.} and $\kappa$ is the seminal Newton's constant.

{\bf 2.} From the minimal action principle we have that

\bqa\label{minEH}
0 = \delta_g S_{EH} = -\frac{1}{16\, \pi \, \kappa} \, \delta_g \, \int d^4x \, \sqrt{|g|}\, \left(g^{\mu\nu} \, R_{\mu\nu} + \Lambda\right) + \delta_g S_M = \nonumber \\ = - \frac{1}{16\, \pi\, \kappa} \, \int d^4x \, \left[\left(\delta \sqrt{|g|}\right) \, \left(R+\Lambda\right) + \sqrt{|g|}\, \left(\delta g^{\mu\nu}\right) \, R_{\mu\nu} + \sqrt{|g|} \, \left(\delta R_{\mu\nu}\right) \, g^{\mu\nu}\right] + \delta_g S_M.
\eqa
Here $\delta_g S \equiv \left[S(g+\delta g) - S(g)\right]_{linear \,\, in \,\, \delta g}$ and in the extremum of $S$ we have that $\delta_g S = 0$.

First, let us find $\delta \sqrt{|g|}$. For that we derive here the generic identity:

\bqa
\delta \log\left|\det \hat{M}\right| \equiv \log \left|\det \left(\hat{M} + \delta \hat{M}\right)\right| - \log \left|\det \hat{M}\right| = \log\frac{\det \left(\hat{M} + \delta \hat{M}\right)}{\det\hat{M}} = \nonumber \\ = \log\det\left[\hat{M}^{-1}\, \left(\hat{M} + \delta\hat{M}\right)\right] = \log \det\left[\hat{\mathbf 1} + \hat{M}^{-1}\, \delta\hat{M}\right] = {\rm Tr} \log \left[\hat{\mathbf 1} + \hat{M}^{-1}\, \delta\hat{M}\right] \approx {\rm Tr} \hat{M}^{-1}\, \delta\hat{M}.
\eqa
In this chain of relations $\hat{M}$ is a generic non--degenerate matrix and we have been keeping trace of only terms which are linear in $\delta\hat{M}$.

Applying the obtained equation for $g_{\mu\nu}$ and its inverse tensor $g^{\mu\nu}$ we obtain:

$$
\delta \log \sqrt{|g|} \equiv \delta \log\sqrt{\left|\det{\left(g_{\mu\nu}\right)}\right|} = - \frac12 \, \delta \log \left|\det\left(g^{\mu\nu}\right)\right| = - \frac12 \, g_{\mu\nu} \, \delta g^{\mu\nu}.
$$
Hence,

\bqa\label{detg}
\delta\sqrt{|g|} = - \frac12 \sqrt{|g|} \, g_{\mu\nu} \, \delta g^{\mu\nu}.
\eqa
Second, let us continue with the term $\int d^4x \, \sqrt{|g|} \, \delta R_{\mu\nu} \, g^{\mu\nu}$ in (\ref{minEH}).
In the LMRS, where $g_{\mu\nu}(x) = \eta_{\mu\nu}$ and $\Gamma^\mu_{\nu\alpha}(x) = 0$, we have that:

\bqa
\delta R_{\mu\nu} = \delta \left(\Gamma^\alpha_{\mu\nu \, ,\alpha} - \Gamma^\alpha_{\mu\alpha \, ,\nu}\right) = \partial_\alpha \left(\delta\Gamma^\alpha_{\mu\nu}\right) - \partial_\nu \left(\delta \Gamma^\alpha_{\mu\alpha}\right) = D_\alpha\left(\delta \Gamma^\alpha_{\mu\nu}\right) - D_\nu \left(\delta \Gamma^\alpha_{\mu\alpha}\right).
\eqa
The last expression here is two tensor. Hence, we have a tensor relation between $\delta R_{\mu\nu}$ and $\delta \Gamma^\mu_{\nu\alpha}$, which is valid in any reference system,
although it was obtained in LMRS. Thus,

\bqa
g^{\mu\nu} \, \delta R_{\mu\nu} = D_\mu \left(g^{\alpha\beta} \, \delta \Gamma^\mu_{\alpha\beta} - g^{\alpha\mu} \, \delta \Gamma^\beta_{\alpha\beta}\right) \equiv D_\mu \delta U^\mu
\eqa
is a total covariant derivative of a four--vector $\delta U^\mu$. As the result,

\bqa
\int_{\cal M} d^4x \, \sqrt{|g|} \, g^{\mu\nu} \, \delta R_{\mu\nu} = \int_{\cal M} d^4x\, \sqrt{|g|} \, D_\mu \delta U^\mu = \oint_{\partial {\cal M}} d\Sigma_\mu \, \delta U^\mu,
\eqa
where ${\cal M}$ is the space--time manifold under consideration and $\partial {\cal M}$ is its boundary. To obtain the last equality, we have used the Stokes' theorem and $d\Sigma_\mu$ is the four--vector normal to $\partial {\cal M}$, whose modulus is the infinitesimal volume element of $\partial {\cal M}$: $d\Sigma_\mu = n_\mu \, \sqrt{g^{(3)}} \, d^3\xi$, where $n_\mu$ is the normal vector to the boundary, $g^{(3)} = \left|\det g_{ij}\right|$ is the determinant of the induced three--dimensional metric, $g_{ij}, \,\, i = 1,2,3$, on the boundary $\partial {\cal M}$ and $\xi$ are the corresponding coordinates parametrizing the boundary.

Consideration of the boundary contributions is a separate interesting subject, but here we are varying the action with such conditions that are as follows $\left.\delta U^\mu\right|_{\partial {\cal M}} = 0$. Hence, $\int d^4x \, \sqrt{|g|} \, g^{\mu\nu} \, \delta R_{\mu\nu} = 0$. Combining in (\ref{minEH}) this fact together with (\ref{detg}), we find:

\bqa\label{preldel}
0 = \delta_g S_M - \frac{1}{16\,\pi\, \kappa} \, \int d^4x \, \sqrt{|g|} \left(R_{\mu\nu} - \frac12 \, g_{\mu\nu} \, R - \frac12 \, g_{\mu\nu} \, \Lambda \right) \, \delta g^{\mu\nu}.
\eqa
What remains to be found is $\delta_g S_M$.

{\bf 3.} Let us assume that the matter action, being invariant under the general coordinate transformations, has the following form $S_M = \int d^4x \,  \sqrt{|g|} \, {\cal L}$, where ${\cal L}$ is an invariant Lagrangian density. For example,

\bqa
\int ds = \int d\tau \, \sqrt{g_{\mu\nu}(z) \, \dot{z}^\mu\, \dot{z}^\nu} = \int d^4x \, \sqrt{|g(x)|} \, \int d\tau \, \frac{\delta^{(4)}\left[x - z(\tau)\right]}{\sqrt{|g(z)|}} \, \sqrt{g_{\mu\nu}(z) \, \dot{z}^\mu\, \dot{z}^\nu}.
\eqa
The other examples will be given below.

Then,

\bqa\label{delSM}
\delta_g S_M = \delta \int d^4x \, \sqrt{|g|} \, {\cal L} = \int d^4x \, \left[\frac{\partial {\cal L}}{\partial g^{\mu\nu}} \, \delta g^{\mu\nu} \, \sqrt{|g|} + {\cal L} \, \delta \sqrt{|g|}\right] = \nonumber \\ = \int d^4x \, \sqrt{|g|} \, \left[\frac{\partial {\cal L}}{\partial g^{\mu\nu}} - \frac12 \, {\cal L} \, g_{\mu\nu}\right] \, \delta g^{\mu\nu} \equiv \frac12 \, \int d^4x \, \sqrt{|g|} \, T_{\mu\nu} \, \delta g^{\mu\nu},
\eqa
where we have introduced a new two--tensor:

\bqa
T_{\mu\nu} = 2 \, \frac{\partial {\cal L}}{\partial g^{\mu\nu}} - {\cal L} \, g_{\mu\nu}, \quad T_{\mu\nu} = T_{\nu\mu}.
\eqa
Let us clarify the physical meaning of $T_{\mu\nu}$. Among the variations of the metric $\delta g^{\mu\nu}$ there are so to say physical ones, which lead to the curvature variations of space--time, i.e. for them $\delta R_{\mu\nu\alpha\beta} \neq 0$. But there are also such variations of the metric tensor which are due to the coordinate changes, i.e. for them $\delta R_{\mu\nu\alpha\beta} = 0$\footnote{While the first type of variations deforms the space--time itself (it changes the actual distances between points), the second type corresponds to the changes between different reference systems in the same space--time.}. Let us specify the form of the latter variations. Under general coordinate transformations the inverse metric tensor transforms as:

\bqa
\bar{g}^{\mu\nu}\left(\bar{x}\right) = g^{\alpha\beta}(x) \, \frac{\partial \bar{x}^\mu}{\partial x^\alpha} \, \frac{\partial \bar{x}^\nu}{\partial x^\beta}.
\eqa
We are looking for the infinitesimal form of this transformation, i.e. when the transformation matrix $\partial \bar{x}^\mu/\partial x^\nu$ is close to the unit matrix.
If $\bar{x}^\mu = x^\mu + \epsilon^\mu(x)$, where $\epsilon^\mu(x)$ is a small vector field, then:

\bqa
\bar{g}^{\mu\nu}\left(\bar{x}\right) \approx g^{\alpha\beta}(x) \, \left(\delta^\mu_\alpha + \partial_\alpha \epsilon^\mu\right) \, \left(\delta^\nu_\beta + \partial_\beta \epsilon^\nu\right) = g^{\mu\nu}(x) + \partial^\mu\epsilon^\nu + \partial^\nu\epsilon^\mu \equiv g^{\mu\nu}(x) + \partial^{(\mu}\epsilon^{\nu)}.
\eqa
Here $\delta^\mu_\nu$ is the Kronecker symbol.
Taking into account that

\bqa
\bar{g}^{\mu\nu}\left(\bar{x}\right) = \bar{g}^{\mu\nu}\left(x + \epsilon\right) \approx \bar{g}^{\mu\nu}(x) + \partial_\alpha \bar{g}^{\mu\nu}(x) \, \epsilon^\alpha,
\eqa
we find that at the linear order in $\epsilon$:

\bqa\label{showeps}
\delta_{\epsilon} g^{\mu\nu} \equiv \bar{g}^{\mu\nu}(x) - g^{\mu\nu}(x) \approx - \partial_\alpha g^{\mu\nu}\, \epsilon^\alpha + \partial^{(\mu} \epsilon^{\nu)} = D^{(\mu}\epsilon^{\nu)}.
\eqa
Under such variations of the metric the action $S_M$ should not change at all, because it is an invariant,
as we agreed above. Hence, using (\ref{delSM}), we obtain

\bqa
0 \equiv \delta_{\epsilon} S_M = \frac12 \, \int_{\cal M} d^4x \, \sqrt{|g|}\, T_{\mu\nu} \, D^{(\mu} \epsilon^{\nu)} = \int_{\cal M} d^4 x\, \sqrt{|g|} \, T_{\mu\nu} \, D^\mu \epsilon^\nu = \nonumber \\ = \int_{\cal M} d^4x\, \sqrt{|g|} \, D^\mu\left(T_{\mu\nu} \, \epsilon^\nu\right) - \int_{\cal M} d^4x \, \sqrt{|g|} \, \epsilon^\nu \left(D^\mu \, T_{\mu\nu}\right) = \nonumber \\ = \oint_{\partial {\cal M}} d\Sigma^\mu \, T_{\mu\nu} \, \epsilon^\nu - \int_{\cal M} d^4x \, \sqrt{|g|}\, \epsilon^\nu \, \left(D^\mu \, T_{\mu\nu}\right),
\eqa
where we have used the symmetry $T_{\mu\nu} = T_{\nu\mu}$, performed the
integration by parts and used the Stokes' theorem.
We assume vanishing variations at the boundary, $\left.\epsilon^\mu\right|_{\partial {\cal M}} = 0$\footnote{The consideration of asymptotic symmetries, i.e. such symmetries which do not vanish at infinity, is a separate interesting and important subject. But it goes beyond the scope of our introductory lectures.}. Then, because $\delta_{\epsilon} S_M$ should vanish at any $\epsilon^\mu$ inside ${\cal M}$, we obtain the identity:

\bqa
D^\mu T_{\mu\nu} = 0,
\eqa
which is a covariant generalization of a conservation law. In fact, in Minkowski space--time it reduces to $\partial^\mu T_{\mu\nu} = 0$, which is a conservation law following from the Nether's theorem applied to space--time translations.
Thus, $T_{\mu\nu}$ is nothing but the \underline{energy momentum tensor}.

{\bf 4.} All in all, from (\ref{delSM}) and (\ref{preldel}) we obtain that

\bqa
0 = - \frac{1}{16 \, \pi \, \kappa} \int d^4x \sqrt{|g|} \left\{R_{\mu\nu} - \frac12 \, R \, g_{\mu\nu} - \frac12 \, \Lambda \, g_{\mu\nu} - 8\, \pi \, \kappa \, T_{\mu\nu}\right\} \, \delta g^{\mu\nu}.
\eqa
This expression should vanish for any infinitesimal value of $\delta g^{\mu\nu}$.
Hence, we obtain the \underline{Einstein equations}:

\bqa\label{Eeq}
R_{\mu\nu} - \frac12 \, R \, g_{\mu\nu} - \frac12 \, \Lambda \, g_{\mu\nu} = 8\, \pi \, \kappa \, T_{\mu\nu},
\eqa
which relate the geometry of space--time (left hand side) to the energy (right hand side).

In the vacuum $T_{\mu\nu}=0$ and $\Lambda = 0$. Hence, we obtain the equation

\bqa
R_{\mu\nu} - \frac12 \, R \, g_{\mu\nu} = 0.
\eqa
Multiplying it by $g^{\mu\nu}$ and using that $g_{\mu\nu}\, g^{\nu\mu} = \delta^\mu_\mu = 4$, we find that it implies $R=0$, i.e. this equation is equivalent to the \underline{Ricci flatness condition}:

\bqa
R_{\mu\nu} = 0.
\eqa
Note that this equation is {\it not} equivalent to the condition of vanishing of the curvature of space--time $R_{\mu\nu\alpha\beta} = 0$. In the following lectures we will find vacuum solutions (with $T_{\mu\nu} = 0$ and $\Lambda = 0$) of the Einstein equations which are Ricci flat, but are {\it not} Riemann flat, i.e. describe curved space--times.

If we apply covariant differential $D^\nu$ to both sides of eq. (\ref{Eeq}) and use the covariant constancy of the metric tensor $D_\alpha g_{\mu\nu} = 0$, we find that

\bqa
R^\nu_{\mu\, ;\nu} - \frac12 \, \partial_\mu R = 8\, \pi\, \kappa \, D^\nu T_{\mu\nu}
\eqa
Using the consequence of the Bianchi identity $R^\nu_{\mu\, ;\nu} = \frac12 \, \partial_\mu R$, which was derived in the previous lecture, we find the energy momentum tensor conservation condition, $D^\nu T_{\mu\nu} = 0$. Thus, even if we do not assume from the very beginning that $T_{\mu\nu}$ is conserved, this condition follows from the Einstein equations and the Bianchi identity.

This situation is similar to the one we encounter in the case of Maxwell equations. In fact, if one applies $\partial^\nu$ derivative to the equation $\partial^\mu F_{\mu\nu} = 4\,\pi\, j_\nu,$
where $F_{\mu\nu} = \partial_\mu A_\nu - \partial_\nu A_\mu$ and $j_\nu$ is four-current, he finds the
continuity equation $\partial^\mu j_\mu = 0$ due to anti--symmetry of the electromagnetic tensor $F_{\mu\nu} = - F_{\nu\mu}$. The continuity equation is just the condition of the charge conservation.
However, from the dynamical point of view the conservation of the energy momentum tensor means much more than the conservation of the electric current. Let us show that conservation of $T_{\mu\nu}$ implies the equations of motion for matter.

Consider, for example, energy momentum tensor for a dust (cloud of free particles which do not create any pressure). As follows from the solution of the problems at the end of this lecture, in this case $T_{\mu\nu}(x) = \rho(x) \, u_\mu(x) \, u_\nu(x)$, here $\rho(x)$ is the energy density of the dust and $u^\mu(x)$ is its four--velocity vector field. (Note that in the comoving reference frame, where $u^\mu = (1, 0, 0, 0)$, we obtain that $\left|\left|T_{\mu\nu}\right|\right| = Diag(\rho, 0, 0, 0)$.) Let us consider the condition of the conservation of such an energy momentum tensor:

\bqa\label{condcont}
0 = D^\mu\left(\rho \, u_\mu u_\nu\right) = \left(D^\mu \rho u_\mu\right)\, u_\nu + \rho \, u_\mu \, D^\mu u_\nu.
\eqa
Multiplying this equation by $u_\nu$ and using that $u_\nu u^\nu = 1$ (hence, $D_\mu u^\nu u_\nu = 2 u^\nu D_\mu u_\nu = 0$), we obtain a covariant generalization, $D^\mu\left(\rho \, u_\mu\right) = 0$, of the ordinary mass continuity equation $\partial^\mu \left(\rho \, u_\mu\right) = 0$. Moreover, as the consequence of this equation from (\ref{condcont}) we obtain that $u_\mu D^\mu u_\nu = 0$. Which means that species of the dust should move along geodesic curves, as the consequence of the energy--momentum tensor conservation. Thus, Einstein equations necessarily imply the dynamical equations of motion for matter. We will frequently encounter the consequences of these observations in the following lectures.

{\bf 5.} Let us describe various simple examples of matter coupling to gravity, i.e. various examples of $S_M$.
Consider e.g. a real scalar field $\phi$. Then, the simplest invariants that one can write are powers of $\phi$. At the same time, the simplest invariant that contains derivatives of $\phi$ is $g^{\mu\nu} \, \partial_\mu \phi \, \partial_\nu \phi$. As the result the simplest action describing the coupling of the scalar field to the gravity is as follows:

\bqa\label{SM2}
S_M = \int d^4x \, \sqrt{|g|} \, \left[g^{\mu\nu} \, \partial_\mu \phi \, \partial_\nu \phi - V(\phi)\right],
\eqa
where $V(\phi)$ is a polynomial in $\phi$.

Let us continue with the curved space--time generalization of the Maxwell theory. The natural covariant generalization of the electromagnetic tensor is:

\bqa
F_{\mu\nu} = D_\mu A_\nu - D_\nu A_\mu = \partial_\mu A_\nu - \partial_\nu A_\mu.
\eqa
As one can see this tensor does not change in passing from the flat space--time to a curved one. As the result, the curved space--time generalization of the Maxwell's action

\bqa\label{SM3}
S_M = \int d^4x \sqrt{|g|} \, F_{\mu\nu}\, F_{\alpha\beta} \, g^{\mu\alpha} \, g^{\nu\beta},
\eqa
is a trivial extension of the flat space action.

\vspace{10mm}

\centerline{\bf Problems}

\vspace{5mm}

\begin{itemize}

\item Find the $T_{\mu\nu}$ tensor for a collection of $N$ free particles (i.e. for the dust):

\bqa
S_M = - \sum_{q=1}^N m_q \, \int d\tau_q \, \sqrt{g_{\mu\nu}(z_q) \, \dot{z}_q^\mu\, \dot{z}_q^\nu}.
\eqa
Find the mass density $\rho$ in this case.

\item Find equations of motion ($\delta_\phi S_M = 0$ and $\delta_A S_M = 0$) and $T_{\mu\nu}$ for (\ref{SM2}) and (\ref{SM3}).

\item  Propose an invariant action for the metric, which is more complicated than the Einstein--Hilbert action. E.g. that which contains higher derivatives of the metric tensor.
({\bf *}) Derive the corresponding equations of motion.

\item Prove the last equality in (\ref{showeps}).

\item Using the experience with the reparametrization invariance from the first lecture, propose an action which contains only scalar field $\Phi$ and does not contain metric tensor $g_{\mu\nu}$, but which is invariant under the general covariant transformations. Note that reparametrization invariance is just one--dimensional general covariance.

\item Propose invariant actions that contain higher powers of derivatives, $\partial_\mu \Phi$, of the scalar field $\Phi$ and/or higher derivatives of the scalar field, $\partial_\mu \partial_\nu \Phi$ and etc..

\end{itemize}

\vspace{10mm}

\centerline{\bf Subjects for further study:}

\vspace{5mm}

\begin{itemize}

\item The minimal action principle for space--times with boundaries (with variations of the metric at the boundary). Boundary terms.

\item Different types of energy conditions for $T^{\mu\nu}$, their origin and meaning.

\item Raychaudhuri equations.

\item Veilbein formalism and spin connection.

\item Three--dimensional gravity and the Chern--Simons theory. (See e.g.	
``Quantum gravity in 2+1 dimensions: The Case of a closed universe'',
S.Carlip, Living Rev.Rel. 8 (2005) 1; arXiv: gr-qc/0409039.)

\item Two--dimensional gravity and Liouville theory. (Gauge fields and strings, A.Polyakov, Harwood Academic Publishers, 1987.)

\end{itemize}

\newpage

\section*{LECTURE IV \\{\it Schwarzschild solution. Schwarzschild coordinates. Eddington--Finkelstein coordinates.}}

\vspace{10mm}

{\bf 1.} Starting with this lecture we look for solutions of the Einstein equations. One of the simplest exact solutions of these equations was found by Schwarzschild. It describes spherically symmetric geometry when the cosmological constant is set to zero, $\Lambda = 0$, and in the absence of matter, $T_{\mu\nu} = 0$.

To find spherically symmetric geometry it is convenient to use spherical spatial coordinates, $x^\mu = (t, r, \theta, \varphi)$, instead of the Cartesian ones, $x^\mu = (t, x, y, z)$, and to use the most general spherically symmetric ansatz for the metric:

\bqa\label{generspher}
ds^2 = g_{tt}(r,t)\, dt^2 + 2\, g_{tr}(r,t) \, dtdr + g_{rr}(r,t) \, dr^2 + k(r,t) \, d\Omega^2, \nonumber \\
{\rm where} \quad d\Omega^2 = d\theta^2 + \sin^2\theta \, d\varphi^2,\quad {\rm and } \quad r \in [0,+\infty) \quad \varphi \in [0,2\pi), \quad \theta \in [0,\pi].
\eqa
Of course if one will choose a different coordinate frame (i.e. if one will make a generic coordinate change to an arbitrary reference system), then the spherical symmetry will be lost, but it is important to stress that there is a reference system in which the metric has the above form. This form of the metric is spherically symmetric because at a given moment of time, $dt = 0$, the space itself is sliced as an onion by spheres of radii following from $g_{rr}(r,t)$ and whose areas are set up by $k(r,t)$. Here $d\Omega^2$ is the metric on the sphere of unit radius.

The form of the metric (\ref{generspher}) is invariant under the following two--dimensional coordinate changes: $r = r\left(\bar{r},\bar{t}\right), \quad t = t\left(\bar{r}, \bar{t}\right)$. In fact, then:

\bqa
\bar{g}_{ab}\left(\bar{x}\right) = g_{cd}(x) \, \frac{\partial x^c}{\partial \bar{x}^a}\, \frac{\partial x^d}{\partial \bar{x}^b}, \quad {\rm and} \quad \bar{k}\left(\bar{x}\right) = k\left[x\left(\bar{x}\right)\right], \nonumber \\ {\rm where} \quad x^a = (r,t), \quad a = 1,2.
\eqa
Using this freedom of choice of two functions, $r\left(\bar{r},\bar{t}\right)$ and $t\left(\bar{r}, \bar{t}\right)$,
one can fix two out of the four functions, $g_{tt}(r,t)$, $g_{tr}(r,t)$, $g_{rr}(r,t)$ and $k(r,t)$. Without loss of generality in the case of nondegenerate metric it is convenient to set $g_{rt} = 0$ and $k(r,t) = - r^2$.

Then, introducing the standard notations $g_{tt} = e^{\nu(r,t)}$ and $g_{rr} = - e^{\lambda(r,t)}$, we arrive at the following convenient ansatz for the Einstein equations:

\bqa\label{dsSchprel}
ds^2 = e^{\nu(r,t)} \, dt^2 - e^{\lambda(r,t)} \, dr^2 - r^2\, d\Omega^2.
\eqa
It is important to note that the form of this metric is invariant under the remaining coordinate transformations $t = t\left(\bar{t}\right)$. In fact, then

\bqa\label{freed}
\bar{\lambda}\left(r,\bar{t}\right) = \lambda\left[r,t\left(\bar{t}\right)\right] \quad {\rm and} \quad \bar{\nu}\left(r,\bar{t}\right) = \nu\left[r,t\left(\bar{t}\right)\right] + \log\left(\frac{dt}{d\bar{t}}\right)^2.
\eqa
{\bf 2.} Thus, we have the following non-zero components of the metric and its inverse tensor:

\bqa
\left|\left|g_{\mu\nu}\right|\right| = Diag\left(e^\nu, - e^\lambda, - r^2, - r^2\, \sin^2\theta\right), \quad
\left|\left|g^{\mu\nu}\right|\right| = Diag\left(e^{-\nu}, - e^{-\lambda}, - \frac{1}{r^2}, - \frac{1}{r^2\, \sin^2\theta}\right).
\eqa
It is straightforward to find that non--zero components of the Christoffel symbols are given by:

\bqa\label{GamSch}
\Gamma^1_{11} = \frac{\lambda'}{2}, \quad \Gamma^0_{10} = \frac{\nu'}{2}, \quad \Gamma^2_{33} = - \sin\theta \, \cos\theta, \quad \Gamma^0_{11} = \frac{\dot{\lambda}}{2}\, e^{\lambda - \nu}, \nonumber \\
\Gamma^1_{22} = - r \, e^{-\lambda}, \quad \Gamma^1_{00} = \frac{\nu'}{2}\, e^{\nu-\lambda}, \quad \Gamma^2_{12} = \Gamma^3_{13} = \frac{1}{r}, \nonumber \\
\Gamma^3_{23} = \cot\theta, \quad \Gamma^0_{00} = \frac{\dot{\nu}}{2}, \quad \Gamma^1_{10} = \frac{\dot{\lambda}}{2}, \quad \Gamma^1_{33} = - r \, \sin^2\theta \, e^{-\lambda}.
\eqa
For the metric ansatz under consideration the other components of $\Gamma^\mu_{\nu\alpha}$ are zero, if they do not follow from (\ref{GamSch}) via the application of the symmetry $\Gamma^\mu_{\nu\alpha} = \Gamma^\mu_{\alpha\nu}$. Here the prime on top of $\nu(t,r)$ and $\lambda(t,r)$ means just the application of $\partial/\partial r$ differential and the dot --- $\partial/\partial t$.

As the result, the non--trivial part of the Einstein equations is as follows:

\bqa\label{einschwarz}
8 \, \pi \, \kappa \, T^1_1 = - e^{-\lambda} \, \left(\frac{\nu'}{r} + \frac{1}{r^2}\right) + \frac{1}{r^2}, \nonumber \\
8 \, \pi \, \kappa \, T^2_2 = 8 \, \pi \, \kappa \, T^3_3 = - \frac12 \, e^{-\lambda} \, \left(\nu'' + \frac{\left(\nu'\right)^2}{2} + \frac{\nu' - \lambda'}{r} - \frac{\nu' \, \lambda '}{2}\right) + \frac12 \, e^{-\nu} \, \left(\ddot{\lambda} + \frac{\dot{\lambda}^2}{2} - \frac{\dot{\lambda} \, \dot{\nu}}{2}\right), \nonumber \\
8 \, \pi \, \kappa \, T^0_0 = - e^{-\lambda} \, \left(\frac{1}{r^2} - \frac{\lambda'}{r}\right) + \frac{1}{r^2}, \nonumber \\ 8 \, \pi \, \kappa \, T^1_0 = - e^{-\lambda} \, \frac{\dot{\lambda}}{r}.
\eqa
Here for the future convenience we have written the Einstein equations in the form $R_\mu^\nu - \frac12 \, R\, \delta_\mu^\nu = 8\, \pi \, \kappa \, T_\mu^\nu$, as if $T_{\mu\nu}$ is not zero. The other part of the Einstein equations, corresponding to $T^3_1$ or $T^2_1$ and etc., is trivially satisfied, if the corresponding components of $T_{\mu\nu}$ are zero.

Now, if $T^\mu_\nu = 0$, as we have assumed from the very beginning, then the equations under consideration reduce to:

\bqa\label{einshw1}
e^{-\lambda} \, \left(\frac{\nu'}{r} + \frac{1}{r^2}\right) - \frac{1}{r^2} = 0, \nonumber \\
e^{-\lambda} \, \left(\frac{\lambda'}{r} - \frac{1}{r^2}\right) + \frac{1}{r^2} = 0, \nonumber \\
\dot{\lambda} = 0.
\eqa
The equations for $T^2_2$ and $T^3_3$ in (\ref{einschwarz}) follow from (\ref{einshw1}).

Now it is straightforward to see that $\nu = \lambda = 0$ solves this equation. This corresponds to the metric $ds^2 = dt^2 - dr^2 - r^2 \, d\Omega^2$, which is just the flat space--time in the spherical spatial coordinates.

From the third equation in (\ref{einshw1}) we immediately see that $\lambda = \lambda(r)$ is time independent. Then, taking the sum of the first and the second equation in (\ref{einshw1}), we obtain the relation $\lambda' + \nu' = 0$, which means that $\lambda + \nu = g(t)$, where $g(t)$ is some function of time only. Using the freedom (\ref{freed}) one can set this function to zero by the appropriate change of $\nu(t,r)$. As the result, we obtain that $\nu = - \lambda(r)$.

Finally, it is not hard to solve the second equation in (\ref{einshw1}) to find that

\bqa
e^{-\lambda} = e^\nu = 1 + \frac{C_2}{r},
\eqa
where $C_2$ is some constant which will be fixed below.
As $r\to \infty$ we restore the flat space--time metric

\bqa\label{C11}
ds^2 = \left(1 + \frac{C_2}{r}\right)\, dt^2 - \frac{dr^2}{1 + \frac{C_2}{r}} - r^2 \, d\Omega^2 \longrightarrow dt^2 - dr^2 - r^2 \, d\Omega^2.
\eqa
In fact, on general physical grounds one can conclude that the geometry in question is created by a spherically symmetric massive body outside itself, i.e. in that part of space--time where $T_{\mu\nu}(x) = 0$. (We discuss these points in grater detail in the lectures that follow.) Then it is natural to expect that there should be flat space at the spatial infinity, where the influence of the gravitating center is negligible. General metrics obeying such a condition are referred to as \underline{asymptotically flat}. (Actually, the precise definition of what is asymptotically flat space--time is more complicated, but for brevity we do not discuss it here.)

Thus, asymptotically flat, spherically symmetric metric in the vacuum (when $T_{\mu\nu} = 0$ and $\Lambda = 0$) is static --- time independent and diagonal. This is the essence of the \underline{Birkhoff theorem}, which is discussed from various perspectives throughout most of the lectures that follow. This theorem is just a more complicated analog of the one stating that in Maxwell's theory spherically symmetric solution, which tends to zero at spatial infinity, is also necessarily static. The latter is the unique seminal solution describing the Coulomb potential created by a point like or spherically distributed charge. But Birkhoff theorem has deeper consequences, which will be discussed in the lectures that follow.

{\bf 3.} To find the value of $C_2$ in (\ref{C11}) consider a probe particle which moves in the background in question. Let the particle be non--relativistic and traveling far away from the gravitating center. Then, as we know from the course of classical mechanics, the action for such a particle should be:

\bqa
S \approx - m \, \int \left[1 - \frac{\dot{\vec{z}}^2}{2} + V\left(\left|\vec{z}\right|\right)\right] \, dt,
\eqa
where $V\left(\left|\vec{z}\right|\right) = V(r)$ is the Newton's potential, i.e. $V(r) = - \frac{\kappa \, M}{r}$, $M$ is the mass of the gravitating center and $\left|\dot{\vec{z}}\right| \ll 1$ is the velocity of the particle.
At the same time we know that $S = - m \, \int ds$. Hence, there is the following approximate relation:

\bqa
ds \approx \left[1 - \frac{\dot{\vec{z}}^2}{2} + V\left(r\right)\right] \, dt, \quad {\rm as} \quad r \to \infty \quad {\rm and} \quad \left|\dot{\vec{z}}\right| \ll 1.
\eqa
Then, at the linear order in $\dot{\vec{z}}^2$ and $V(r)$, we have

\bqa
ds^2 \approx \left[1 + 2\, V(r)\right]\, dt^2 - \dot{\vec{z}}^2 \, dt^2 = \left[1 + 2\, V(r)\right]\, dt^2 - d\vec{z}^2, \quad {\rm as} \quad r \to \infty.
\eqa
As the result for the weak field, $r\to \infty$, we ought to obtain

\bqa
g_{tt} \approx 1 + 2\, V(r) = 1 - \frac{2\,\kappa \, M}{r},
\eqa
i.e. from (\ref{C11}) we deduce that $C_2 = - 2\, \kappa \, M$.

All in all, we have found the so called \underline{Schwarzschild solution} of the Einstein equations:

\bqa\label{Schwarzschild}
ds^2 = \left(1 - \frac{r_g}{r}\right)\, dt^2 - \frac{dr^2}{1 - \frac{r_g}{r}} - r^2 \, d\Omega^2, \quad {\rm where} \quad r_g \equiv 2\, \kappa \, M.
\eqa
$r_g$ is referred to as the \underline{gravitational radius} for the mass $M$.
We will discuss the physical meaning of this metric in greater detail in the next lecture. But let us point out here a few relevant points concerning the geometry under consideration.

{\bf 4.} The metric (\ref{Schwarzschild}) is invariant under time translations $t\to t + const$ and time reversal $t\to - t$ transformation. Time slices, $dt = 0$, of this metric are themselves sliced by static spheres. As the  result, curves corresponding to $dr = d\theta = d\varphi = 0$ are world--lines of non--inertial observers that are fixed over the gravitating center. (We say that Schwarzschild metric is seen by non--inertial observers, which are fixed at various radii $r$ and angles $\theta, \varphi$ over the gravitating body.) Note that physical distance between $(r_1,\theta, \varphi)$ and $(r_2, \theta, \varphi)$ is given by $\int_{r_1}^{r_2} \frac{dr}{\sqrt{1 - \frac{r_g}{r}}} \neq r_2 - r_1$. There is the approximate equality in the latter equation only in the limit as $r_{1,2} \to \infty$.

The metric (\ref{Schwarzschild}) degenerates as $r \to r_g$: in fact, then $g_{tt} \to 0$ and $g_{rr}\to \infty$. As we will see in a moment, this singularity of the metric is a coordinate (unphysical) one. It is similar to the singularity of the Rindler's metric at $\rho = 0$, which was discussed in the first lecture. Hence, the Schwarzschild coordinates $r$ and $t$ are applicable only for $r > r_g$.

In fact, none of the invariants, which one can build from this metric are singular at $r=r_g$. For example,
the simplest invariant --- the volume form --- is equal to $d^4 x \sqrt{|g|} = r^2 \sin\theta \, dr\, d\theta \, d\varphi \, dt$ and is regular at $r=r_g$. The Ricci scalar is zero, as follows from the Einstein equations. One can build up other invariants using the Riemann tensor. The non--zero components of this tensor for the metric in question are given by:

\bqa\label{RiemSchwtens}
R_{0101} = \frac{r_g}{r^3}, \quad R_{0202} = \frac{R_{0303}}{\sin^2\theta} = - \frac{r_g\, \left(r-r_g\right)}{2\, r^2}, \nonumber \\ R_{1212} = \frac{R_{1313}}{\sin^2\theta} = \frac{r_g}{2\, \left(r - r_g\right)}, \quad R_{2323} = - r_g \, r \, \sin^2\theta.
\eqa
Other nonzero components of $R_{\mu\nu\alpha\beta}$ are obtained by permutations of the indexes of (\ref{RiemSchwtens}) according to the symmetries of this tensor.
Then it is straightforward to find one of the invariants:

\bqa\label{Rieminv}
R_{\mu\nu\alpha\beta} \, R^{\mu\nu\alpha\beta} = \frac{12 \, r_g^2}{r^6},
\eqa
which is regular at $r=r_g$.

\begin{figure}
\begin{center}
\includegraphics[scale=0.5]{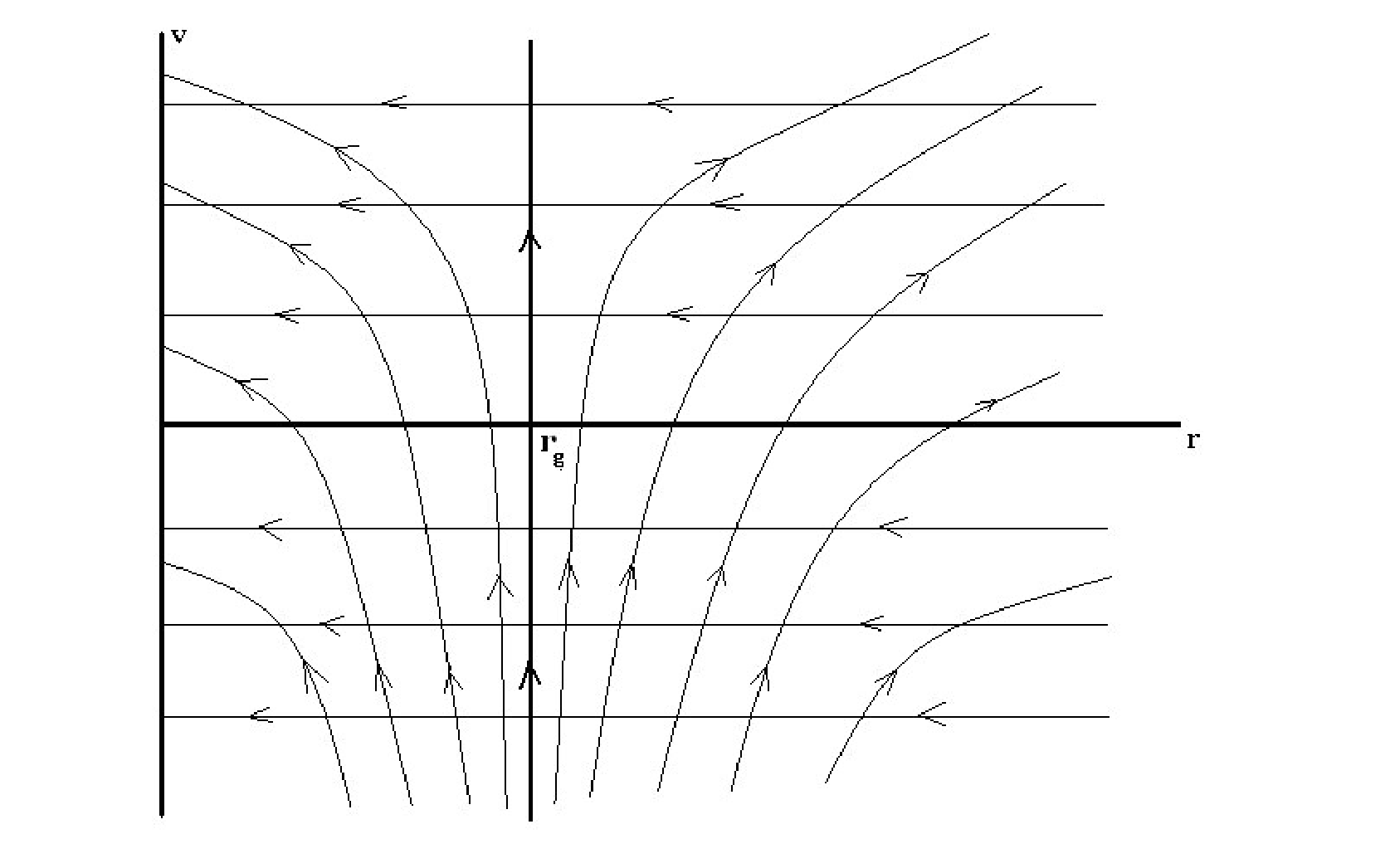}\caption{}\label{fig5}
\end{center}
\end{figure}

{\bf 5.} Another way to see that the space--time (\ref{Schwarzschild}) is regular at $r=r_g$ is to make a coordinate transformation to such a metric tensor which is regular at this surface. Let us perform the following transformations:

\bqa\label{EFC}
ds^2 = \left(1 - \frac{r_g}{r}\right)\, dt^2 - \frac{dr^2}{1 - \frac{r_g}{r}} - r^2 \, d\Omega^2 = \left(1 - \frac{r_g}{r}\right)\, \left[dt^2 - \frac{dr^2}{\left(1 - \frac{r_g}{r}\right)^2}\right] - r^2 \, d\Omega^2 = \nonumber \\ = \left(1 - \frac{r_g}{r}\right)\, \left[dt^2 - dr_*^2\right] - r^2 \, d\Omega^2.
\eqa
Here we have introduced the so called \underline{tortoise coordinate} $r_*$:

\bqa
dr_* = \frac{dr}{1 - \frac{r_g}{r}}, \quad {\rm and} \quad r_* = r + r_g \, \log\left(\frac{r}{r_g} - 1\right).
\eqa
We have fixed an integration constant in the relation between this coordinate, $r_*$, and the radius $r$. So that $r_* \approx r$, as $r\to\infty$ and $r_*\to -\infty$ as $r\to r_g$.

Now, if one introduces $v = t + r_*$ and transforms from $(t,r)$ to $(v,r)$ coordinates in (\ref{EFC}), then he finds the Schwarzschild space--time in the so called \underline{ingoing Eddington--Finkelstein coordinates}:

\bqa\label{EFC1}
ds^2 = \left(1 - \frac{r_g}{r}\right) \, dv^2 - 2\, dv \, dr - r^2 \, d\Omega^2.
\eqa
The obtained form of the metric tensor is not singular at $r=r_g$ and can be extended to $r\leq r_g$.

In the Eddington--Finkelstein coordinates the invariant (\ref{Rieminv}) has the same from. Hence, we see that the Schwarzschild space--time has a physical singularity at $r=0$, i.e. the space--time in question is meaningful only beyond the point $r=0$. Moreover, it is natural to expect that Einstein's theory brakes down as one approaches the point $r=0$, where the curvature becomes enormous. The reason for that can be understood after the solution of the problem for the previous lecture, which addresses modifications of the Einstein--Hilbert action.

Let us consider the behavior of radial light--like geodesics in the metric (\ref{EFC1}). In flat space--time light rays travel according to the law that $ds=0$. In a vicinity of any point a curved space--time looks almost as flat. Hence, in curved space--time light rays also travel according to the law that $ds=0$. Thus, for radial light rays we have that $ds=0$ and $d\theta = d\varphi = 0$. Then from (\ref{EFC1}) we find

\bqa
\left[\left(1 - \frac{r_g}{r}\right) \, dv - 2 \, dr\right]\, dv = 0.
\eqa
From $dv = 0$ we obtain the ingoing light rays $v \equiv t + r_* = const$. They are ingoing because as $t\to +\infty$ we have to take $r_*\to - \infty$ to keep $v=const$. At the same time, from $\left(1 - \frac{r_g}{r}\right) \, dv = 2 \, dr$ we obtain ``outgoing'' light rays. They are actually outgoing ($r\to +\infty$ as time goes by) only when $r>r_g$. These light rays evolve towards $r\to 0$ when $r<r_g$. The resulting picture is shown on the fig. (\ref{fig5}). The thin lines are light--like geodesics. The arrows on these lines show the directions of the light propagation as one advances forward in time. The vertical line $r=r_g$, $dr=0$, on the fig. (\ref{fig5}) is also one of the light--like geodesics.

\vspace{10mm}

\centerline{\bf Problems}

\vspace{5mm}

\begin{itemize}

\item Show that from the variational equation $0 = \delta \int d\tau \, g_{\mu\nu}[z(\tau)] \, \dot{z}^\mu \, \dot{z}^\nu$ follows the equation $\ddot{z}^\mu + \Gamma^\mu_{\nu\alpha} \, \dot{z}^\nu \, \dot{z}^\alpha = 0$. This observation frequently gives a practical way to calculate Christoffel symbols. Using this method find (\ref{GamSch}) from (\ref{dsSchprel}).

\item Derive components of $R_{\mu\nu}$.

\item Derive (\ref{Rieminv}) from (\ref{RiemSchwtens}).

\item Show that if in (\ref{EFC}) one will introduce $u = t - r_*$ and transform from $(t,r)$ to $(u,r)$, he would find the Scwarzschild space--time in the so called
    \underline{outgoing Eddington--Finkelstein coordinates}:

\bqa
ds^2 = \left(1 - \frac{r_g}{r}\right) \, du^2 + 2\, du \, dr - r^2 \, d\Omega^2.
\eqa
Show that this metric is mapped to (\ref{EFC1}) under the time reversal $t \to - t$.

\item Show that if one will make the change $\rho = \sqrt{1 - \frac{r_g}{r}}$ in the Schwarzschild space--time, he would obtain the metric, which in the vicinity of $r=r_g$ looks like:

\bqa
ds^2 \approx \rho^2 \, dt^2 - \left(2\, r_g\right)^2 \, d\rho^2 - r_g^2 \, d\Omega^2, \quad {\rm as} \quad \rho \to 0.
\eqa
Which is very similar to the Rindler space--time.

\item Find the metric for the Schwarzschild space--time as seen by the free falling observers. (See the corresponding paragraph in Landau--Lifshitz.)
For the free falling observers coordinate time coincides with the proper one. Hence, the corresponding metric should have $g_{00} = 1$.

\end{itemize}

\vspace{10mm}

\centerline{\bf Subjects for further study:}

\vspace{5mm}

\begin{itemize}

\item Black hole and black brane solutions in higher and lower dimensional Einstein theories. (E.g. in ``String Theory'', by J.Polchinski, Cambridge University Press, 2005.)

\item Reissner--Nordstrom solution of the Einstein--Maxwell theory.

\item Wormhole solutions. (See e.g. ``Wormholes in spacetime and their use for interstellar travel: A tool for teaching general relativity'', M.S.Morris and Kip S.Thorne, Am. J. Phys. 56(5), May 1988.)

\item Stability of the Schwarzschild solution under linearized perturbations. (See e.g. ``The mathematical theory of black holes'', S. Chandrasekhar, Oxford University Press, 1992)

\end{itemize}

\newpage

\section*{LECTURE V \\{\it Penrose--Carter diagrams. Kruskal--Szekeres coordinates. Penrose--Carter diagram for the Schwarzschild black hole.}}

\vspace{10mm}

{\bf 1.} Let us discuss now the properties of the Schwarzschild solution on the so called \underline{Penrose--Carter diagram}. The idea of such a diagram is to select a relevant two--dimensional part of a space--time under consideration and to make its stereographic projection on a compact space. For two--dimensional spaces such a projection always possible via a conformal map. In fact, a two--dimensional metric tensor, being symmetric $2\times 2$ matrix, has 3 independent components. Two of them can be fixed using transformations of two coordinates. As the result any two--dimensional metric on $R^{1,1}$ can be transformed to the following form $g_{ab} = \omega^2(x) \, \eta_{ab}, \quad a =1,2$. Here $\omega^2(x)$ is a space--time dependent function, which is referred to as \underline{conformal factor}. One just has to make sure that the corresponding coordinates, $x^a$, take values in a compact range. The reason for that will be clear in a moment.

The main point behind the Penrose--Carter diagrams is that under conformal maps (when one drops off the conformal factor) light--like world--lines and angles between them do not change. As the result one can clearly see causal properties of the original space--time on a compact diagram. The disadvantage of such diagram is that to draw it one has to know the whole space--time throughout its entire history, which is frequently impossible in generic physical situations. Moreover Penrose--Carter diagrams are sensitive only to global structure of space--time.

{\bf 2.} To illustrate these points let us draw the Penrose--Carter diagram for Minkowski space--time: $ds^2 = dt^2 - dx^2 - dy^2 - dz^2$. Select e.g. $(t,x)$ part of this space--time and make the following transformation $t\pm x = \tan\left(\frac{\psi \pm \xi}{2}\right)$. Here if $t,x \in (-\infty,+\infty)$, then $\psi,\xi \in [-\pi,\pi]$. Under such a coordinate transformation the Minkowskian metric changes as follows:

\bqa\label{114}
dt^2 - dx^2 = \frac{1}{\left[2\, \cos\left(\frac{\psi + \xi}{2}\right) \, \cos\left(\frac{\psi - \xi}{2}\right)\right]^2} \, \left[d\psi^2 - d\xi^2\right].
\eqa
The conformal factor of the new metric, $\left[\frac{1}{2\, \cos\left(\frac{\psi + \xi}{2}\right) \, \cos\left(\frac{\psi - \xi}{2}\right)}\right]^2$, blows up at $\left|\psi \pm \xi\right| = \pi$, which
makes the boundary of the compact $(\psi,\xi)$ space--time infinitely far away from any its internal point. This fact allows one to map the compact $(\psi,\xi)$ space--time onto the non--compact $(t,x)$ space--time.

\begin{figure}
\begin{center}
\includegraphics[scale=0.5]{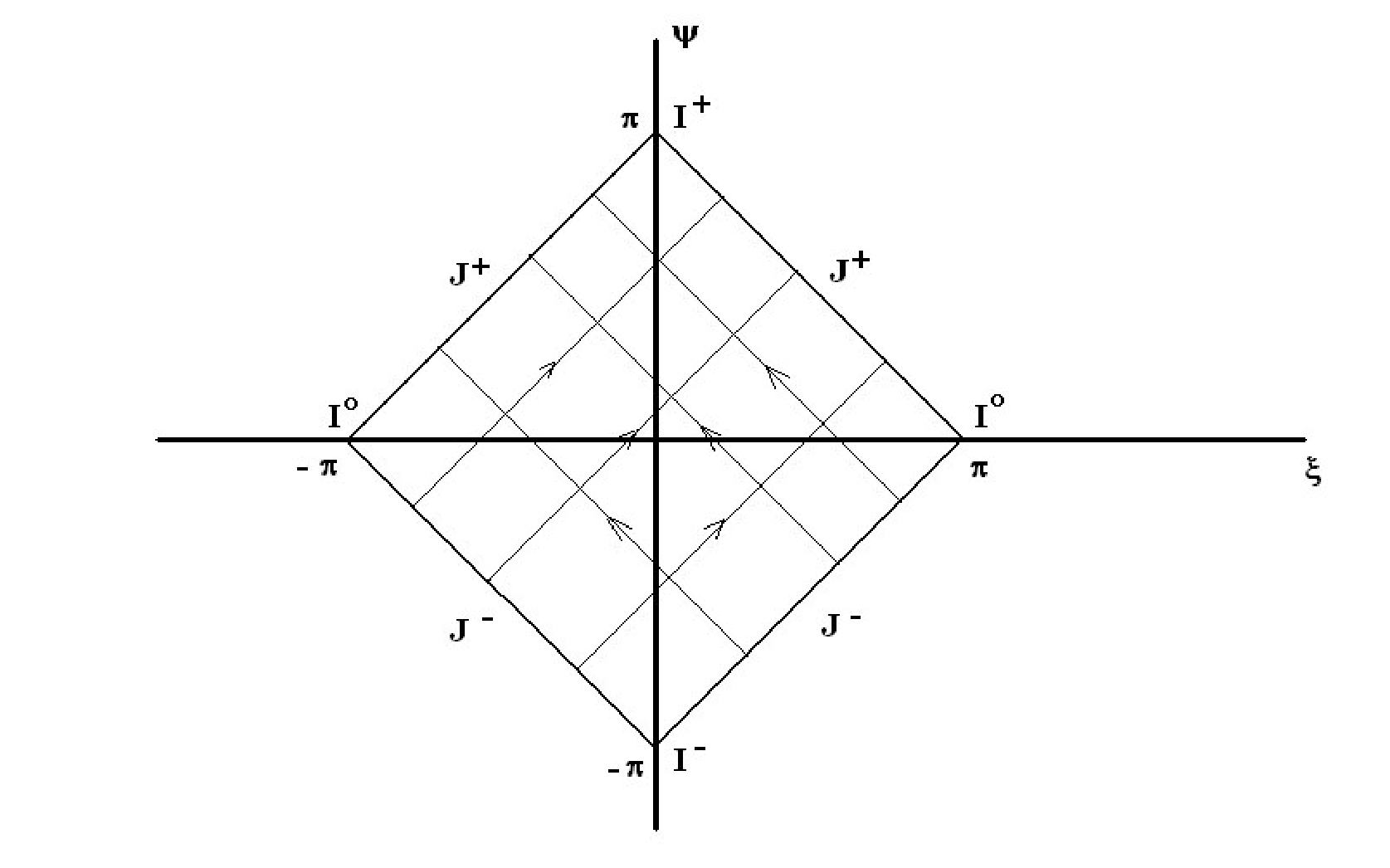}\caption{}\label{fig6}
\end{center}
\end{figure}

Furthermore, it is not hard to see that equality $dt^2 - dx^2 = 0$ implies also that $d\psi^2 - d\xi^2 = 0$, and vise versa. Hence, conformal factor is irrelevant in the study of the properties of the light--like world--lines --- those which obey, $ds^2 = 0$.
The latter in $(\psi,\xi)$ space--time are also straight lines making $45^o$ angles with respect to the $\psi$ and $\xi$ axes. Then, let us just drop off the conformal factor and draw the compact $(\psi,\xi)$ space--time.
It is shown on the fig. (\ref{fig6}).

On this diagram we show light--like rays by thin straight lines. The arrows on them show the direction of the light propagation, as $t$ is changing from past to the future. Furthermore, on this diagram $I^\pm$ represent the entire space, $x\in (-\infty, +\infty)$, at $t = \pm \infty$. These are space--like past and future infinities. Also $I^0$ is the entire time line, $t\in (-\infty, +\infty)$, at $x=\pm \infty$, i.e. this is time--like space infinity. And finally $J^\pm$ are light--like past and future infinities, i.e. these are the curves on which light--like world--lines originate and terminate, correspondingly.

The reason why after the stereographic projection of a two--dimensional plane we obtain the square rather than the sphere is its Minkowskian signature. While with Euclidian signature all points at infinity of the plane are indistinguishable and, hence, map to the single northern pole of the sphere, in the case of Minkowskian signature different parts at infinity have different properties. They can be either of space--like, time--like or light--like type.

\begin{figure}
\begin{center}
\includegraphics[scale=0.5]{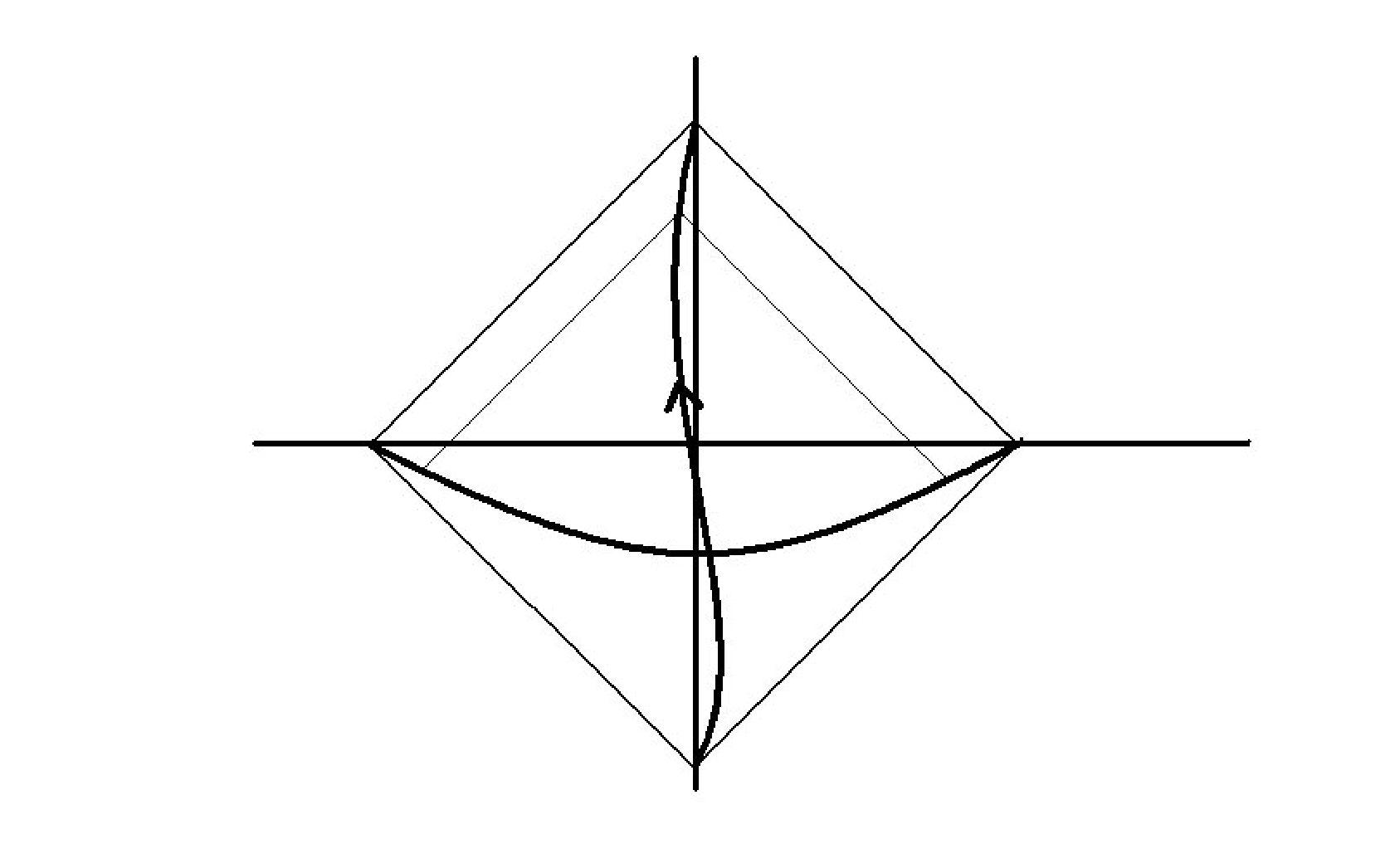}\caption{}\label{fig7}
\end{center}
\end{figure}

Let us discuss now causal properties of the Minkowski space--time on the obtained diagram. Consider the fig. (\ref{fig7}). Here we show a time--like world--line of an observer or of a massive particle. It is the bold curly vertically directed line with the arrow. Also we depict on the fig. (\ref{fig7}) a space--like Cauchy surface --- a surface of a fixed time slice, $t=const$. It is the bold curly horizontally directed line. From the picture under consideration it is not hard to see that any observer in this space--time can access (view) whole space--like sections as he reaches future infinity, $I^+$. Hence, in Minkowski space--time there are no regions which are causally disconnected from each other. Below we will see that the situation in the case of Schwarzschild space--time is quite different.

{\bf 3.} Before drawing the Penrose--Carter diagram for the Schwarzschild space--time one should find coordinates which cover it completely. As we have explained at the end of the previous lecture, Schwarzschild coordinates $(t,r)$ should cover only a part of the entire space--time. In this respect they are similar to the Rindler coordinates, which were introduced in the first lecture.

To find coordinates which cover entire space--time one has to embed it as a hyperplane into a higher--dimensional Minkowski space. Then one, in principle, can find such coordinates which cover the hyperplane completely and, hence, they cover the entire original curved space--time.

At every point of a four--dimensional space--time its metric, being a symmetric two--tensor, has $\frac{4(4+1)}{2} = 10$ independent components. From this we can subtract four degrees of freedom according to the four coordinate transformations,$\bar{x}^\mu(x)$. Thus, we have six independent degrees of freedom at every point. Hence, an arbitrary four--dimensional space--time can be embedded locally as a four--dimensional hyperplane into the $(4+6)$--dimensional Minkowski space--time, with a map which has appropriate properties.

However, if a curved space--time has extra symmetries, then it can be embedded into a flat space of a dimensionality less than ten. For example, Schwarzschild space--time, being quite symmetric, can be embedded into six--dimensional flat space. This embedding is done with the use of the so called \underline{Kruskal--Szekeres coordinates}. The machinery of such an embedding is beyond the scope of our concise lectures. We will present the Kruskal--Szekeres coordinates from a different perspective. At this point the reader just has to believe us that the coordinates in question cover the Schwarzschild space--time completely.

So let us start with the metric

\bqa\label{EFagain}
ds^2 = \left(1 - \frac{r_g}{r}\right) \, \left(dt^2 - dr_*^2\right) - r^2 \, d\Omega^2,
\eqa
which is written in terms of the tortoise coordinate,

\bqa\label{tortagain}
r_* = r + r_g\, \log\left(\frac{r}{r_g} - 1\right).
\eqa
It was introduced in the previous lecture. Our goal here is to get rid of the singularity of (\ref{EFagain})
at $r=r_g$, but in a way which is different from the one that was used in the previous lecture.

Let us introduce light--like coordinates $u = t - r_*$ and $v = t + r_*$. Then the metric acquires the following form:

\bqa
ds^2 = \left[1 - \frac{r_g}{r(u,v)}\right] \, du \, dv - r^2(u,v) \, d\Omega^2,
\eqa
where $r(u,v)$ is understood as an implicit function of $u$ and $v$ following from the relation (\ref{tortagain}) between $r_*$ and $u,v$:

\bqa
r(u,v) + r_g \, \log \left(\frac{r(u,v)}{r_g} - 1\right) = \frac12 \, \left(v - u\right).
\eqa
The coordinate singularity of this metric in the new coordinates is now placed at $r_* = - \infty$, or at $v-u = - \infty$.

It is not hard to see that from (\ref{tortagain}) we obtain the approximate relation $r_* \approx r_g \, \log \left|\frac{r}{r_g} - 1\right|$, as $r\to r_g$. Hence, in the vicinity of $r = r_g$ we find that $\left(1 - \frac{r_g}{r}\right) \approx e^{\frac{v-u}{2\, r_g}}$ and

\bqa
ds^2 \approx e^{\frac{v-u}{2r_g}}\, du \, dv - r^2_g \, d\Omega^2 = \left(e^{-\frac{u}{2r_g}} \, du\right)\, \left(e^{\frac{v}{2r_g}} \, dv\right) - r^2_g \, d\Omega^2.
\eqa
Thus, if one makes a change to the new coordinates $U = - 2 r_g \, e^{- \frac{u}{2r_g}}$ and $V = 2 r_g \, e^{\frac{v}{2r_g}}$, then the metric reduces to $ds^2 \approx dU\, dV - r^2_g\, d\Omega^2$ in the vicinity of $r = r_g$, i.e. the metric becomes flat rather than singular in the $(U,V)$ plane.

As the result, if from the very beginning one has made the following coordinate transformation:

\bqa\label{KScoord}
U = - 2r_g \, \exp\left[-\frac{t - r_*}{2r_g}\right] = - 2 r_g \, e^{-\frac{t-r}{2r_g}}\, \left(\frac{r}{r_g} - 1\right)^\frac12, \nonumber \\
V = 2r_g \, \exp\left[\frac{t + r_*}{2r_g}\right] = 2 r_g \, e^{\frac{t+r}{2r_g}}\, \left(\frac{r}{r_g} - 1\right)^\frac12,
\eqa
in the Schwarzschild space--time, he would obtain the metric

\bqa\label{KSmetr}
ds^2 = \frac{r_g}{r\left(U,V\right)}\, e^{-\frac{r\left(U,V\right)}{r_g}} \, dU \, dV - r^2\left(U,V\right) \, d\Omega^2.
\eqa
Here $r\left(U,V\right)$ is an implicit function of $U$ and $V$, which is given by the relation:

\bqa\label{rUVrel}
\left(\frac{r\left(U,V\right)}{r_g} - 1\right) \, e^{\frac{r\left(U,V\right)}{r_g}} = - \frac{U\,V}{\left(2r_g\right)^2}
\eqa
following from (\ref{KScoord}). The eq. (\ref{KScoord}) defines the Kruskal--Szekeres coordinates. The obtained metric (\ref{KSmetr}) is regular at $r=r_g$ and covers entire Schwarzschild space--time.

\begin{figure}
\begin{center}
\includegraphics[scale=0.5]{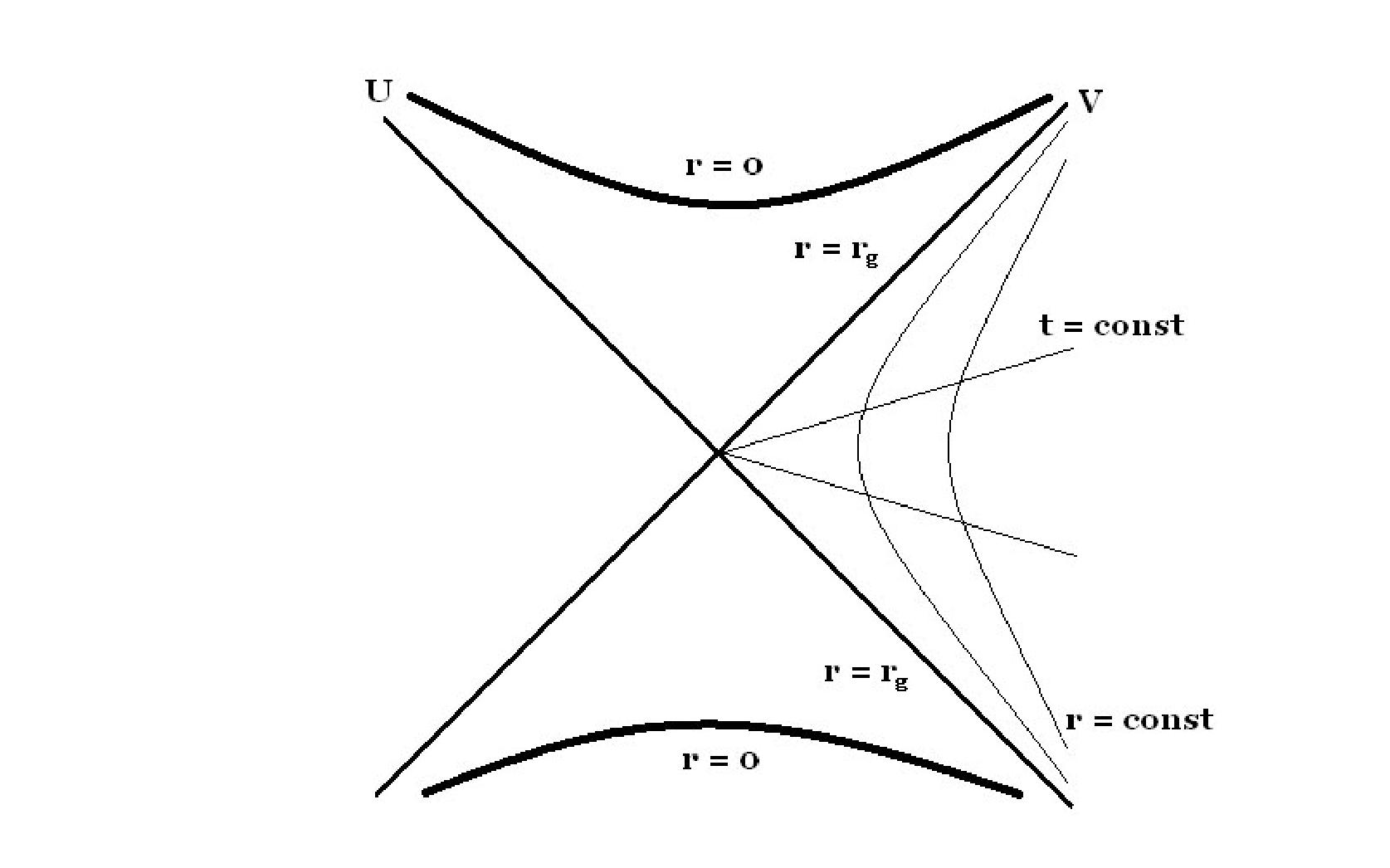}\caption{}\label{fig8}
\end{center}
\end{figure}

{\bf 4.} Let us describe the coordinate lattice in the new coordinates. Here we will discuss only the relevant two--dimensional, $(U,V)$, part of the space--time under consideration. From (\ref{rUVrel}) one can see that curves of constant $r$ are hyperbolas in the $(U,V)$ plane. At $r=r_g$ these hyperbolas degenerate to $UV = 0$, i.e. into two straight lines $U=0$ and $V=0$. At the same time from

\bqa\label{UV}
\frac{V}{U} = - e^{\frac{t}{r_g}}
\eqa
one can deduce that curves of constant $t$ are just straight lines.

As the result the relation between $(t,r)$ and $(U,V)$ coordinates is similar to the one we have had in the first lecture between Rindler's $(\tau,\rho)$ and Minkowskian $(t - x, t + x)$ coordinates. Note that $U$ and $V$ are light--like coordinates, because equations $dV = 0$ and $dU = 0$ describe light rays. Having these relations in mind, one can understand the picture shown on the fig. (\ref{fig8}).

As we have explained at the end of the pervious lecture the Schwarzschild space--time has a physical singularity at $r=0$. The two sheets of the corresponding hyperbola in (\ref{rUVrel}) are depicted by the bold lines on the fig. (\ref{fig8}). The Kruskal--Szekeres coordinates and space--time itself are not extendable beyond these curves. At the same time the Schwarzschild metric and $(t,r)$ coordinates cover only quarter of the fig. (\ref{fig8}), namely --- the right quadrant.

\begin{figure}
\begin{center}
\includegraphics[scale=0.5]{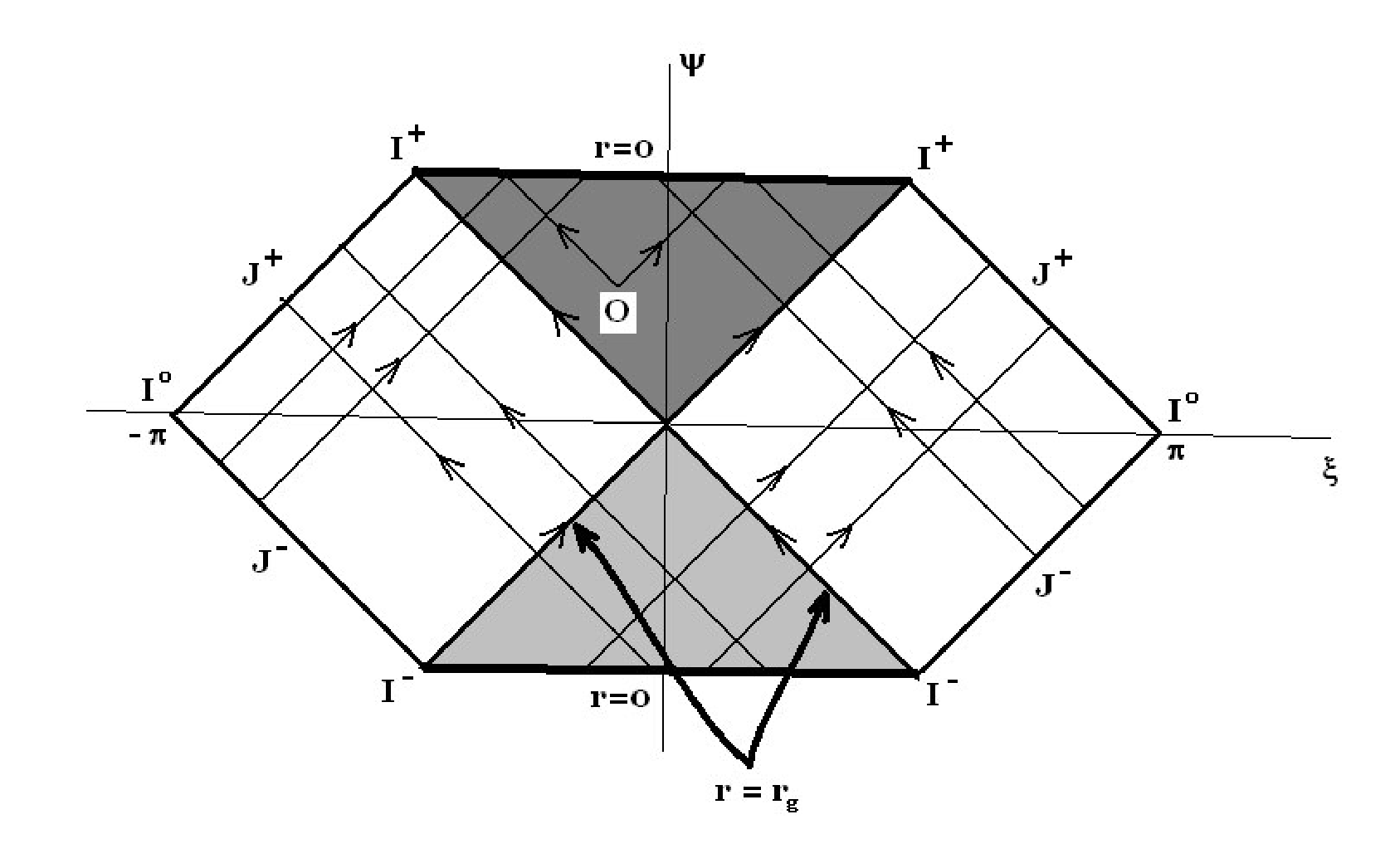}\caption{}\label{fig9}
\end{center}
\end{figure}

{\bf 5.} To draw the Penrose--Carter diagram for the Schwarzschild space--time let us do exactly the same transformation as at the beginning of this lecture. Let us transfrom from $U \equiv T - X$ and $V \equiv T + X$ to the $\psi \pm \xi$ coordinates and drop off the conformal factor. The resulting compact $(\psi,\xi)$ space--time is shown on the fig. (\ref{fig9}). This is basically the same picture as is shown on the fig. (\ref{fig6}), but with chopped off two triangular pieces from the top and from the bottom.

By the thin curves on the fig. (\ref{fig9}) we depict the light--like world--lines. The arrows on these curves show directions of light propagations, as $T \equiv \left(U + V\right)/2$ changes from the past towards future. Two bold lines corresponding to $r=r_g$ are also light--like. The singularity at $r=0$ is depicted on the fig. (\ref{fig9}) as two bold lines on the top and at the bottom of this picture. Actually these lines should be curved after the conformal map under discussion, but we draw them straight, because to have such a picture one can always adjust the conformal factor.

Now one can see that if an observer finds himself at a point like $O$ in the upper dark grey region, he inevitably will fall into the singularity $r=0$, because his world--line has to be within the light--cone emanating from the point $O$. For such an observer to avoid falling into the singularity is the same as to avoid the next Monday. Similar picture can be seen from the fig. (\ref{fig5}) of the previous lecture.

Thus, there is no way for such an observer to get out to the right or left quadrant if he found himself in the dark grey region. Also light rays from this region cannot reach future light--like infinity $J^+$. This region, hence, is referred to as \underline{black hole}. Its right boundary, $r=r_g$ or $U=0$, is referred to as future event horizon.

At the same time, the lower light--grey region has the opposite properties. Nothing can fall into this region
and everything escapes from it. This is the so called \underline{white hole}. Its right boundary is the past event horizon. Actually, there is nothing surprising in the appearance of this region for the Schwarzschild space--time.
In fact, General Relativity can be formulated in a Hamiltonian form. (This subject is beyond the scope of our lectures.) The Hamiltonian evolution is time--reversal. Hence, for every solution of the Einstein equations, its time reversal also should be a solution. As the result, in the static situation we simultaneously have the presence of both solutions.

\begin{figure}
\begin{center}
\includegraphics[scale=0.5]{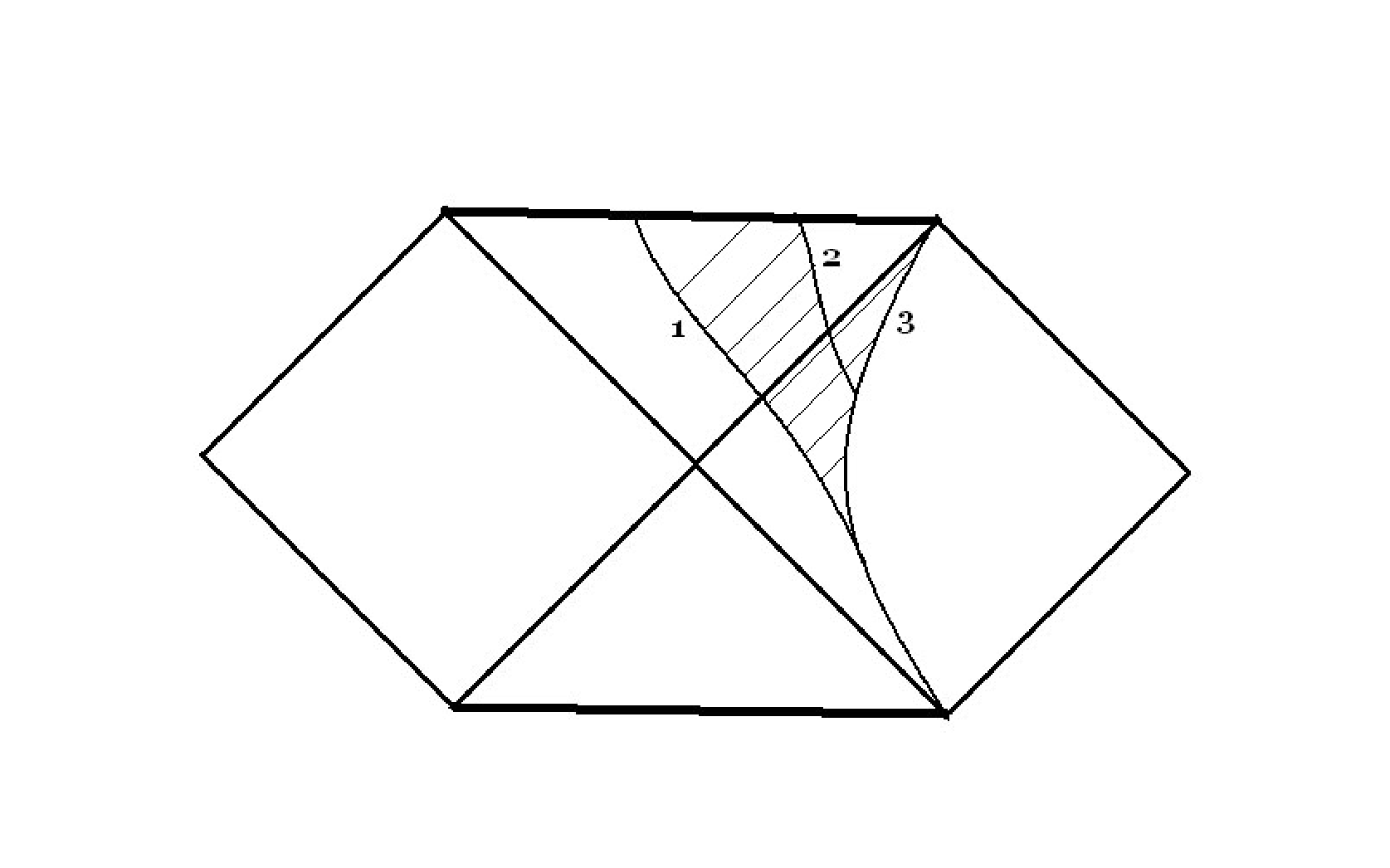}\caption{}\label{fig10}
\end{center}
\end{figure}

{\bf 6.} In the lectures that follow we will quantitatively describe the properties of the black holes in grater details. But let us discuss some of these properties qualitatively here on the Penrose--Carter diagram. Consider the fig. (\ref{fig10}). On this diagram an observer, depicted as having number 3, is fixed on some radius above the black hole. (Number 3 also can rotate around the black hole on a circular orbit: Penrose--Carter diagram cannot distinguish these two types of behavior, because it is not sensitive to the change of spherical angles $\theta$ and $\varphi$.) Thus, the observer number 3 always stays outside the black hole.

Then at some point another observer, e.g. number 1, starts his fall into the black hole from the same orbit. As we will show in the lectures that follow he crosses the event horizon, $r=r_g$, or even reaches the singularity, $r=0$, within finite proper time. At the same time from the diagram shown on the fig. (\ref{fig10}), one can deduce that the observer number 3 never sees how the number 1 crosses the event horizon. In fact, the last light ray which is scattered off by the number 1 goes along the horizon and reaches the number 3 only at $I^+$, i.e. at the future infinity. The situation is completely similar to the one which we have encountered in the first lecture for the case of Rindler's metric. In fact, similarly to that case, the relation between the proper and coordinate time is as follows:

\bqa
ds = \left(1 - \frac{r_g}{r}\right)^\frac12 \, dt = \sqrt{g_{00}}\, dt.
\eqa
Hence, fixed portions of the proper time, $ds$, correspond to the longer portions of the coordinate time, $dt$, if one resides closer to the horizon, $r\to r_g$. Recall also that, as follows from (\ref{UV}), $t=-\infty$ corresponds to $V=0$ (past event horizon) and $t=+\infty$ corresponds to $U=0$ (future event horizon).

But the picture which is seen by another falling observer, shown as the number 2 on the fig. (\ref{fig10}), which starts his fall after the number 1, is quite different. He never looses the number 1 from his sight and sees him crossing the horizon. But the light rays which are scattered off by the number 1 before the crossing of the horizon are received by the number 2 also before he himself crosses the horizon. At the same time, only
after crossing the horizon the number 2 starts to receive those light rays which are scattered off by the number 1 after he crossed the horizon.

\vspace{10mm}

\centerline{\bf Problems}

\vspace{5mm}

\begin{itemize}

\item Derive (\ref{114}).

\item Show the world line of an eternally accelerating observer/particle on the Penrose--Carter diagram of the Minkowskian space--time, i.e. on the fig. (\ref{fig6}).

\item Which part of the black hole Penrose--Carter diagram is covered by the ingoing Edington--Finkelstein coordinates? Why?

\item Which part of the black hole Penrose--Carter diagram is covered by the outgoing Edington--Finkelstein coordinates? Why?

\item Draw the Penrose--Carter diagram for $(t,r)$ part of the Minkowskian space--time in the spherical coordinates: $ds^2 = dt^2 - dr^2 - r^2 \, d\Omega^2$.

\end{itemize}

\vspace{10mm}

\centerline{\bf Subjects for further study:}

\vspace{5mm}

\begin{itemize}

\item Fronsdal--Kruskal's embedding of the Schwarzschild solution into the six dimensional Minkowski space--time. (``Maximal extension of Schwarzschild metric'', M.D. Kruskal, Phys. Rev., Vol. 119, No. 5 (1960) 1743.)

\item Cauchy problem in General Relativity and Hamiltonian formulation of the Einstein's theory. Time reversal of the Hamiltonian evolution.

\item Apparent horizon and other types of black hole horizons.

\item Penrose and Hawking singularity theorems.

\item Positive energy theorem, Penrose bound and their proof. (See e.g. ``A new proof of the positive energy theorem'', E. Witten, Commun. Math. Phys. 80, 381-402 (1981);``The inverse mean curvature flow and the Riemannian Penrose inequality'', G. Huisken and T. Ilmanen, J. Differential Geometry 59 (2001) 353-437;
    ``Proof of the Riemannian Penrose inequality using the positive mass theorem'', H.L. Bray, J. Differential Geometry 59 (2001) 177-267)

\item Asymptotic conformal infinity and asymptotically flat space--times.

\item Newman--Penrose formalism.

\end{itemize}

\newpage

\section*{LECTURE VI \\{\it Killing vectors and conservation laws. Test particle motion on Schwarzschild black hole background. Mercury perihelion rotation. Light ray deviation in the vicinity of the Sun.}}

\vspace{10mm}

{\bf 1.} In this lecture we provide a quantitative approval of several qualitative observations that have been made in the previous lectures. Furthermore, we derive from the General Theory of Relativity some effects that have been approved by classical experiments.

We will find geodesics in the Schwarzschild space--time. To do that it is convenient to find integrals of motion. Now we will provide them.
As we have shown in one of the previous lectures, under an infinitesimal transformation,

$$
\bar{x}^\mu = x^\mu + \epsilon^\mu(x),
$$
the inverse metric tensor transforms as

$$
\bar{g}^{\mu\nu}(x) = g^{\mu\nu}(x) + D^{(\mu} \epsilon^{\nu)}(x).
$$
If for some of the transformations, $\epsilon^\mu = k^\mu$, the metric tensor does not change,
i.e.

\bqa
D^{(\mu} k^{\nu)} \equiv D^\mu k^\nu + D^\nu k^\mu = 0,
\eqa
then the corresponding vector field, $k^\mu$, is referred to as \underline{Killing vector} and the transformations are called \underline{isometries} of the metric. For example, the Schwarzschild metric tensor does not depend on time, $t$, and angle, $\varphi$. Hence, it's isometries include at least the translations in time
$t \to t + a$ and rotations $\varphi \to \varphi + b$, for some constants $a$ and $b$. The corresponding Killing vectors, $k^\mu = \left(k^t, k^r, k^\theta, k^\varphi\right)$, have the following form: $k^\mu = (1,0,0,0)$ and $k^\mu = (0,0,0,1)$.

{\bf 2.} Here we show that if there is a Killing vector, then there also should be a conserved quantity as a revelation of the Noerther theorem. Consider a particle moving along a world--line $z^\mu(s)$ with the four--velocity $u^\mu(s) \equiv dz^\mu/ds$. Then, let us calculate the following derivative:

$$
\frac{d}{ds} k^\mu u_\mu = \partial_\nu\left(k^\mu u_\mu\right) \, \frac{dz^\nu}{ds} = k^\mu \, u^\nu \, D_\nu u_\mu + u^\nu \, u^\mu \, D_\nu k_\mu = k^\mu \, \left(u^\nu \, D_\nu u_\mu\right) + \frac12 \, u^\nu \, u^\mu \, \left(D_\nu \, k_\mu + D_\mu \, k_\nu\right).
$$
If the particle moves along a geodesic, then $u^\nu \, D_\nu u_\mu = 0$ and, if $k^\mu$ is the Killing vector, then
$D_\mu k_\nu + D_\nu k _\mu = 0$. As the result, $\frac{d \left(k^\mu u_\mu\right)}{ds} = 0$, i.e. the corresponding quantity is conserved, $k^\mu u_\mu = const$ for the motion along a geodesic.

{\bf 3.} Consider the Schwarzschild space--time and the Killing vector $k^\mu = (1,0,0,0)$. Then, the conserved quantity is given by

\bqa\label{dtds}
k^\mu u_\mu = u_0 = g_{00} \, u^0 = \left(1 - \frac{r_g}{r}\right) \, \frac{dt}{ds} \equiv \frac{E}{m},
\eqa
where $m$ is the mass of the particle and the physical meaning of $E$ will be specified in a moment.

Similarly for the Killing vector $k^\mu = (0,0,0,1)$ we obtain the following conserved quantity:

\bqa
r^2 \, \sin^2\theta \, \frac{d\varphi}{ds} \equiv \frac{L}{m}.
\eqa
The physical meaning of $L$ will also be defined in a moment. As can be seen from this conservation law and similarly to the Newton's theory, in radially symmetric space--times a trajectory of a particle is restricted to a plane\footnote{Note that $r^2 \, \sin^2\theta \, \frac{d\varphi}{ds}$ is the area swept by the radius--vector of the orbiting point during a unit time. Hence, the equation under consideration is just one of the Kepler's laws.}. We choose such a plane to be at $x_3 = 0$, i.e. at $\theta = \frac{\pi}{2}$. Hence, we can represent the last conserved quantity as $r^2 \, \frac{d\varphi}{ds} = \frac{L}{m}$.

Furthermore, in the background of the Schwarzschild metric the four--velocity should obey:

\bqa
g_{\mu\nu} \, u^\mu \, u^\nu = \left(1 - \frac{r_g}{r}\right)\, \left(\frac{dt}{ds}\right)^2 - \frac{\left(\frac{dr}{ds}\right)^2}{1 - \frac{r_g}{r}} - r^2 \, \left(\frac{d\varphi}{ds}\right)^2 = 1,
\eqa
under the assumption that $\theta = \frac{\pi}{2}$ and, hence, $d\theta = 0$. Using the two conservation laws that have been derived above, we find that the world--line of a massive particle in Schwarzschild space--time obeys the following equation:

\bqa\label{eqmas}
\left(\frac{dr}{ds}\right)^2 = \left(\frac{E}{m}\right)^2 - \left(1 - \frac{r_g}{r}\right)\, \left(1 + \frac{L^2}{m^2 \, r^2}\right).
\eqa
To clarify the physical meaning of $E$ and $L$, let us consider the Newtonian, $r_g \ll r$, and non--relativistic, $dr/dt \ll 1$, limit of this equation. Then $dr/ds \approx dr/dt$ and:

\bqa
\frac{E^2 - m^2}{2 \, m} \approx \frac{m}{2}\, \left(\frac{dr}{dt}\right)^2 + \frac{L^2}{2\, m\, r^2} - \frac{m \, r_g}{2 r}.
\eqa
If $E^2 - m^2 = (E-m)\, (E+m) \approx 2 \, m \, {\cal E}$, where ${\cal E} = (E-m) \ll m$ is the non--relativistic total energy, and $L$ is the angular momentum, then this equation defines the trajectory of a massive particle in the Newtonian gravitational field. Thus, as it should be, the conservation of energy, $E$, follows from the invariance under time translations and the conservation of the angular momentum, $L$, follows form the invariance under rotations.

{\bf 4.} For the radial infall into the black hole, $L=0$,  eq. (\ref{eqmas}) reduces to:

\bqa\label{77}
\left(\frac{dr}{ds}\right)^2 = \left(\frac{E}{m}\right)^2 - \left(1 - \frac{r_g}{r}\right).
\eqa
Let us assume that $dr/ds \to 0$, as $r\to \infty$, i.e. the particle starts its free fall at infinity with the zero velocity. Then, as follows from (\ref{77}), $E=m$ and this equation simplifies to:

\bqa
\left(\frac{dr}{ds}\right)^2 = \frac{r_g}{r}.
\eqa
From here one can deduce that the proper time of the particle's free fall from a radius $r=R$ to the horizon $r=r_g$ is equal to:

\bqa
s\left(R \to r_g\right) = - \int_R^{r_g} \left(\frac{r}{r_g}\right)^\frac12 \, dr = \frac{2\, r_g}{3}\, \left[\left(\frac{R}{r_g}\right)^{\frac{3}{2}} - 1\right].
\eqa
The minus sing in front of the integral here is due to the fact that for the falling in trajectory ---
$dr < 0$ as $ds> 0$. Hence, as we have mentioned in the previous lectures, it takes a finite proper time for a particle or an observer to cross the black hole horizon.

At the same time, from (\ref{dtds}) it follows that $dt/ds = 1\left/\left(1 - \frac{r_g}{r}\right)\right.$, if $E=m$. Then, the ratio of $dr/ds$ by $dt/ds$ leads to

\bqa
\frac{dr}{dt} = - \left(\frac{r_g}{r}\right)^\frac12 \, \left(1 - \frac{r_g}{r}\right).
\eqa
As the result, the Schwarzschild time, which is necessary for a particle to fall from a radius $r=R \gg r_g$ to a radius $r = r_g + \epsilon$ in the vicinity of the horizon, $\epsilon \ll r_g$, is given by:

\bqa
t\left(R \to r_g + \epsilon\right) = - \int_R^{r_g + \epsilon} \, \left(\frac{r}{r_g}\right)^{\frac12} \, \frac{r \, dr}{r - r_g} \sim r_g \log \frac{R}{\epsilon}, \quad {\rm as} \quad \epsilon \to 0.
\eqa
Hence, $t\to +\infty$, as $\epsilon \to 0$, and a particle cannot approach the horizon within finite time, as measured by an observer fixed over the black hole. This again coincides with our expectations from the previous lectures.

{\bf 5.} Let us continue with the classical experimental approvals of the General Theory of Relativity.
From the seminal Newton's solution of the Kepler's problem it is known that planets orbit along ellipsoidal trajectories around stars. How does this behavior change, if relativistic corrections become relevant?

Using the above mentioned conservation laws, we can find that:

\bqa
\frac{dr}{ds} = \frac{dr}{d\varphi} \, \frac{d\varphi}{ds} = \frac{dr}{d\varphi}\, \frac{L}{m\, r^2} = - \frac{L}{m} \,\frac{du}{d\varphi},
\eqa
if the notation $u = 1/r$ is introduced. Then, the equation (\ref{eqmas}) acquires the following form:

\bqa
\left(\frac{du}{d\varphi}\right)^2 + u^2 \, \left(1 - r_g \, u\right) = \frac{E^2 - m^2}{L^2} + \frac{m^2}{L^2}\, r_g \, u.
\eqa
Differentiating both sides of this equation by $d/d\varphi$ and dividing by $du/d\varphi$, we obtain:

\bqa\label{pertN}
\frac{d^2 u}{d\varphi^2} + u = \frac{m^2 \, r_g}{2\, L^2} + \frac{3\, r_g}{2}\, u^2.
\eqa
In the Newtonian limit, $r_g/r = r_g\, u \ll 1$, the last term on the right hand side of the obtained equation is negligible. Here we discuss small deviations from the Newtonian gravitation and, hence, consider the term $\frac{3\, r_g}{2}\, u^2$, on the right hand side of (\ref{pertN}), as a perturbation.

Thus, in the Newtonian limit (\ref{pertN}) reduces to

\bqa\label{15}
\frac{d^2 u_0}{d\varphi^2} + u_0 = \frac{m^2 \, r_g}{2\, L^2}.
\eqa
The subscript ``$0$'' corresponds to the zero approximation in the perturbation over $u\, r_g \ll 1$.
The solution of this oscillator type equation is given by:

\bqa
u_0 \equiv \frac{1}{r_0} = \frac{m^2 \, r_g}{2\, L^2} \, \left(1 + e \, \cos\varphi\right).
\eqa
If the eccentricity, $e$, is less than one, then the resulting trajectory is ellipse. Substituting $u \approx u_0 + u_1$ into (\ref{pertN}) and using (\ref{15}) one can find that the first correction, $u_1$, obeys the following equation

\bqa\label{pertN1}
\frac{d^2 u_1}{d\varphi^2} + u_1 = \frac{3\, r_g}{2}\, u_0^2.
\eqa
Here:

\bqa\label{complm}
\frac{3\, r_g}{2}\, u_0^2 = \frac{3\, r_g^3 \, m^4}{8\, L^4} \, \left(1 + e \, \cos\varphi\right)^2 = \frac{3\, r_g^3 \, m^4}{8\, L^4} + \frac{3\, r_g^3 \, m^4}{4\, L^4} \, e \, \cos\varphi + \frac{3\, r_g^3 \, m^4}{8\, L^4} \, e^2 \, \cos^2\varphi = \nonumber \\ = \frac{3\, r_g^3 \, m^4}{8\, L^4}\, \left(1 + \frac{e^2}{2}\right) + \frac{3\, r_g^3 \, m^4}{4\, L^4} \, e \, \cos\varphi + \frac{3\, r_g^3 \, m^4}{16\, L^4} \, e^2 \, \cos(2\varphi).
\eqa
The contribution $\frac{3\, r_g^3 \, m^4}{8\, L^4} \, \left(1 + \frac{e^2}{2}\right)$ to $u_1$ leads to a small correction to the $\frac{m^2 \, r_g}{2\, L^2}$ factor in $u_0$, i.e. in (\ref{15}). This is just a small deviation of the length of the main axis of the ellipse, which is not a very interesting correction, because the main axis of the planetary ellipse in the Sun system cannot be measured with such an accuracy.

We are looking for the largest correction to $u \approx u_0 + u_1$ coming from $u_1$. This correction is provided by a resonant solution of (\ref{pertN1}). In this respect the last term, $\sim \cos(2\varphi)$, in (\ref{complm}) leads to a suppressed resonant contribution to $u_1$ in comparison with the term $\sim \cos\varphi$, due to the mismatch of the frequency of this ``external force'' with the frequency of the oscillator in (\ref{pertN1}).

All in all, to find the biggest correction $u_1$ we just have to find the resonant solution of the equation:

\bqa\label{19}
\frac{d^2 u_1}{d\varphi^2} + u_1 \approx \frac{3\, r_g^3 \, m^4}{4\, L^4} \, e \, \cos\varphi.
\eqa
We look for the solution of this equation in the following form:

\bqa
u_1 = A(\varphi)\, \sin\varphi + B(\varphi)\, \cos \varphi,
\eqa
where $A(\varphi)$ and $B(\varphi)$ are slow functions. Such a $u_1$ function solves (\ref{19}), if:

\bqa
A'' = B'' = 0, \quad {\rm and} \quad 2\, A' \, \cos\varphi - 2 \, B' \, \sin\varphi = \frac{3\, r_g^3 \, m^4}{4 \, L^4} \, e \, \cos \varphi.
\eqa
Wherefrom we find that $B'=0$ and $A' = \frac{3\, r_g^3 \, m^4}{8 \, L^4} \, e$. Hence, $B = const$ and the corresponding correction is not resonant. As the result, the relevant for our considerations part of $u_0 + u_1$ is as follows

\bqa\label{22}
u \approx u_0 + u_1 \equiv \frac{1}{r_{0+1}} \approx \frac{m^2 \, r_g}{2 \, L^2} \, \left(1 + e\, \cos\varphi\right) + \frac{3\,r_g^3 \, m^4 \, e}{8\, L^4}\, \varphi \, \sin\varphi.
\eqa
The last term here was obtained under the assumption that it is just a small correction, i.e. $\frac{3 \, r_g^2 \, m^2}{4 \, L^2} \, \varphi \ll 1$. Hence, eq. (\ref{22}) can be rewritten as:

\bqa
\frac{1}{r_{0+1}} \approx \frac{m^2 \, r_g}{2 \, L^2} \, \left[1 + e\, \cos\left(\varphi - \frac{3 \, r_g^2 \, m^2}{4 \, L^2} \, \varphi\right)\right].
\eqa
Now one can see that while the unperturbed trajectory is periodic in the standard sense, $r_0(\varphi) = r_0(\varphi + 2\pi)$, a particle moving along the corrected trajectory returns back after the rotation over an angle, which is different from $2\pi$

$$
r_{0+1}(\varphi) \approx r_{0+1}\left(\varphi + 2\pi + \frac{3 \, \pi \, r_g^2 \, m^2}{2 \, L^2}\right).
$$
Hence, due to relativistic effects the perihelion of a planet is rotated by the following angle

\bqa
\delta \varphi = \frac{3 \, \pi \, r_g^2 \, m^2}{2 \, L^2},
\eqa
in one its period.
Let us estimate this quantity for a circular orbit. In the latter case $L = m \, v \, R$ and $v^2 = \frac{r_g}{2\,R}$, where $R$ is the radius of the orbit and $v$ is the planet's velocity.
Hence, $\delta \varphi \approx 3\, \pi \, \frac{r_g}{R}$. At the same time for an elliptic orbit the same quantity is equal to $\delta \varphi \approx 3\, \pi \, \frac{r_g}{R(1 - e^2)}$.

If we substitute here the parameters of the Mercury orbit, which is the closest planet to the Sun and, hence, is the most sensitive to the relativistic corrections, we obtain that $\delta \varphi \approx 0,1''$. However, this is the secular effect, i.e. it grows with the number of times the planet rotates around the Sun. During one hundred years the gained angle is equal to $43''$. This angle perfectly agrees with the observational data.

\begin{figure}
\begin{center}
\includegraphics[scale=0.5]{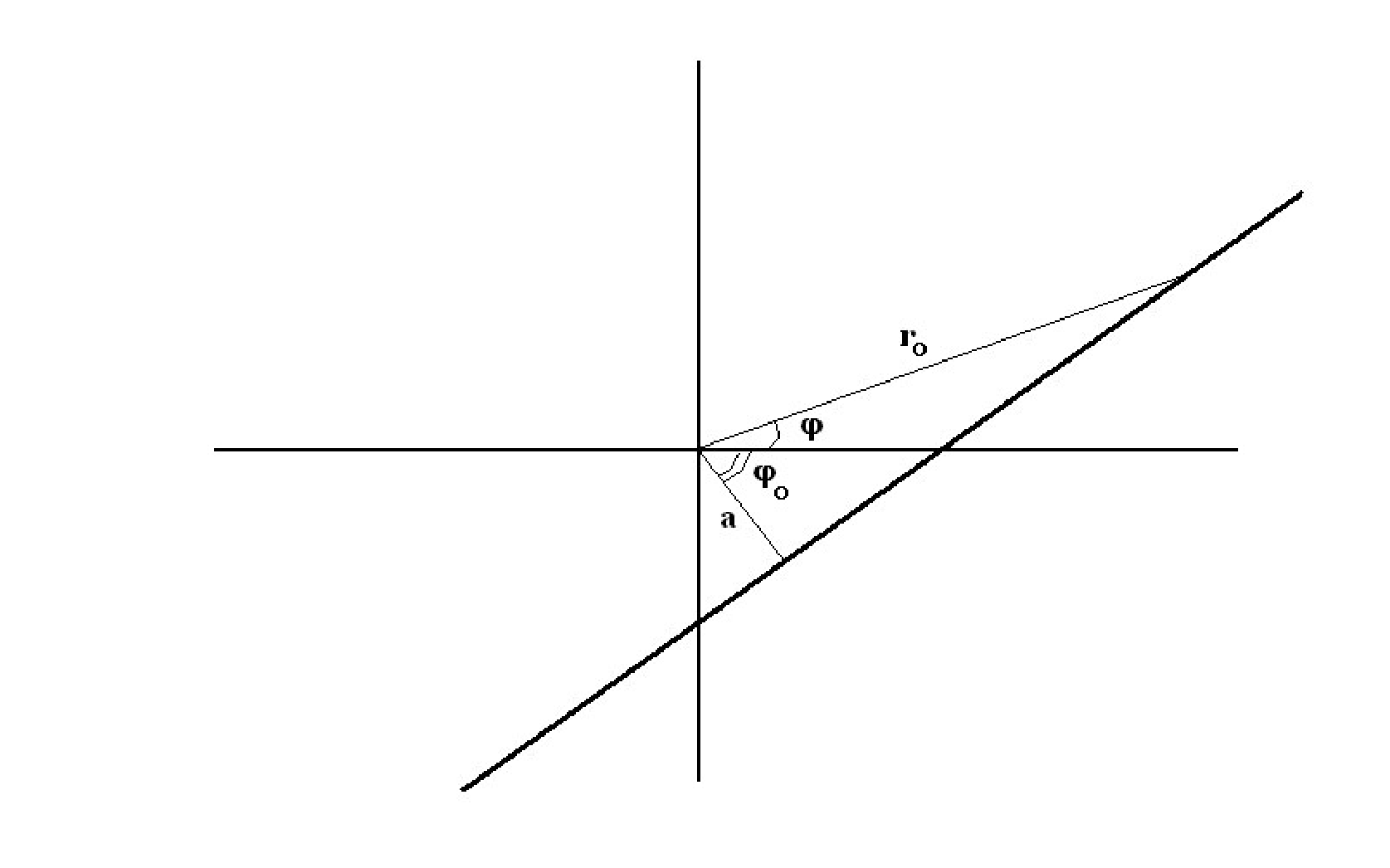}\caption{}\label{fig15}
\end{center}
\end{figure}

{\bf 6.} Let us find now the deviation angle of a light ray passing in the vicinity of a massive body. For the light $m=0$ and the equation (\ref{pertN}) reduces to:

\bqa
\frac{d^2 u}{d\varphi^2} + u = \frac{3\, r_g}{2}\, u^2.
\eqa
As before we consider the right hand side of this equation as a perturbation, $u\, r_g \ll 1$. Then, at zero approximation the equation has the following form:

\bqa
\frac{d^2 u_0}{d\varphi^2} + u_0 = 0.
\eqa
Its solution is:

\bqa
u_0 \equiv \frac{1}{r_0} = \frac{1}{a} \, \cos(\varphi + \varphi_0),
\eqa
for constant $a$ and $\varphi_0$. This expression defines a line $a = r_0 \, \cos(\varphi + \varphi_0)$ in the $(r_0, \varphi)$ plane. See the fig. (\ref{fig15}).

Acting as in the case of the massive particle, to find the first correction, $r_1$, we have to solve the equation as follows:

\bqa
\frac{d^2 u_1}{d\varphi^2} + u_1 = \frac{3\, r_g}{2\,a^2}\, \cos^2\varphi,
\eqa
where we put $\varphi_0 = 0$ for the simplicity. This equation has the following solution

\bqa
\frac{1}{r_1} = \frac{r_g}{2\, a^2} \, \left(1 + \sin^2 \varphi\right).
\eqa
Then,

\bqa
\frac{1}{r_{0+1}} = \frac{1}{a} \, \cos\varphi + \frac{r_g}{2\, a^2} \, \left(1 + \sin^2 \varphi\right).
\eqa
As the result, if the unperturbed solution has such a property that $r_0\left(\varphi = \pm\frac{\pi}{2}\right) = \infty$, the perturbed one has the feature that $r_{0+1}\left[\varphi = \pm \left(\frac{\pi}{2} + \alpha\right)\right] = \infty$, where $\alpha$ solves the following equation:

\bqa
-\frac{1}{a} \, \sin\alpha + \frac{r_g}{2\, a^2}\, \left(1 + \cos^2\alpha\right) = 0.
\eqa
Assuming that $\alpha$ is very small, we find that $\alpha \approx \frac{r_g}{a}$.

\begin{figure}
\begin{center}
\includegraphics[scale=0.5]{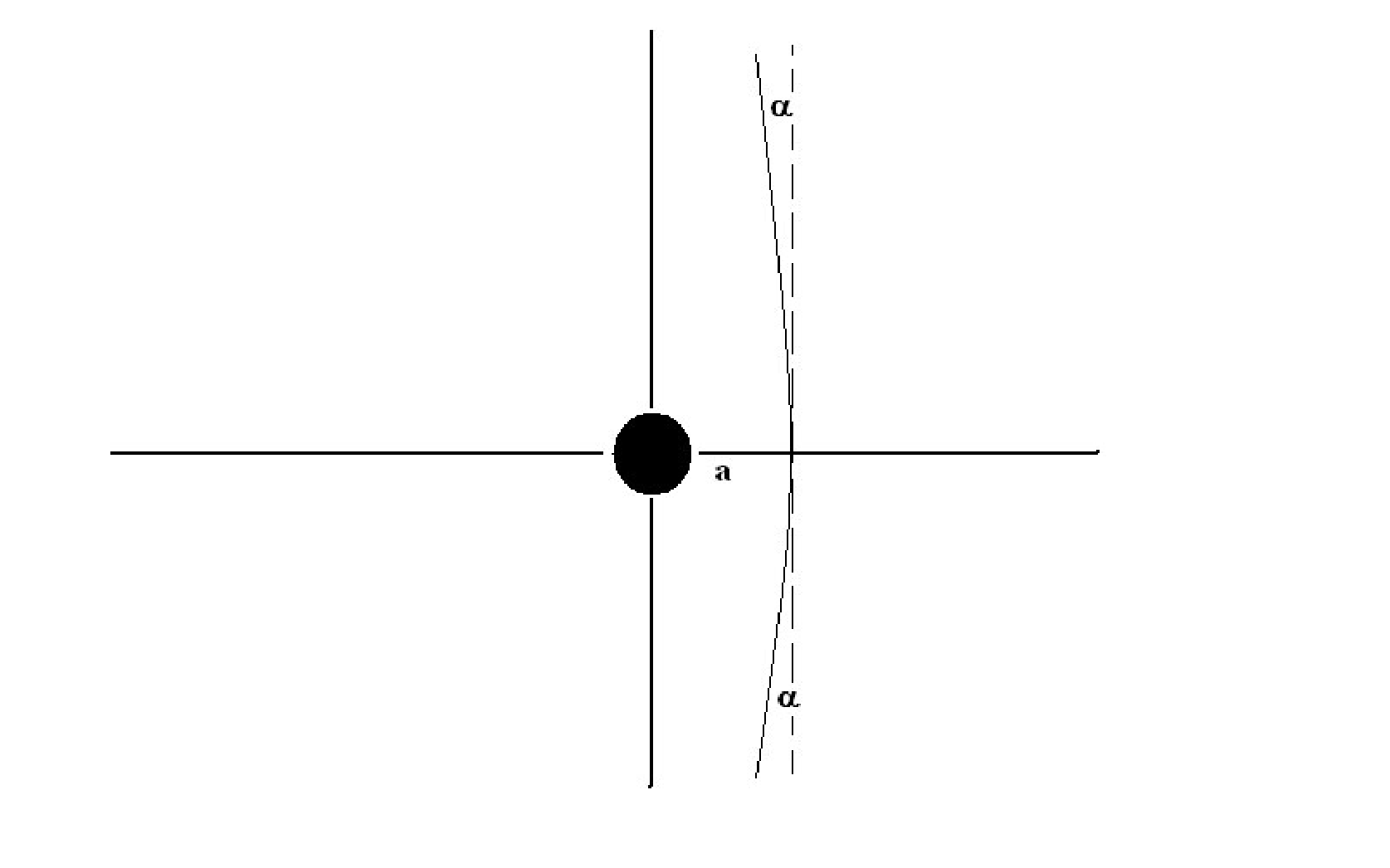}\caption{}\label{fig16}
\end{center}
\end{figure}

From the fig. (\ref{fig16}) one can deduce that the total light ray deviation angle is equal to $\delta\varphi = 2\,\alpha \approx \frac{2\, r_g}{a}$. For the case of the Sun its actual radius is $a \approx 7 \times 10^8 m$ and its gravitational radius is $r_g \approx 3 \times 10^3 m$. As the result, $\delta \varphi \approx 1,75''$, which perfectly agrees with the measured value.

\vspace{10mm}

\centerline{\bf Problems:}

\vspace{5mm}

\begin{itemize}

\item Show on formulas that after the crossing of the horizon $r=r_g$ an observer or a particle inevitably will fall into the singularity $r=0$. Use the ingoing Eddington--Finkelstein coordinates.

\item Show that in the Schwarzschild metric there are circular light like orbits at $r = 3r_g/2$.

\item Show that from eq. (\ref{pertN}) it follows that the deviation angle does not depend on $m$.

\end{itemize}

\vspace{10mm}

\centerline{\bf Subjects for further study:}

\vspace{5mm}

\begin{itemize}

\item Pound--Rebka experiment and other classic experiments in General Relativity.

\item How does Global Positioning System or Glonass system works? (``Relativity in the Global Positioning System'', Neil Ashbey, Living Reviews in Relativity, 6 (2003) 1.)

\item Particle orbits and absorbtion cross--sections for black holes.

\end{itemize}

\newpage

\section*{LECTURE VII \\ {\it Energy--momentum tensor for a perfect relativistic fluid. Interior solution. Kerr's rotating black hole. Proper time. Gravitational redshift. Concise comments on Cosmic Censorship Hypothesis and on the No Hair Theorem.}}

\vspace{10mm}

{\bf 1.} In this lecture we find a solution of Einstein equations which describes a stable star --- an ideal ball of matter surrounded by vacuum. This is referred to as \underline{interior solution}. Inside the ball we assume that the energy--momentum tensor is not zero, $T_{\mu\nu} \neq 0$, but it is homogeneous and spherically symmetric. At the same time outside the ball there is vacuum, i.e. $T_{\mu\nu} = 0$. And everywhere $\Lambda = 0$.

To have a stable star, there should be a matter inside it, which is capable to create a strong enough pressure to prevent its collapse. Hence, $T_{\mu\nu}$ of a dust, which was considered in the previous lectures, is not sufficient. The appropriate energy momentum tensor is given by

\bqa\label{EMTpf}
T_{\mu\nu}(\underline{x}) = \rho(\underline{x}) \, u_\mu(\underline{x}) \, u_\nu(\underline{x}) + p(\underline{x}) \, \left[u_\mu(\underline{x})^{\phantom{\frac12}} u_\nu(\underline{x}) - g_{\mu\nu}(\underline{x})\right],
\eqa
where $g_{\mu\nu}$ is the metric tensor and $\underline{x} = \left(t, \vec{x}\right)$ is the coordinate four--vector. We start with providing arguments that this is the \underline{energy momentum tensor} \underline{for a relativistic perfect fluid} and $\rho$ is the energy density, while $p$ is the pressure. In fact, in the comoving reference system, where $g_{\mu\nu} \approx \eta_{\mu\nu}$ and the four--velocity  is $u^\mu = \left(1,0,0,0\right)$, one obtains that

\bqa
\left|\left|T^\mu_\nu\right|\right| = Diag\left(\rho, - p, - p, - p\right), \quad T^\mu_\nu = g^{\mu\alpha} \, T_{\alpha\nu}.
\eqa
Already this form of the tensor can reveal the physical meaning of $\rho$ and $p$, as follows from the course of classical electrodynamics.

Furthermore, the condition of the conservation of this energy momentum tensor is as follows:

\bqa\label{cons1}
0 = D^\mu\, T_{\mu\nu} = \left(\rho + p\right) \, u_\mu \, D^\mu u_\nu + D^\mu\left[(\rho + p)\, u_\mu\right]\, u_\nu - D_\nu p.
\eqa
Multiplying this equality by $u^\nu$ and using that $u^\nu \, u_\nu = 1$ (and, hence, $u^\nu \, D^\mu u_\nu = 0$), we arrive at the following equation:

\bqa\label{cont1}
D^\mu\left[\left(\rho + p\right)\, u_\mu\right] - u^\mu \, D_\mu p = D^\mu\left[\rho \, u_\mu\right] + p\, D^\mu\, u_\mu = 0.
\eqa
The last equality is just a covariant generalization of the continuity equation in the flat space--time: $\partial^\mu \left[\rho \, u_\mu\right] = \dot{\rho} + div \left[\rho \vec{v}\right] = 0.$ Note that in the non--relativistic limit the pressure is always much less than the energy density, $p \ll \rho$, for the same reason as why non--relativistic kinetic energy is much smaller than the rest energy.

Using (\ref{cons1}) and (\ref{cont1}), we find the following equation:

\bqa
\left(\rho + p\right)\, u_\mu \, D^\mu u_\nu = \left(\delta^\alpha_\nu - u^\alpha \, u_\nu\right)\, D_\alpha p,
\eqa
Here $\left|\left|\delta^\alpha_\nu - u^\alpha \, u_\nu\right|\right| = Diag(0,1,1,1)$ --- is the projector on spatial directions. Hence, this is a relativistic, covariant extension of the Euler equation for a liquid without viscosity:

\bqa
\rho \left[\partial_t \vec{v} + \left(\vec{v}, \vec{\nabla}\right)\, \vec{v}\right] = - \vec{\nabla}p.
\eqa
The last equation is just the Newton's second law, $m \, \vec{a} = \vec{F}$, both sides of which are attributed per unit volume. In fact,

$$\vec{a}\left[t,\vec{x}(t)\right] \equiv \frac{d \vec{v}\left[t,\vec{x}(t)\right]}{dt} = \partial_t \vec{v} + \left(\partial_i \vec{v}\right) \, \left(\partial_t x_i\right) = \partial_t \vec{v} + \left(\vec{v}, \vec{\nabla}\right)\, \vec{v},$$
and recall that the pressure $p$ is just a force attributed per unit area. The Euler equation can be rewritten in the following form:

\bqa
\partial_t \left(\rho v_i\right) + \partial_j \, T_{ij} = 0
\eqa
with the use of the tensor $T_{ij} = \rho \, v_i \, v_j + p \, \delta_{ij}$.

All in all, the given above arguments should convince reader that (\ref{EMTpf}) is the energy momentum tensor for a relativistic perfect fluid. To specify the fluid one just has to choose an \underline{equation of state} --- $p = p(\rho)$, i.e. the relation between the pressure and density of the fluid.

{\bf 2.} Thus, we are looking for a time--independent, spherically symmetric and regular everywhere solution of the Einstein equations with the energy--momentum tensor, which is given by (\ref{EMTpf}) with $\rho = \rho(r)$ and $p = p(r)$ for $r \leq R$, and $\rho=0=p$ for $r > R$, where $R$ is the radius of the star. Then the appropriate ansatz for the metric should be as in the lecture IV:

\bqa\label{starsol}
ds^2 = e^{\nu(r)}\, dt^2 - e^{\lambda(r)} \, dr^2 - r^2\, d\Omega^2,
\eqa
where $\nu$ and $\lambda$ are functions of $r$ only, because we are looking for a static and spherically symmetric solution. For such a metric we obtain the Einstein equations that are following from those found in the lecture IV:

\bqa\label{Eeqr}
8 \, \pi \, \kappa \, \rho(r) = \frac{e^{-\lambda}}{r} \, \lambda' + \frac{1}{r^2}\, \left(1 - e^{-\lambda}\right), \nonumber \\
- 8 \, \pi \, \kappa\, p(r) = - \frac{e^{-\lambda}}{r} \, \nu' + \frac{1}{r^2}\, \left(1 - e^{-\lambda}\right), \nonumber \\
- 8 \, \pi \, \kappa\, p(r)  = - \frac{e^{-\frac{\nu + \lambda}{2}}}{2}\, \frac{d}{dr} \left[ e^{\frac{\nu - \lambda}{2}} \, \nu'\right] - \frac{e^{-\lambda}}{2\, r} \, \nu' + \frac{e^{-\lambda}}{2\,r} \, \lambda'.
\eqa
Here $\nu' \equiv d\nu/dr$ and $\lambda' \equiv d\lambda/dr$. The rest of the components of the Einstein equations either follow from (\ref{Eeqr}) or lead to the trivial relations, stating that $0=0$.

The first equation in (\ref{Eeqr}) contains only $\lambda$ and can be rewritten as:

\bqa
\frac{1}{r^2} \, \frac{d}{dr}\left[r\, \left(1 - e^{-\lambda}\right)\right] = 8\, \pi \, \kappa \, \rho(r).
\eqa
Its solution is as follows:

\bqa\label{M(r)}
e^{- \lambda(r)} = 1 - \frac{2 \, \kappa \, M(r)}{r}, \quad {\rm where} \nonumber \\
M(r) = 4\,\pi \, \int_0^r dx \, x^2 \, \rho(x) + c,
\eqa
and $c$ is an integration constant. To obtain a solution, which is regular at $r=0$, we have to choose $c=0$. Also to avoid having a horizon --- a surface, beyond which coordinate $r$ ceases to be space--like and becomes time--like --- we have to have that everywhere $e^{-\lambda(r)} > 0$. Hence, there should be $r > 2\, \kappa \, M(r)$, which imposes a restriction on the density $\rho(r)$.

As we have assumed above, $\rho(r) = 0$, if $r>R$. In this region we expect that the solution (\ref{starsol}) acquires the form of the Schwarzschild one. And indeed it does that with the mass equal to $M \equiv M(R) = 4\, \pi \, \int_0^R dr \, r^2 \, \rho(r)$. But the real mass of the star should be

$$M_{real} = \int_{B} d^3V \, \sqrt{g^{(3)}} \, \rho(r) = \int_0^R dr \, \int_0^{\pi}d\theta \, \int_0^{2\pi} d\varphi \, \sin\theta \, r^2 \, e^{\lambda/2} \, \rho(r) = 4 \, \pi \, \int_0^R \frac{dr \, r^2 \, \rho(r)}{\sqrt{1 - \frac{2\, \kappa \, M(r)}{r}}},$$
where $B$ is the body of the star, $d^3V\sqrt{g^{(3)}}$ is the volume form of the three--dimensional time--slice, following from the metric (\ref{starsol}). Correspondingly $g^{(3)}$ is the determinant of the three--dimensional spatial part of the metric (\ref{starsol}). Thus, the difference $E_{bin} = M_{real} - M > 0$ is the gravitational binding energy.

Coming back to (\ref{Eeqr}), from the second equation we find that

\bqa\label{Newtpoten}
\frac{d\nu}{dr} = \frac{2\, \kappa \, M(r) + 8\,\pi \, r^3 \, \kappa \, p}{r \, \left[r - 2\, \kappa\, M(r)\right]}.
\eqa
In the Newtonian limit, $p\ll \rho$ and $\kappa \, M(r) \ll r$, this equation reduces to
$d\nu/dr \approx 2\, \kappa\, M(r)/r^2$, i.e. to the spherically symmetric version of the Poisson equation for the gravitational potential. Thus, in the case of the static, spherically symmetric gravitational field, $\nu/2$ has the meaning of the relativistic analog of the Newtonian potential.

Instead of the last equation in (\ref{Eeqr}) it is convenient to use the condition of the conservation of the energy--momentum tensor, $D^\mu \, T_{\mu\nu} = 0$\footnote{As we have explained in the Lecture III this equation and the Einstein equations are not independent.}. In our case it reduces to:

\bqa
2\, \frac{dp}{dr} = - \left(p + \rho\right)\, \frac{d\nu}{dr}.
\eqa
Hence, using eq. (\ref{Newtpoten}) here, we obtain:

\bqa\label{ToOpVo}
\frac{dp}{dr} = - \left(p + \rho\right)\, \frac{\kappa \, M(r) + 4\,\pi \, r^3 \, \kappa \, p}{r \, \left[r - 2\, \kappa\, M(r)\right]},
\eqa
which is the so called \underline{Tolman--Oppenheimer--Volkoff equation} for the hydrodynamic equilibrium of the star. In the Newtonian limit, $p \ll \rho$ and $\kappa\, M(r) \ll r$, this equation reduces to:

\bqa
\frac{dp}{dr} \approx - \rho \, \frac{\kappa \, M(r)}{r^2}.
\eqa

{\bf 3.} In conclusion, we have found that the metric for a homogeneous, static, spherically symmetric and non--rotating star is as follows:

\bqa
ds^2 = e^{\nu(r)} \, dt^2 - \frac{dr^2}{1 - \frac{2\, \kappa\, M(r)}{r}} - r^2\, d\Omega^2,
\eqa
where $M(r)$ is defined in (\ref{M(r)}), while $\nu(r)$ follows from (\ref{Newtpoten}). At the same time, the necessary and sufficient condition of the star equilibrium is given by the equation (\ref{ToOpVo}).

Outside the star body $p=0$ and $\rho=0$, hence, one can solve (\ref{Newtpoten}) to find that $e^\nu = 1 - \frac{2\,\kappa\, M(R)}{r}$. As the result, outside the star surface, $r\geq R$, the solution under consideration reduces to the Schwarzschild one, as expected due to the Birkhoff theorem.

To solve the equations (\ref{M(r)}), (\ref{Newtpoten}) and (\ref{ToOpVo}) explicitly one has to specify the equation of state $p = p(\rho)$. The simplest model corresponds to the incompressible fluid, when $\rho = \rho_0 = const$. Then $M(r) = \frac{4\, \pi}{3}\, r^3 \, \rho_0$ for $r<R$ and the solution of (\ref{ToOpVo}) is as follows:

\bqa
p(r) = \rho_0 \, \frac{\sqrt{1 - \frac{2\, \kappa\, M}{R}} - \sqrt{1 - \frac{2\, \kappa \, M \, r^2}{R^3}}}{\sqrt{1 - \frac{2\, \kappa\, M \, r^2}{R^3}} - 3 \, \sqrt{1 - \frac{2\, \kappa \, M}{R}}}, \quad M = \frac{4\pi}{3}\, R^3 \, \rho_0.
\eqa
From this equation it is not hard to see that the central pressure $p(r=0)$ blows up to infinity if $R = \frac{9}{4}\, \kappa \, M$. That simply means that stars with $M > M_{max} = \frac{4}{9\kappa}\, R$ cannot exist, if their matter consists of the incompressible fluid.

\begin{figure}
\begin{center}
\includegraphics[scale=0.5]{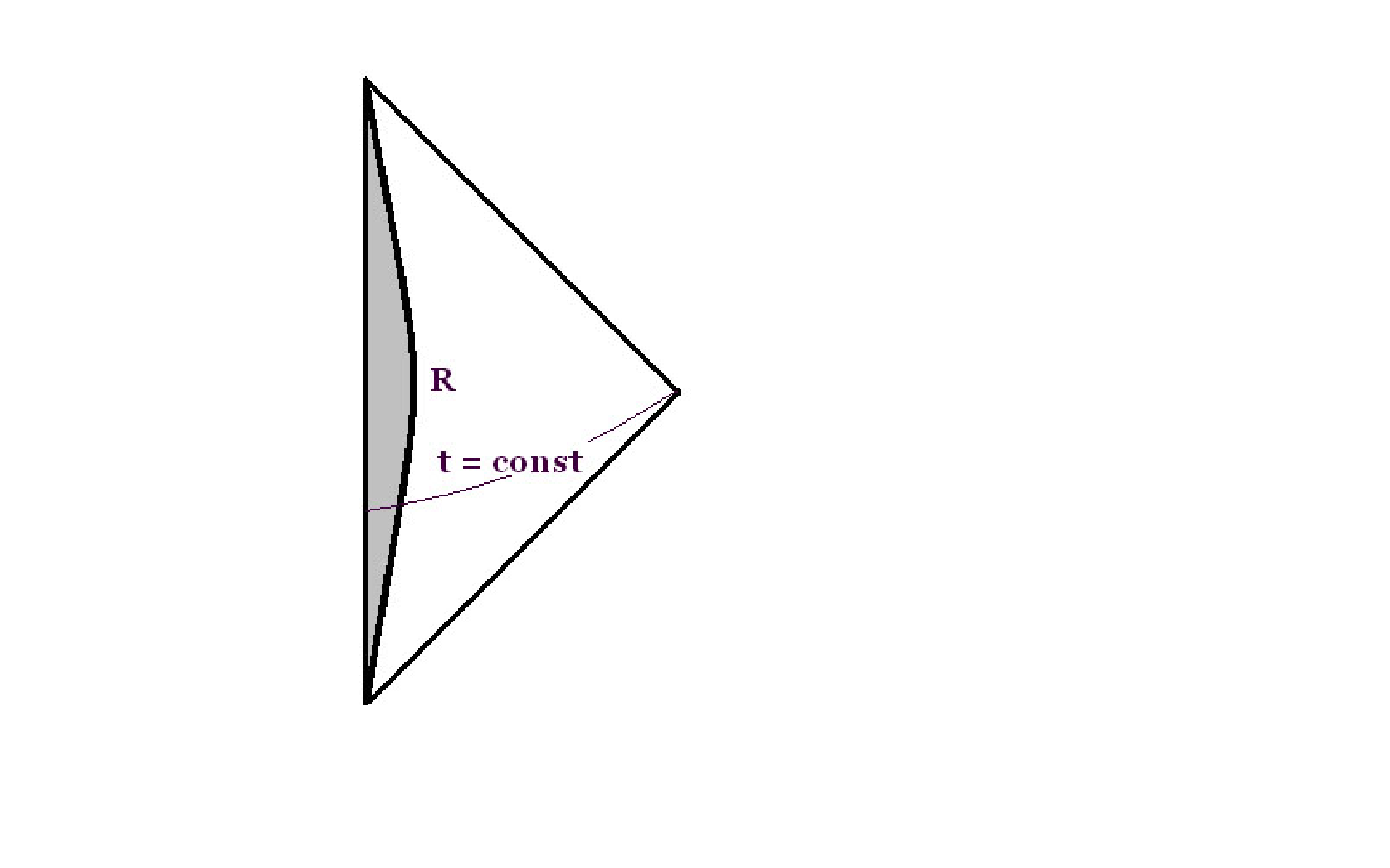}\caption{}\label{fig13}
\end{center}
\end{figure}

{\bf 4.} To draw the corresponding Penrose--Carter diagram it is convenient to choose the relevant $(t,r)$ part of the space--time and to make the following coordinate change:

\bqa
e^{\nu(r)} \, dt^2 - \frac{dr^2}{1 - \frac{2\, \kappa\, M(r)}{r}} = e^{\nu(r)} \, \left[dt^2 - \frac{dr^2}{e^{\nu(r)}\, \left[1 - \frac{2\, \kappa\, M(r)}{r}\right]}\right] =  e^{\nu(r)} \, \left[dt^2 - d\bar{r}^2_*\right], \nonumber \\
{\rm where} \quad d\bar{r}_*^2 \equiv \frac{dr^2}{e^{\nu(r)}\, \left[1 - \frac{2\, \kappa\, M(r)}{r}\right]}.
\eqa
Here $\bar{r}_*$ is ranging from $0$ to $+\infty$, unlike tortoise $r_*$, which is ranging from $-\infty$ to $+\infty$. That is because the denominator in the definition of $d\bar{r}_*$ does not have a pole for
any $r$, due to the regularity of the solution under consideration. The range of values of $\bar{r}_*$ can be adjusted to $[0,+\infty)$ by a suitable choice of the integration constant in the relation between $\bar{r}_*$ and $r$.

As the result, if we make the same transformation for $t\pm \bar{r}_*$ as the one which was done at the beginning of the lecture V, we arrive at the diagram shown on the fig. (\ref{fig13}). The grey region on this picture shows the star's body. The resulting picture is just a half of the Minkowskian Penrose--Carter diagram, which was shown at the  beginning of the lecture V. It is similar to the diagram of the $(t,r)$ part of the Minkowski space--time in the spherical coordinates.

{\bf 5.} We continue our consideration of collapse and star solutions with the concise discussion of the most general stationary,  vacuum solution of the Einstein equations. This is the so called \underline{Kerr's rotating black hole}. With the use of the so called \underline{Boyer--Lindquist coordinates} its metric can be represented in the following form:

\bqa\label{KerrBL}
ds^2 = \frac{\Delta - a^2 \, \sin^2 \theta}{\rho^2}\, dt^2 &+& \frac{4\, \kappa \, M \, a}{\rho^2} \, r\, \sin^2\theta \, d\varphi \, dt - \frac{\rho^2}{\Delta}\, dr^2 - \rho^2 \, d\theta^2 - \frac{A\, \sin^2\theta}{\rho^2}\, d\varphi^2, \nonumber \\
{\rm where} \quad \Delta &=& r^2 - 2\,\kappa \, M \, r + a^2, \nonumber \\
\rho^2 &=& r^2 + a^2 \, \cos^2\theta, \nonumber \\
A &=& \left(r^2 + a^2\right)^2 - a^2 \, \Delta \, \sin^2\theta.
\eqa
This metric is stationary but not static. In fact, components of the metric tensor are time independent. However, the line element (\ref{KerrBL}) is not invariant under time inversion $t\to - t$, due to the presence of the non--diagonal term $dt\, d\varphi$.

While Schwarzschild black hole depends on one parameter $M$, the Kerr solution depends on two parameters $M$ and $a$. It reduces to the Schwarzschild black hole if $a=0$. In the asymptotic spatial infinity, $r \to \infty$, Kerr's metric simplifies to:

\bqa
ds^2 \approx \left(1 - \frac{2\, \kappa \, M}{r}\right)\, dt^2 + \frac{4\, \kappa \, M \, a \, \sin^2\theta}{r}\, dt \, d\varphi - \left(1 + \frac{2\, \kappa\, M}{r}\right)\, dr^2 - r^2 d\Omega^2.
\eqa
This is the week gravitational field, which describes massive rotating body with the angular momentum $J = M \, a$, which explains the physical meaning of the parameter $a$. In fact, the presence of the $dt d\varphi$ term in this line element is responsible for rotation, which can be understood even from the transformation to the rotating reference system in flat space. (See the corresponding paragraph of Landau--Lifshitz for more details.)

It is not hard to notice that $g_{11}$ component of metric is singular when $\Delta = 0$. The corresponding quadratic equation has two solutions $r_\pm = \kappa \, M \pm \sqrt{\left(\kappa\, M\right)^2 - a^2}$. As $a \to 0$ one can find that $r_+ \to r_g$, while $r_- \to 0$. This is the hint that $r_+$ is the event horizon. Explicitly this fact can be seen, e.g., from the corresponding Penrose--Carter diagram, drawing of which goes beyond the scope of our lectures. Another option is to see that the normal vector, $n_\mu$, to the surface $r=r_+$ is light--like, $n_\mu n^\mu = 0$ (which is left as an exercise for the reader). If the normal vector is light--like then one has to exceed the speed of light to cross the corresponding surface in one of the directions. as can be seen on the example of the light--cone.

\begin{figure}
\begin{center}
\includegraphics[scale=0.5]{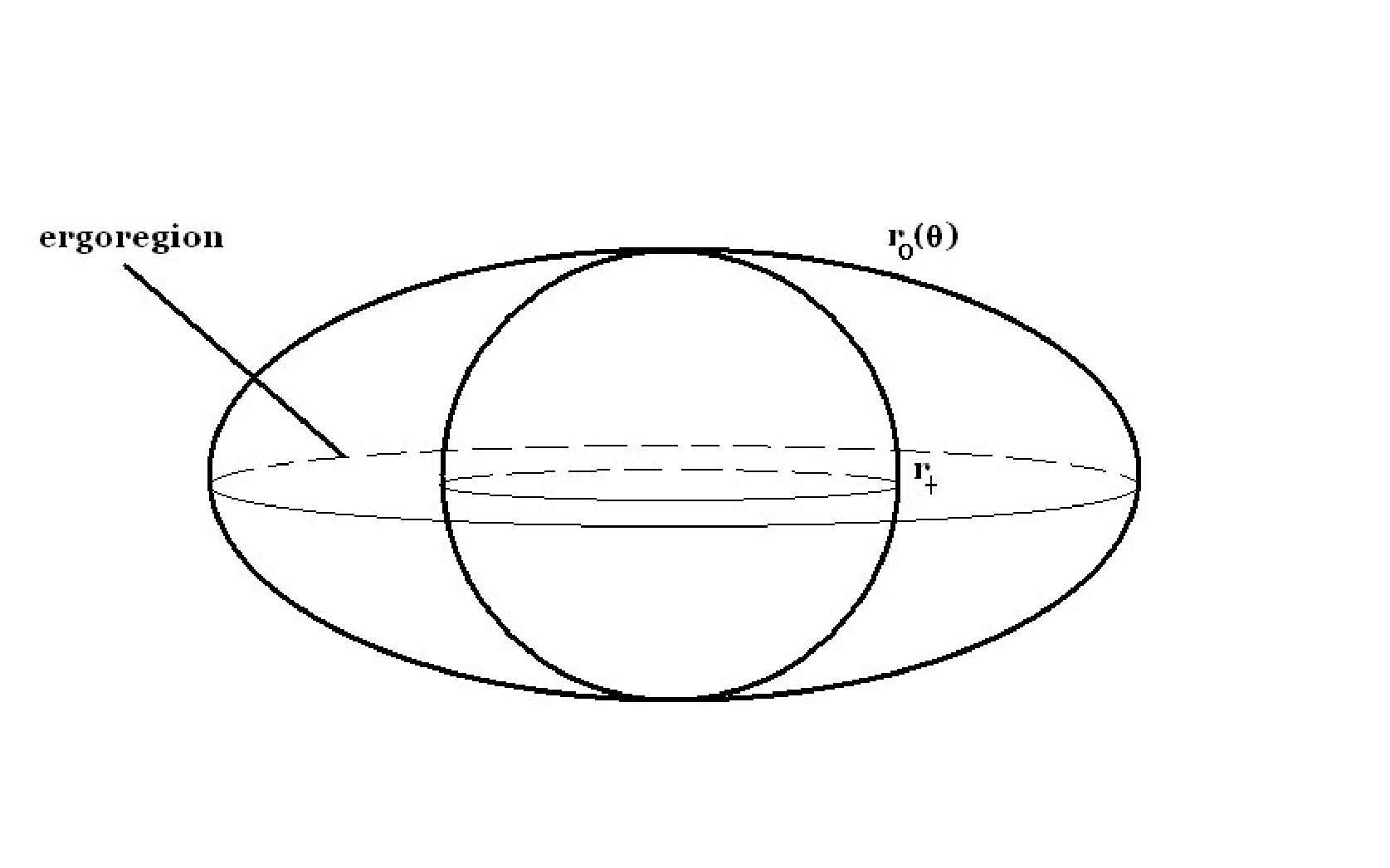}\caption{}\label{fig14}
\end{center}
\end{figure}

Furthermore, the condition $g_{00} = 0$ has two solutions. The outer one, $r_{0}(\theta) = \kappa \, M + \sqrt{\left(\kappa \, M\right)^2 - a^2 \, \cos^2\theta}$, defines the boundary of the so called \underline{ergoregion}.
As the result one encounters the picture shown on the fig. (\ref{fig14}): the event horizon is surrounded by the ergoregion. If an observer finds himself inside the ergoregion, he in principle can escape to spatial infinity. This can be seen, e.g., from the corresponding Penrose--Carter diagram or from the fact that the normal vector to the ergosphere $r_0(\theta)$ is {\it not} light--like.

{\bf 6.} Now it is worth pointing out that equations $\Delta = 0$ and $g_{00} = 0$ have real solutions only if $\kappa \, M > a$. Otherwise the singularity of the space--time under consideration is not hidden behind the event horizon. This is so called \underline{naked singularity}. In the latter case the singularity will be reachable within finite time as measured by any distant observer. Moreover, it can be shown that in this case there are closed time--like curves. Such a peculiar situation seems to be physically unacceptable. As the result scientific community believes in the so called \underline{Cosmic censorship hypothesis}.

Vague formulation of this hypothesis is as follows. Einstein equations can have many spurious (unphysical) solutions, e.g., having naked singularities (not surrounded by event horizons) or closed time--like loops. But physicists believe that in real situations black holes are results of collapse processes. Hence, it is natural to expect that during these processes any physically measurable quantity remains regular. The question is, however, whether one can achieve this within the theory of General Relativity or one has to modify it? In any case we expect that naked singularities cannot be created and, e.g., the Kerr solution is physically meaningful only if $\kappa \, M \geq a$. The solution with $\kappa \, M = a$ is referred to as \underline{critical}.

Another important property of black holes is provided by the so called \underline{No Hair Theorem}. The proof of this theorem goes beyond the scope of our lectures, but in simple terms it can be explained as follows. It states that the final stage of a generic collapse process is given by a stationary Kerr solution, if the total electric charge of the black hole is zero.

The point is that black hole horizon is a surface of the infinite redshift --- from any frequency to its zero value. To explain this fact, let us define what means proper time for the curved space--times. If $c=1$, proper time spend by an observer is equal to the length of its world--line: $z^\mu(t) = \left[t, \vec{z}(t)\right]$ and $d\tau = \sqrt{g_{00} dt^2 + g_{0i} dt dz^i + g_{ij} dz^i dz^j}$. If the observer is at rest, then $d\tau = \sqrt{g_{00}} dt$. Consider now two stationary observers, which are placed at positions $r_1$ and $r_2$ in a time independent metric. Suppose one of the observers emits two signals towards the other one. The coordinate time separation between the emission of the signals is equal to $\Delta t$. Because the metric is time independent, the coordinate time separation between the arrival of the signals is also $\Delta t$. Then, the proper time separations of the signals for the two observers are as follows $\frac{\Delta \tau_1}{\sqrt{g_{00}(r_1)} } = \Delta t = \frac{\Delta \tau_2}{\sqrt{g_{00}(r_2)}}$. But the measured frequency is related to the clock rate as $\omega \sim 1/ \Delta \tau$. Hence,

\bqa
\omega_2 = \omega_1\, \sqrt{\frac{g_{00}(r_1)}{g_{00}(r_2)}}.
\eqa
This effect can be understood on general physical grounds. In fact, a photon performs
a work to climb out from the gravitational attraction of the massive center. Then, if $r_2 > r_1$ in Scwharzschild metric, photon's energy and, hence, frequency should reduce, $\omega_2 < \omega_1$.

Now if $r_1 \to r_g$, then $\omega_2 \to 0$ independently of the value of $r_2$ and $\omega_1$.
Namely, if someone creates a radiation just behind the horizon, it cannot be seen even just outside the horizon. In fact, whatever happens behind the horizon, observers outside it, resting at any radius, still see only stationary fields due to the infinite redshift. And anything that is radiated under the horizon, remains under it.

This observation, probably, will help to acquire an intuitive understanding of the following picture. Obviously during a collapse process (which is discussed in grater details in the next lecture), if there is no any spherical symmetry, there are plenty of radiation processes. However, as the matter of the collapsing star, the source of the radiation under discussion, approaches the corresponding Schwarzschild radius, this radiation experience higher and higher redshift. As the result, independently of the frequency of the initial radiation, which was created in the vicinity of the horizon surface, distant observers see fields which are closer and closer to the stationary ones, as the collapse process is approaching its termination. Such a situation reveals itself via the following effect: All intensities for multi--pole radiations are falling as inverse powers of time.

All in all, during the collapse process, everything which can be radiated is indeed radiated away and the outside observes eventually detect only stationary fields. As we will see in the lectures IX and X, to have a gravitational radiation there should be at least quadruple moment changing in time (or higher momenta). Also it is known that to have an electromagnetic radiation there should be at least dipole moment changing in time. Thus, monopole electric moment (i.e. electric charge), monopole gravitational moment (i.e. mass) and dipole gravitational moment (i.e. angular momentum) are capable to provide only stationary fields. It is this content that can remain for the gravitational and electromagnetic fields at the end of the collapse process. As the result, the most generic black hole, in the presence of the electromagnetic fields, can carry only three parameters --- mass, angular momentum and electric charge. Note that to create a radiation of scalar fields it is sufficient to have a monopole moment (i.e. the corresponding charge) changing in time. As the result black holes cannot carry ``scalar hairs''. This completes the intuitive explanation of the essence of the No Hair theorem.

\vspace{10mm}

\centerline{\bf Problems:}

\vspace{5mm}

\begin{itemize}

\item Show that for the Kerr solution the normal vector to the surface $r=r_+$ is light--like. (See the corresponding paragraph in Landau--Lifshitz.)

\item Show that one can escape from the ergoregion.

\end{itemize}

\vspace{10mm}

\centerline{\bf Subjects for further study:}

\vspace{5mm}

\begin{itemize}

\item Criteria for the star stability. Chandrasekhar limit. (Stars and relativity,
Y.B. Zel'dovich and I.D. Novikov, University of Chicago Press, 1971)

\item Penrose--Carter diagram for Kerr solution.

\item No Hair Theorem.

\item Cosmic censorship hypothesis.

\item Membrane paradigm. Black hole horizon as a viscous liquid having electric conductivity. Black hole thermodynamics

\item Zeldovich--Penrose effect. Energy extraction from the ergoregion.

\item Multi--pole radiation during the collapse process. (``Nonspherical perturbations of relativistic gravitational collapse. I. Scalar and gravitational perturbations'', R.H. Price, Phys. Rev. D 5 (1972) 2419)

\end{itemize}

\newpage

\section*{LECTURE VIII \\ {\it Oppenheimer--Snyder collapse. Concise comments on the origin of the thermal nature of the Hawking's radiation
and on black hole creation.}}

\vspace{10mm}

{\bf 1.} In this lecture we continue the discussion of the physics underlying the Schwarzschild geometry. Namely we find here the so called \underline{Oppenheimer--Snyder} solution, which describes the collapse process of a spherically symmetric, non--rotating body into the black hole.

Consider a static star (a ball filled with a matter) surrounded by empty space. The star is static due to an internal pressure until some moment of time $t=0$ and then the pressure is switched off. Say before $t=0$ there were some thermonuclear processes inside this star, which were producing the internal pressure. But by the moment $t=0$ the entire thermonuclear fuel was
used out.

To model the collapse process after $t=0$ we assume that inside the star there is a homogeneous {\it pressureless} dust. Also we assume that the original star was ideal ball with ideal spherical surface and that the collapse process goes in such a way that homogeneity of the matter inside the star and the spherical symmetry is never violated. This is a highly unstable situation because any its perturbation violating these symmetries will grow in time due to the tidal forces. We neglect such perturbations.

Thus, inside the ball the energy momentum tensor is $T_{\mu\nu} = \rho(\tau) u_\mu u_\nu$, where the density $\rho(\tau)$ is just a function of time $\tau$ because of the spatial homogeneity. Outside the ball we have vacuum, $T_{\mu\nu} = 0$.

Such a massive ball of matter should create a spherically symmetric and asymptotically flat gravitational field in vacuum. Thus, due to Birkhoff theorem outside the ball the metric has to be the Schwarzschild one. Any violation of the spherical symmetry will lead to time--dependent gravitational fields in vacuum, i.e. to a creation of gravitational waves, which are discussed in the lectures IX and X. Neglecting such processes is exactly the approximation that we use here.

To give an intuitive explanation why spherically symmetric collapse does not create gravitational radiation, let us discuss the following situation. Consider a ball which is electrically charged and the charge is homogeneously
distributed over its volume. Suppose now that for some reason this ball rapidly shrinks in such a way that the homogeneity and spherical symmetry are respected. It is not hard to see that independently of the radius of the ball it creates the same Coulomb field outside itself. This is related to the uniqueness of the Coulomb solution of the Maxwell equations, which was mentioned in the lecture IV.

Thus, the magnetic field outside the ball is vanishing and such an accelerated motion of the charge does not create an electromagnetic radiation. The point is that to have a radiation there should be at least dipole moment, which is changing in time, while in the case of the ball all momenta are zero with respect to its center. It happens that to have a gravitational radiation, as we will see in the lectures IX and X, there should be even quadruple moment, which changes in time. Dipole moment changing in time is not enough for the generation of the gravitational radiation. Also from the Newton's gravitation we know that an ideal spherical massive ball creates the same potential outside itself independently of its radius. The form of the potential just depends on the mass of the ball. In general relativity the situation is similar due to the Birkhoff's theorem.

{\bf 2.} All in all, the metric outside the ball is

\bqa
ds_+^2 = \left(1 - \frac{r_g}{r}\right) \, dt^2 - \frac{dr^2}{1 - \frac{r_g}{r}} - r^2 \, d\Omega^2.
\eqa
Here $r_g = 2\kappa M$, where $M$ is the mass of the ball, which remains constant during the collapse process, because of the absence of the radiation. Intuitively it should be clear that then there is no energy which fluxes away to infinity with the radiation and the energy of the ball remains constant.

This metric is valid outside of the surface of the ball $\Sigma$, which radially shrinks down during the collapse process. Then the world--hypersurface of $\Sigma$ is $z^\mu(\tau) = \left[T(\tau), R(\tau)\right]$ and at every given time slice it is an ideal sphere. (Note that in this lecture we denote by the same $\Sigma$ both the three--dimensional world--hypersurface and its two--dimensional time--slices.) Hence, $\Sigma$ occupies all the values of the spherical angles $\theta$ and $\varphi$. Here $R(\tau)$ decreases as the proper time $\tau$ goes by. The initial value of the $R(\tau_0) = R_0$ is grater than $r_g$. Otherwise at the initial stage we would have had a black hole rather than a star.

Inside the ball we have a spatially homogeneous metric whose time--slices are compact and are decreasing in size as time goes by. The suitable metric is:

\bqa\label{ballmetr1}
ds_-^2 = d\tau^2 - a^2(\tau) \, \left[d\chi^2 + \sin^2\chi \, d\Omega^2\right], \quad d\Omega^2 = d\theta^2 + \sin^2\theta \, d\varphi^2.
\eqa
We discuss the physics and the origin of such metrics in the lecture XI in grater details. Its spatial section $d\tau = 0$ is the ball or three--dimensional disc, which is a part of the three--sphere, whose metric is $a^2(\tau) \, \left[d\chi^2 + \sin^2\chi \, d\Omega^2\right]$. The latter line element would represent the three--sphere if $\chi \in [0,\pi]$, but in the case under consideration $\chi \in [0, \chi_0]$ for some $\chi_0 < \pi$. Note that one can obtain the three--sphere by gluing to each other two three--discs for $\chi \in [0, \chi_0]$ and $\chi \in [\chi_0, \pi]$. The radius of the three--disc under consideration, $a(\tau)$, is decreasing as the time $\tau$ goes by.

Thus, we have

\bqa\label{ballmetr}
g_{\tau\tau} = 1, \quad g_{\chi\chi} = - a^2, \quad g_{\theta\theta} = - a^2 \, \sin^2\chi, \quad {\rm and} \quad g_{\varphi\varphi} = - a^2 \, \sin^2\chi \, \sin^2\theta.
\eqa
The resulting non--zero components of the Christoffel symbols are $\Gamma^0_{ij} = \dot{a} \, a \tilde{g}_{ij}$ and $\Gamma^i_{0j} = \frac{\dot{a}}{a} \, \delta^i_j$. Here $\dot{a} \equiv da/d\tau$, $i = 1,2,3$ and $\tilde{g}_{ij}$ is the metric of the three--sphere of unit radius: $dl^2 = \tilde{g}_{ij} \, dx^i dx^j$. As the result $\Gamma^i_{jk}$ components of the Christophel symbols also are not zero and are proportional to those of the three--sphere, but we do not need their explicit form in this lecture.

One can find from these Christoffel symbols that the ``$00$'' part of the \underline{Einstein tensor}, $G_{\mu\nu} \equiv R_{\mu\nu} - \frac12 \, g_{\mu\nu} \, R$, has the following form:

\bqa\label{G00}
G_{00} = \frac{3}{a^2}\, \left(\dot{a}^2 + 1\right).
\eqa
At the same time the energy--momentum tensor inside the ball is $T_{\mu\nu} = \rho(\tau) \, u_\mu \, u_\nu$.
In the reference frame of (\ref{ballmetr1}) the dust remains stationary, i.e. $d\chi = d\theta = d\varphi = 0$. Hence, in this frame $u_\mu = (1,0,0,0)$ and the only non--zero component of $T_{\mu\nu}$ is $T_{00} = \rho(\tau)$. As the result from (\ref{G00}) we find that one of the Einstein equations is as follows:

\bqa\label{frideq}
\dot{a}^2 + 1 = \frac{8\pi\kappa}{3} \, \rho \, a^2.
\eqa
Equations for $T_{0j}$ do {\it not} lead to non--trivial relations: they just give relations stating that $0=0$. As we have explained at the end of the lecture III, instead of some of the remaining equations one can use the energy--momentum tensor conservation condition,

\bqa
0 = D_\mu T^\mu_\nu = \partial_\mu \left(\rho \, u^\mu \, u_\nu\right) + \Gamma^\mu_{\beta\mu} \, \rho \, u^\beta \, u_\nu - \Gamma^\beta_{\nu\mu} \, \rho \, u^\mu \, u_\beta.
\eqa
Here we have four equations --- one for each value of $\nu$. If $\nu = 1,2,3$ then again we get trivial relations $0=0$. However, if $\nu = 0$ there is the following equation:

\bqa
\dot{\rho} + 3\, \frac{\dot{a}}{a} \, \rho = 0, \quad {\rm where} \quad \dot{\rho} = \frac{d\rho}{d\tau}.
\eqa
Hence, $\partial_\tau \left(\rho \, a^3\right) = 0$ and we obtain the obvious result that if the volume of spatial sections decreases as $a^3(\tau)$ the density of the dust is increasing as $\rho(\tau) = \frac{const}{a^3(\tau)}$. Let us choose here such a constant that $\rho \, a^3 = \frac{3}{8\pi\kappa} \, a_0$. Then the solution of (\ref{frideq}) has the following parameterized form:

\bqa\label{difsol}
a(\eta) = \frac{a_0}{2} \, \left(1 + \cos \eta\right), \quad {\rm and} \quad \tau(\eta) = \frac{a_0}{2} \, \left(\eta + \sin\eta\right).
\eqa
We discuss the physical meaning of the parameter $\eta$ and of such solutions in the lecture XI in grater details.
Here we just check that (\ref{difsol}) indeed solves (\ref{frideq}). In fact, $\frac{d\tau}{d\eta} = \frac{a_0}{2} \, \left(1 + \cos\eta\right) = a(\eta)$. Then,
$\frac{da}{d\tau} = \frac{da}{d\eta} \, \frac{d\eta}{d\tau} = \frac{da}{d\eta} \, \frac{1}{a}$. As the result from
(\ref{frideq}) we derive the simple oscillator type equation

\bqa
\left(\frac{da}{d\eta}\right)^2 + a^2 = a_0 \, a.
\eqa
Obviously $a(\eta)$ from (\ref{difsol}) solves this equation. The integration constants in (\ref{difsol}) are chosen to fulfill the initial conditions.

From (\ref{difsol}) we see that the collapse starts at $\eta = 0$, which corresponds to $\tau = 0$. In Schwarzschild coordinates this corresponds to $t=0$, when the thermonuclear fuel inside the ball was completely spent. At this moment $a=a_0$. The collapse ends as $a \to 0$. That happens as $\eta \to \pi$, i.e. when the proper time reaches $\tau = \frac{\pi a_0}{2}$. Thus, the collapse process takes the finite proper time: the shell crosses the horizon $r=r_g$ and even reaches the singularity within the finite proper time.

{\bf 3.} What remains to be done is to glue the metrics $ds_+^2$ and $ds_-^2$ and their first derivatives across the surface $\Sigma$ of the ball. (Their second derivatives, which are related to the Ricci tensor, are fixed by Einstein equations, i.e. by the form of $T_{\mu\nu}$ in the corresponding region, and we have used this fact by fixing $ds_+^2$ and $ds_-^2$ above.) These gluing conditions follow from the least action principle for the Einstein--Hilbert action\footnote{We do not derive these gluing conditions from first principles (from the least action principle), because it demands some space and time and elements of differential geometry. That goes beyond our introductory course, but at the same time the conditions themselves are pretty much obvious.}.

In terms of the metric $ds_-^2$ the surface $\Sigma$ is just a two sphere at
some value $\chi_0$ of the angle $\chi$. Then the induced metric on the surface is

\bqa
\left. ds^2_-\right|_\Sigma = d\tau^2 - a^2(\tau)\sin^2 \, \chi_0 \, d\Omega^2.
\eqa
This metric has to be related to the one on the world--hypersurface $z^\mu(\tau) = \left[T(\tau), \, R(\tau)\right]$ in the Schwarzschild background:

\bqa
\left. ds_+^2 \right|_\Sigma = \left[\left(1 - \frac{r_g}{R}\right)\, \left(\frac{dT}{d\tau}\right)^2 - \frac{1}{1 - \frac{r_g}{R}}\, \left(\frac{dR}{d\tau}\right)^2\right] \, d\tau^2 - R^2(\tau)\, d\Omega^2.
\eqa
To have that $\left. ds^2_-\right|_\Sigma = \left. ds_+^2 \right|_\Sigma$, there should be relations as follows:

\bqa
R(\tau) = a(\tau) \, \sin\chi_0, \nonumber \\
\left(1 - \frac{r_g}{R}\right)\, \left(\frac{dT}{d\tau}\right)^2 - \frac{1}{1 - \frac{r_g}{R}}\, \left(\frac{dR}{d\tau}\right)^2 = 1.
\eqa
These relations allow us to find $R(\tau)$ from (\ref{difsol}) and, then, to solve for $T(\tau)$ the second equation here:

\bqa
\dot{T} = \frac{\sqrt{\dot{R}^2 + 1 - \frac{r_g}{R}}}{1 - \frac{r_g}{R}}, \quad {\rm where} \quad \dot{T} = \frac{dT}{d\tau}, \quad {\rm and} \quad \dot{R} = \frac{dR}{d\tau}.
\eqa
This is the world--hypersurface of the boundary of the ball as it is seen by outside observers.

From the last equation one can see that as $R \to r_g$ in the collapse process, we can neglect $1 - \frac{r_g}{R}$
in comparison with $\dot{R}^2$ under the square root. Then,

\bqa
dT \approx - \frac{dR}{1 - \frac{r_g}{R}},
\eqa
and the minus sign appears here because during the collapse process we have that $dR < 0$, while $dT > 0$. Thus, one obtains that

\bqa\label{RT}
R(T) \approx r_g \left(1 + e^{-\frac{T}{r_g}}\right),
\eqa
i.e. from the point of view of an observer, which is fixed at some $r > R(\tau)$, the fall of the star's matter through the surface $r=r_g$ never happens. The matter of the star just asymptotically approaches its gravitational radius as $t \to + \infty$. That is true although the star falls behind its gravitational radius within the above mentioned finite {\it proper} time.

The next step is to glue the first derivatives of $ds_-^2$ and $ds_+^2$ across $\Sigma$. This demands some straightforward calculations with the use of differential geometry for surfaces in curved space--times. This goes beyond the scope of our lectures. But the result of the calculation is very simple and can be predicted on general physical grounds. In fact, from the gluing conditions in question one finds that

\bqa
r_g = 2 \, \kappa \, \frac{4\pi}{3} \, \rho(\tau) \, R^3(\tau) = const.
\eqa
Which just means that the mass of the star/black hole remains constant during the ideal spherical collapse. Moreover, the mass is appropriately related to the gravitational
radius, $r_g$, of the Schwarzschild geometry.

\begin{figure}
\begin{center}
\includegraphics[scale=0.5]{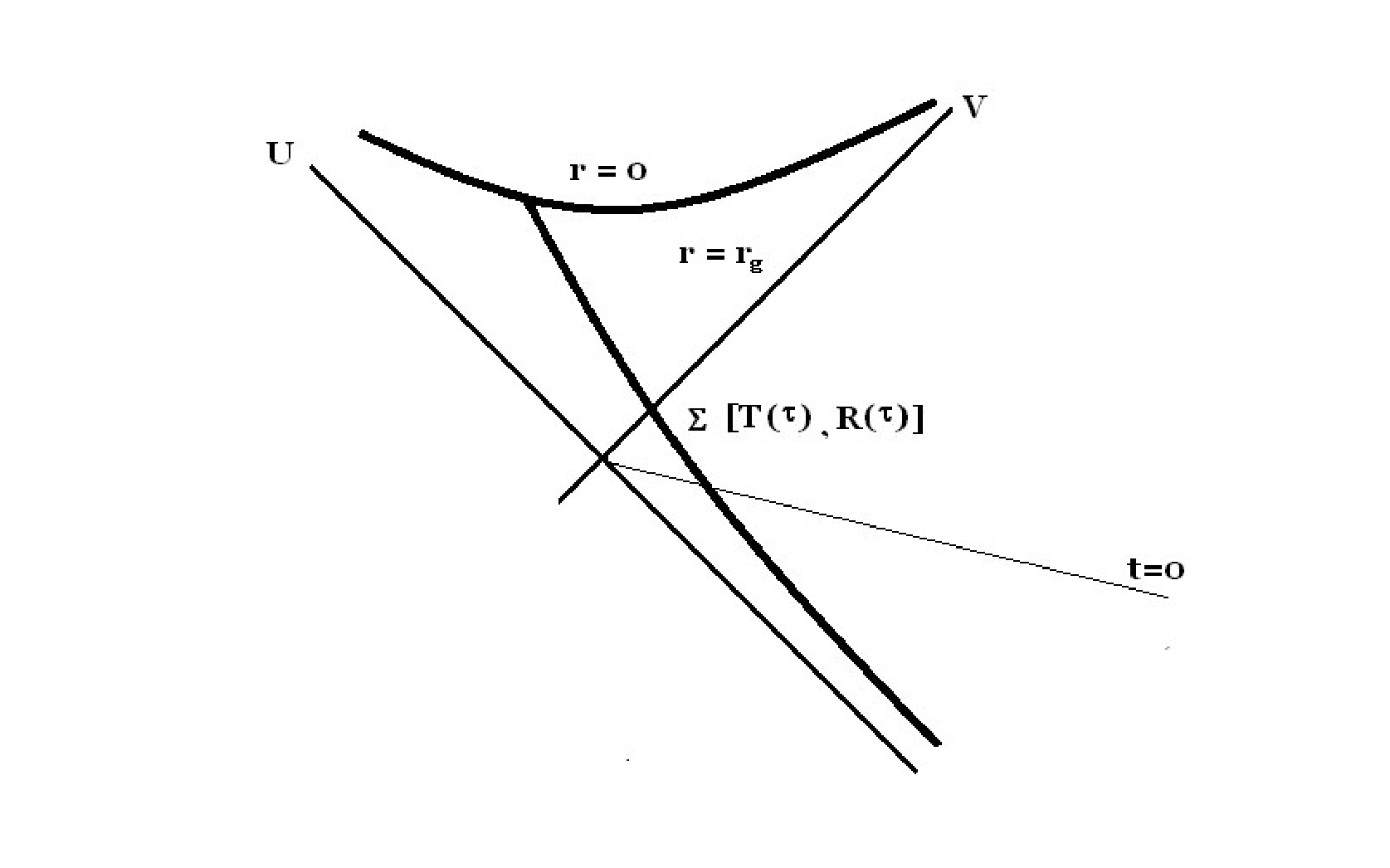}\caption{}\label{fig11}
\end{center}
\end{figure}

\begin{figure}
\begin{center}
\includegraphics[scale=1]{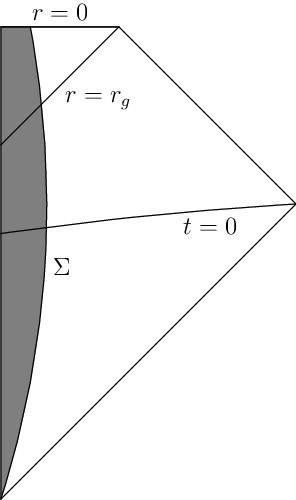}\caption{}\label{fig12}
\end{center}
\end{figure}

{\bf 4.} Let us draw the Penrose--Carter diagram for the Oppenheimer--Snyder collapsing solution. To do that we have to draw separately the diagrams for $ds_+^2$ and $ds_-^2$ and to glue them across $\Sigma$. For the Schwarzschild part, $ds_+^2$, the diagram follows from the fig. (\ref{fig11}). This is just the same diagram as in the previous lectures, but it is valid only beyond $\Sigma$ whose world--surface is $[T(\tau), R(\tau)]$ after $t=0$. Before $t=0$ the surface $\Sigma$ remains stationary at some radius $r=R_0 > r_g$.

To draw the Penrose--Carter diagram for $ds_-^2$ let us represent this metric as follows:

\bqa
ds_-^2 = d\tau^2 - a^2(\tau) \, \left[d\chi^2 + \sin^2\chi \, d\Omega^2\right] = a^2(\eta) \, \left[d\eta^2 - d\chi^2 -\sin^2\chi \, d\Omega^2\right],
\eqa
where $\eta \in [0,\pi]$ is defined in (\ref{difsol}). Now we should drop off the conformal factor $a^2(\eta)$ and choose the relevant two--dimensional part i.e. $(\eta, \chi)$. Then we obtain just a part of the square for $0 \leq \chi \leq \chi_0$ and $0 \leq \eta \leq \pi$. This has to be glued to the diagram for the fig. (\ref{fig11}) after $t=0$. Before this moment of time the Penrose--Carter diagram of the space--time is the same as for the star, which was found in the previous lecture.

All in all, this way one finds the total Penrose--Carter diagram for the Oppenheimer--Snyder collapsing solution which is shown on the fig. (\ref{fig12}). It can be adjusted to the shown here form by a suitable conformal transformation. It is worth stressing at this point that in doing this gluing of diagrams we drop off different conformal factors for different parts of the diagram.

As one can see, the obtained diagram does not contain the white hole part.
But it is not hard to find the time reversal of the collapsing solution. It is given by the same equation as
(\ref{difsol}) for $\eta \in [\pi, 2\pi]$ or by

\bqa
a(\eta) = \frac{a_0}{2} \, \left(1 - \cos \eta\right), \quad {\rm and} \quad \tau(\eta) = \frac{a_0}{2} \, \left(\eta - \sin\eta\right),
\eqa
if one chooses $\eta \in [0,\pi]$. This solution describes explosion starting from $\eta = 0$, when $\tau = 0$ and $a=0$. Then as $\eta$ reaches $\pi$ the conformal factor inflates to $a=a_0$. If something stops the explosion at $a=a_0$, then the corresponding Penrose--Carter diagram is just the time reversal of the one shown on the fig. (\ref{fig12}), i.e. it is the flip of this picture over a horizontal line.

{\bf 5.} We continue this lecture with the discussion of two complicated, but very important questions.
Let us see now what happens with waves which are created in the vicinity of a black hole horizon. The arguments, which are presented here, are borrowed from the book ``General Relativity'' by I. Khriplovich. Consider say an electromagnetic excitation, which is created at a radius $r_0 = r_g + \epsilon$, $\epsilon \ll r_g$ in the vicinity of the horizon. Radial propagation time of this excitation from $r_0$ to $r \gg r_g$ follows from the equation:

\bqa
0 = ds^2 = \left(1 - \frac{r_g}{r}\right)dt^2 - \frac{dr^2}{1 - \frac{r_g}{r}},
\eqa
and is equal to

\bqa\label{teps}
t = \int_{r_0}^r \frac{dr}{1 - \frac{r_g}{r}} = r - r_0 + r_g \, \log \frac{r - r_g}{r_0 - r_g} \approx r + r_g \log\frac{r}{\epsilon}.
\eqa
If the frequency of the excitation at $r_0$ is equal to $\omega_0$, then, as we explained at the end of the previous lecture, at $r \gg r_0$ the frequency is reduced to

\bqa
\omega = \omega_0\, \sqrt{\frac{g_{00}(r_0)}{g_{00}(r)}} \approx \omega_0 \sqrt{\frac{\epsilon}{r_g}}.
\eqa
As follows from (\ref{teps}),

\bqa
\frac{\epsilon}{r} \approx e^{-\frac{t-r}{r_g}}.
\eqa
Hence, the frequency of the excitation, as it crawls out from the vicinity of the horizon, depends on time as follows:

\bqa\label{omegat}
\omega(t) \approx \omega_0 \, \sqrt{\frac{r}{r_g}} \, e^{-\frac{t-r}{2r_g}}.
\eqa
Furthermore, the phase of the electromagnetic excitation is changing as

\bqa
\Psi(t) = \int_0^t dt' \omega(t') \approx - 2 \, \omega_0 \, \sqrt{r\, r_g}\, e^{-\frac{t-r}{2\, r_g}} + const.
\eqa
This way we have found \underline{eikonal} $\Psi$, i.e. $e^{i \Psi}$ is the solution of the Maxwell equations in the geometric optic approximation, $\Box(g) e^{i \Psi} = 0 \rightarrow g^{\mu\nu} \partial_\mu \Psi \partial_\nu \Psi = 0$, and with the given above initial conditions.
Then the spectrum of frequencies contributing to this wave--packet is as follows:

\bqa
f(\omega) \sim \int_0^\infty dt \, e^{i \, \omega \, t} \, \exp\left[- 2 \, i \, \omega_0 \, \sqrt{r\, r_g}\, e^{-\frac{t-r}{2\, r_g}}\right] \sim \left(2\, \omega_0 \, \sqrt{r\, r_g}\right)^{2 \, i \, \omega \, r_g} \,
e^{-\pi \, \omega \, r_g} \, \Gamma\left(- 2 \, i \, \omega \, r_g\right),
\eqa
where $\Gamma(x)$ is the $\Gamma$--function. Here we have dropped off the factors that do not depend on $\omega$. As the result, the spectral density of any wave--like excitation that was created in the vicinity of the horizon is given by:

\bqa
\left|f(\omega)\right|^2 \sim e^{-2\,\pi\,\omega \, r_g}\left|\Gamma\left(- 2 \, i \, \omega \, r_g\right)\right|^2 = \frac{\pi}{\omega \, r_g}\, \frac{1}{e^{4\, \pi \, \omega \, r_g} - 1} \approx \frac{\pi}{\omega \, r_g} \, e^{- 4 \, \pi \, \omega \, r_g}.
\eqa
The last approximation is used because our estimates are valid in the geometric optic approximation, when $\omega r_g \gg 1$.

This calculation is absolutely classical. However, if in the last expression we multiply $\omega$ by $\hbar$, to obtain the energy, $E = \hbar \omega$, then the exponent will acquire the form of the Boltzman's thermal factor:

\bqa
w(E) \sim e^{- \frac{E}{T}},
\eqa
where $T = \frac{\hbar}{4\, \pi \, r_g}$ is the so called \underline{Hawking's temperature}. \underline{Hawking effect} is that black holes are decaying via creation of particles with the thermal spectrum. Formally, Hawking radiation of black holes follows from similar equations. However, conceptually the effect in question is much more complicated and its discussion goes beyond the scope of our lectures. To stress the difference let us just point out here that Hawking effect follows from a change of the ground state of quantum fields due to a collapse process and, unlike classical radiation process, does not have a source so to say. At the same time, in this section we just have made an observation that if there is a radiation, which is sourced in the vicinity of the horizon, it will thermalize as it climbs out to spatial infinity.

{\bf 6.} Another seemingly unrelated question is as follows. Does actually the creation of the black hole ever happens from the point of view of those observers who always stay outside it? Or do they just see such an eternal asymptotically slowing down process which was described above? Although this is just an academic question, because there seems to be no device which will be always sensitive to the exponentially suppressed factor $e^{-t/r_g}$ in (\ref{RT}) or in (\ref{omegat}), as $t\to + \infty$, we still would like to address it here. We think that the answer to this question may be relevant for the deeper understanding of the Hawking radiation and backreaction on it. Note that, as pointed out above, formally this effect appears due to the same exponential factor.

Now we present some intuitive speculations, which still need to get some solid mathematical approval.
Let us see what happens with the light rays which are scattered off the surface of the star or radiated by it.
If one has absolutely sensitive device and takes the above picture seriously, he is expecting to see that the signal will be eternally coming out from the star.

However, this picture is valid only if we assume that light rays are going along light--like geodesics on the Schwarzschild background. That is true, if one neglects that an electromagnetic wave itself carries energy and, hence, also curves space--time. The latter effect is very small, but the question is if one can neglect it, once he addresses the issues of the exponentially suppressed contributions.

To understand what we are actually up to here, consider the fig. (\ref{fig5}) of the lecture IV. From the corresponding picture one can see that the horizon is just one of the light--like geodesics. It is a boundary between two families of ``outgoing'' geodesics. Moreover if a light ray was going along the horizon it will remain there eternally.

However, if one takes into account that electromagnetic field also does curve space--time, he has to draw real light--like world--lines rather than geodesics. The picture of the fig. (\ref{fig5}) from the lecture IV is not applicable anymore, because it is static. The new picture will not be static. But certain its relevant features will remain unchanged. Namely, we still expect to have two families of ``outgoing'' light--like world--lines --- those which escape to infinity and those which are directed towards the singularity. To see this one can just study short parts of the world--lines in question, i.e. seeds of light--like lines during small periods of time. But the boundary separating these two families of curves will not belong to the class of the light--like world--lines. The light ray cannot anymore eternally stay on the fixed radius $r=r_g$.

Rephrasing this, we expect that among the photons, which are emitted by a collapsing star, there will be a last one that will reach outside observers. The next photon after that will just participate into the creation of the black hole as a part of its matter content. As the result, the outside observer sooner or latter will stop receiving signals sourced by the matter of the collapsing star. And that will happen objectively rather than due to a lack of the sensitivity of his device.

One of the disadvantages of the picture that we have described here is that, if it is true, then the moment of black hole formation depends on the energy of the last escaping photon. But at this point for us it is important to see that the black hole is actually created during a finite time as measured by outside observers. In any case we just qualitatively described some phenomenon which remains to be described quantitatively somehow.

\vspace{10mm}

\centerline{\bf Problems}

\vspace{5mm}

\begin{itemize}

\item Calculate the Christoffel symbols and the Ricci tensor for the metric $ds_-^2$.

\item Show that equation $\hbar^2 \, \Box \, e^{i \, \frac{\Psi}{\hbar}} = 0$, where $\Box = \frac{1}{\sqrt{|g|}} \, \partial_\mu \sqrt{|g|} \, g^{\mu\nu} \partial_\nu $, reduces to $g^{\mu\nu} \, \partial_\mu \Psi \, \partial_\nu \Psi = 0$ in the limit $\hbar \to 0$.

\end{itemize}

\vspace{10mm}

\centerline{\bf Subjects for further study:}

\vspace{5mm}

\begin{itemize}

\item Thin--shell collapse and Vaidya space--time.

\item Gluing conditions for metric tensors, extrinsic curvature and differential geometry of embedded hypersurfaces.

\item Derivation of the Hawking radiation (see e.g. ``Hawking radiation and secularly growing loop corrections'',
Emil T. Akhmedov, Hadi Godazgar and Fedor K. Popov;
Published in Phys.Rev. D93 (2016) 2, 024029; e-Print: arXiv:1508.07500)

\item Black hole thermodynamics.

\item Eikonal, geometric optic or quasiclassical approximation for wave equations. Hamiltonian--Jacoby equations in various situations.

\end{itemize}

\newpage

\section*{LECTURE IX \\ {\it Energy--momentum pseudo--tensor for gravity. Weak field approximation. Energy--momentum pseudo--tensor in the weak field limit on the Minkowskian background. Free gravitational waves.}}

\vspace{10mm}

{\bf 1.} To define what means gravitational radiation one has to specify the meaning of energy flux in the presence of gravity. That is the problem we are going to address in this lecture.

In Minkowski space--time the energy--momentum conservation has the following form $\partial^\mu T_{\mu\nu} = 0$. After the integration over a time--slice $\Sigma$ and application of the Stokes theorem, this equation expresses the actual conservation of a quantity referred to as four momentum:

\bqa
\frac{dP^0}{dt} + \oint_{\partial \Sigma} d^2\sigma^{i} \, T_{i 0} = 0, \quad {\rm where} \quad P^\mu(t) \equiv \int_{\Sigma} T^{\mu 0}\left(t,\vec{x}\right) \, d^3\Sigma\left(\vec{x}\right).
\eqa
Here $d^3\Sigma\left(\vec{x}\right)$ is the volume element of a Cauchy surface, $\Sigma$, or of a time--slice $t = const$; $t$ is time and $d^2\sigma^{i}$ is the area element of the boundary $\partial \Sigma$ of the Cauchy surface in question. Having such a quantity, one can define what means energy inflow or outflow, $\oint_{\partial\Sigma} d^2\sigma^i \, T_{i0}$, in a system. In the presence of gravity, however, the equation, that we have been referring to as energy--momentum conservation law in the previous lectures, has the form as follows --- $D^\mu T_{\mu\nu} = 0$. It can be transformed into:

\bqa\label{conslaw}
0 = D_\mu T^\mu_\nu = \partial_\mu T^\mu_\nu + \Gamma^\mu_{\beta\mu} \, T^\beta_\nu - \Gamma^\beta_{\nu\mu}\, T^\mu_\beta = \frac{1}{\sqrt{|g|}} \, \partial_\mu \left(T^\mu_\nu \, \sqrt{|g|}\right) - \frac12 \, \left(\partial_\nu g_{\mu\alpha}\right)\, T^{\mu\alpha},
\eqa
where we have used that $T^{\mu\nu} = T^{\nu\mu}$ and the expression $\Gamma^\mu_{\nu\mu} = \frac{1}{\sqrt{|g|}} \, \partial_\nu \sqrt{|g|}$, which follows from the definition of the Christoffel symbols.

Unlike the equation $\partial^\mu T_{\mu\nu} = 0$ the obtained relation (\ref{conslaw}) does not express any actual conservation of some quantity. This is natural to expect in the presence of gravity, because the energy and momentum carried only by matter, $T_{\mu\nu}$, should not be conserved alone. In fact, there can be energy--momentum transfer between matter and gravity. But then what is a quantity that describes energy--momentum for gravity itself? What is the total energy--momentum tensor for gravity and matter together, which we expect to be conserving?

Let us choose such a reference system at a point $x_0$ that $\partial_\alpha g_{\mu\nu}(x_0) = 0$. (Note that this choice of the frame does {\it not} mean that $g_{\mu\nu}(x_0) = \eta_{\mu\nu}$.) From (\ref{conslaw}) we obtain that $\partial_\mu T^\mu_\nu(x_0) = 0$, because in such a reference frame also $\partial_\mu \sqrt{|g(x_0)|} = 0$. How does this equation change in an arbitrary reference system?

From the Einstein equations of motion it follows that $T^{\mu\nu} = \frac{1}{8\, \pi \, \kappa} \, \left(R^{\mu\nu} - \frac12 \, g^{\mu\nu} \, R\right)$. At the same time in the above defined gauge we have that $\Gamma^\mu_{\nu\alpha}(x_0) = 0$, but $\partial_\beta \Gamma^\mu_{\nu\alpha}(x_0) \neq 0$ and, hence, from the definition of the Ricci tensor we have that

\bqa\label{203}
R^{\mu\nu}(x_0) = \frac12 \, g^{\mu\alpha} \, g^{\nu\gamma} \, g^{\sigma \delta}\, \left[\partial_\alpha \partial_\delta g_{\sigma\gamma} + \partial_\sigma \partial_\gamma g_{\alpha\delta} - \partial_\alpha \partial_\gamma g_{\sigma\delta} - \partial_\sigma \partial_\delta g_{\alpha\gamma}\right].
\eqa
Plaguing this expression into the Einstein equations and performing straightforward transformations, we obtain that:

\bqa\label{defeta}
T^{\mu\nu}(x_0) = \partial_\alpha \left\{\frac{1}{16\, \pi \, \kappa} \, \frac{1}{|g|} \, \partial_\beta\left[|g|\, \left(g^{\mu\nu} \, g^{\alpha\beta} - g^{\mu\alpha} \, g^{\nu\beta}\right)\right]\right\} \equiv \frac{1}{|g|} \, \partial_\alpha \eta^{\mu\nu\alpha},
\eqa
in the reference frame under consideration. Here

\bqa\label{defeta1}
\eta^{\mu\nu\alpha} = \frac{1}{16 \, \pi\, \kappa} \, \partial_\beta\left[|g|\, \left(g^{\mu\nu} \, g^{\alpha\beta} - g^{\mu\alpha} \, g^{\nu\beta}\right)\right].
\eqa
In eq.(\ref{defeta}) we have used that $\partial_\alpha \left|g(x_0)\right| = 0$ and took $|g|$ out from the derivative $\partial_\alpha$. To make all the above transformations one has to keep in mind that in the vicinity of $x_0$ we have that $g_{\mu\nu}(x) = g_{\mu\nu}(x_0) + \partial_\alpha \partial_\beta g_{\mu\nu}(x_0) \,(x-x_0)^\alpha \, (x-x_0)^\beta$ and we do not distinguish terms that coincide at the leading order in $(x-x_0)$.

This quantity, $\eta^{\mu\nu\alpha}$, has obvious properties, which follow from its definition:

\bqa\label{propeta}
\eta^{\nu\mu\alpha} = \eta^{\mu\nu\alpha} = - \eta^{\mu\alpha\nu}.
\eqa
Thus, in the gauge $\partial_\alpha g_{\mu\nu} (x_0) = 0$ we have the following relation: $\partial_\alpha \eta^{\mu\nu\alpha}(x_0) = |g| \, T^{\mu\nu}(x_0)$. In an arbitrary gauge this relation is not true. Hence, let us define

\bqa\label{deft}
\partial_\alpha \eta^{\mu\nu\alpha} - |g|\, T^{\mu\nu} \equiv |g| \, t^{\mu\nu}.
\eqa
The newly defined quantity $t^{\mu\nu}$ is symmetric, $t^{\mu\nu} = t^{\nu\mu}$, as follows from the properties of $T^{\mu\nu}$ and $\eta^{\mu\nu\alpha}$. Expressing $T^{\mu\nu}$ through $R^{\mu\nu}$ via Einstein equations and using (\ref{defeta1}), we obtain that

\bqa
|g| \, t^{\mu\nu} = \frac{1}{16 \, \pi\, \kappa} \, \partial_\alpha  \, \partial_\beta \, \left[|g|\, \left(g^{\mu\nu} \, g^{\alpha\beta} - g^{\mu\alpha} \, g^{\nu\beta}\right)\right] - \frac{|g|}{8\, \pi\,\kappa} \, \left(R^{\mu\nu} - \frac12 g^{\mu\nu} \, R\right).
\eqa
From here, after a rather tedious calculation, one can find that:

\bqa\label{tmunu}
|g| \, t^{\mu\nu} = \frac{1}{16\, \pi \kappa} \, \left[{G^{\mu\nu}}_{,\alpha} \, {G^{\alpha\beta}}_{,\beta} - {G^{\mu\alpha}}_{,\alpha} \, {G^{\nu\beta}}_{,\beta} + \frac12 \, g^{\mu\nu} \, g_{\alpha\beta} \, {G^{\alpha\gamma}}_{,\sigma} \, {G^{\sigma\beta}}_{,\gamma} - \right. \nonumber \\ - \left. \left(g^{\mu\gamma} \, g_{\alpha\beta} \, {G^{\nu\beta}}_{,\sigma} \, {G^{\alpha\sigma}}_{,\gamma} + g^{\nu\alpha} \, g_{\beta\gamma} \, {G^{\mu\gamma}}_{,\sigma} \, {G^{\beta\sigma}}_{,\alpha}\right) + g_{\alpha\beta} \, g^{\gamma\sigma} \, {G^{\mu\alpha}}_{,\gamma} \, {G^{\nu\beta}}_{,\sigma} \right. +\nonumber \\ + \left. \frac18 \, \left(2\, g^{\mu\alpha} \, g^{\nu\beta} - g^{\mu\nu} \, g^{\alpha\beta}\right)\, \left(2\, g_{\gamma\sigma} \, g_{\delta\xi} - g_{\sigma\delta} \, g_{\gamma\xi}\right) \, {G^{\gamma\xi}}_{,\alpha} \, {G^{\sigma\delta}}_{,\beta}\right],
\eqa
where $G^{\mu\nu} \equiv \sqrt{|g|} \, g^{\mu\nu}$ and as usual ${G^{\mu\nu}}_{,\alpha} \equiv \partial_\alpha G^{\mu\nu}$. One can straightforwardly check that this quantity is vanishing in the gauge $\partial_\alpha g_{\mu\nu}(x_0) = 0$.

It is important to note that $t^{\mu\nu}$ is not a tensor, as follows already from its definition (\ref{deft}). In fact, $\partial_\alpha \eta^{\mu\nu\alpha}$ contains the ordinary rather than the covariant derivative.
The reason why following Landau and Lifshitz we have introduced the \underline{gravity energy--momentum pseudo--tensor} $t^{\mu\nu}$ is because then we have the following conservation law:

\bqa
\partial_\mu \left[|g|\, \left(T^{\mu\nu\phantom{\frac12}} + \,\, t^{\mu\nu}\right)\right] = 0,
\eqa
for the total energy--momentum of gravity and matter together.
This equation follows from (\ref{deft}) because $\partial_\nu \partial_\alpha \eta^{\mu\nu\alpha} = 0$, as
the consequence of (\ref{propeta}). Thus, there is the conserved momentum:

\bqa
P^\mu \equiv \int_\Sigma d^3\Sigma_\nu \, |g|\, \left(T^{\mu\nu} + t^{\mu\nu}\right),
\eqa
where $d^3\Sigma_\nu$ is a four--vector, which is perpendicular to the Cauchy surface $\Sigma$ and whose modulus is equal to the elementary three--volume form on $\Sigma$. In the absence of the gravitational field this $P^\mu$ reduces to the above defined four--momentum in Minkowski space--time.

{\bf 2.} To understand the essential idea behind the introduction of $t_{\mu\nu}$ let us consider the weak gravitational field approximation. Namely, consider the metric tensor of the form:

\bqa
g_{\mu\nu} \approx \eta_{\mu\nu} + h_{\mu\nu}, \quad {\rm where} \quad \left|h_{\mu\nu}\right| \ll 1,
\eqa
which describes small perturbations on top of the flat background. Here $\eta_{\mu\nu}$ is the background Minkowskian metric tensor, while $h_{\mu\nu}$ is a small perturbation on top of it. In principle all the formulas below in this lecture can be extended to more general (non--flat) background metrics, but for the illustrative reasons we consider here only the Minkowskian background space--time.

Let us restrict ourselves to the linear order in $h_{\mu\nu}$. Then, the inverse metric tensor $g^{\mu\nu}$ is defined from the equation: $\left(\eta_{\mu\alpha} + h_{\mu\alpha}\right)\, g^{\alpha\nu} = \delta^\nu_\mu$,
and at the linear order is equal to:

\bqa
g^{\mu\nu} \approx \eta^{\mu\nu} - h^{\mu\nu},
\eqa
where from now on in this lecture we higher and lower indexes with the use of the background metric: e.g., $h^\mu_\nu \equiv \eta^{\mu\alpha}\, h_{\alpha \nu}$. With the same precision:

\bqa
|g| \approx 1 + h, \quad {\rm where} \quad h \equiv h^\mu_\mu.
\eqa
As follows from the generic infinitesimal transformation law $\bar{g}_{\mu\nu}(x) = g_{\mu\nu}(x) + D_{(\mu} \epsilon_{\nu)}(x)$ at the linearized order:

\bqa\label{transh}
\bar{h}_{\mu\nu}(x) = h_{\mu\nu}(x) + \partial_{(\mu} \epsilon_{\nu)}.
\eqa
Furthermore:

\bqa
\Gamma^\mu_{\nu\alpha} \approx \frac12 \, \eta^{\mu\beta} \, \left(\partial_\nu h_{\alpha\beta} + \partial_\alpha h_{\beta\nu} - \partial_\beta h_{\nu\alpha}\right).
\eqa
Then, the terms that are $\sim \Gamma^2$ should be of the order of $h^2$. Hence, only terms $\sim \partial \Gamma$ do contribute to the Riemann tensor at the linear order, which then has the following form:

\bqa\label{217}
R^{(1)}_{\mu\nu\alpha\beta} = \eta_{\mu\gamma} \, \left[\partial_\alpha \Gamma^\gamma_{\nu\beta} - \partial_\beta \Gamma^\gamma_{\nu\alpha}\right] \approx \frac12\, \left[\partial_\nu \partial_\alpha h_{\mu\beta} + \partial_\mu \partial_\beta h_{\nu\alpha} - \partial_\mu \partial_\alpha h_{\nu\beta} - \partial_\nu \partial_\beta h_{\mu\alpha}\right].
\eqa
As the result the Ricci tensor is equal to:

\bqa\label{ricci1}
R^{(1)}_{\mu\nu} \approx g^{\alpha\beta} \, R_{\alpha\mu\beta\nu} \approx \eta^{\alpha\beta} \, R_{\alpha\mu\beta\nu} \approx \frac12 \, \left[- \Box h_{\mu\nu} + \partial_\nu \partial_\alpha h^\alpha_\mu + \partial_\mu \partial_\alpha h^\alpha_\nu - \partial_\mu \partial_\nu h \right],
\eqa
where $\Box \equiv \eta^{\alpha\beta} \partial_\alpha \partial_\beta$ and as before $h \equiv h^\alpha_\alpha$.

Using the transformations (\ref{transh}) one can fix the gauge:

\bqa\label{gauge}
\partial_\mu \psi^\mu_\nu = 0, \quad {\rm where} \quad \psi^\mu_\nu = h^\mu_\nu - \frac12 \, \delta^\mu_\nu \, h.
\eqa
In fact, $\partial^\mu \bar{\psi}_{\mu\nu} \equiv \partial^\mu \bar{h}_{\mu\nu} - \frac12 \, \partial_\nu \bar{h} = \partial^\mu \psi_{\mu\nu} + \Box \epsilon_\nu$. Hence, if we choose $\Box \epsilon_\nu = - \partial^\mu \psi_{\mu\nu}$, then $\partial^\mu \bar{\psi}_{\mu\nu} = 0$. It is important to note, however, that this gauge is fixed only up to the remnant transformations of the same form as (\ref{transh}), but with such a $\epsilon_\nu(x)$ that $\Box \epsilon_\nu(x) = 0$. In fact, the equation $\Box\epsilon_\nu = - \partial^\mu \psi_{\mu\nu}$ can be solved only up to the ambiguity of the addition of the harmonic vector--function $\Box \epsilon_\nu = 0$.

In the gauge (\ref{gauge}) the Ricci tensor (\ref{ricci1}) reduces to

\bqa\label{220}
R^{(1)}_{\mu\nu} = - \frac12 \, \Box h_{\mu\nu}.
\eqa
Thus, in the linearized approximation Einstein equations have the following form:

\bqa\label{linEin}
- \Box \psi_{\mu\nu} = 16 \, \pi \, \kappa \, T_{\mu\nu} + O\left(h^2\right).
\eqa
Let us consider now the terms that are of higher orders in $h_{\mu\nu}$, which are denoted here as $O\left(h^2\right)$. To find them we have to expand the relevant quantities at least to the second order in $h_{\mu\nu}$, e.g.:

\bqa
g^{\mu\nu} \approx \eta^{\mu\nu} - h^{\mu\nu} + h^\mu_\alpha \, h^{\alpha\nu}, \nonumber \\
|g| \approx 1 + h + \frac12 h^2 - \frac12 h^\mu_\nu \, h^\nu_\mu.
\eqa
Also the second order contribution to the Ricci tensor is equal to:

\bqa
R^{(2)}_{\mu\nu} = \frac12 \, h^{\rho\sigma} \partial_\mu \partial_\nu h_{\rho\sigma} - h^{\rho\sigma} \partial_\rho \partial_{(\mu} h_{\nu)\sigma} + \frac14 \, \left(\partial_\mu h_{\rho\sigma}\right) \, \left(\partial_\nu h^{\rho\sigma}\right) + \left(\partial^\sigma h^\rho_\nu\right)\, \left(\partial_{[\sigma} h_{\rho]\mu}\right) + \nonumber \\ + \frac12 \partial_\sigma\left(h^{\rho\sigma} \partial_\rho h_{\mu\nu}\right) - \frac14 \, \left(\partial_\rho h_{\mu\nu}\right) \, \left(\partial^\rho h\right) - \left(\partial_\sigma h^{\rho\sigma} - \frac12 \, \partial^\rho h\right) \, \partial_{(\mu} h_{\nu)\rho}.
\eqa
Here $A_{(\mu} \, B_{\nu)} \equiv A_\mu \, B_\nu + A_\nu \, B_\mu$ and $A_{[\mu} \, B_{\nu]} \equiv A_\mu \, B_\nu - A_\nu \, B_\mu$.
Using this expression and fixing the gauge (\ref{gauge}) one can find the term $O\left(h^2\right) \approx R^{(2)}_{\mu\nu} - \frac12 \, \eta_{\mu\nu} \, R^{(2)} - \frac12 h_{\mu\nu} \, R^{(1)}$ up to the second order in eq. (\ref{linEin}). It is straightforward to check that, as follows from (\ref{tmunu}), this term coincides with $|g| t_{\mu\nu}$ up to the same order. In particular, (\ref{tmunu}) is vanishing at the linear order in $h_{\mu\nu}$. That means that eq.(\ref{linEin}) actually has the form:

\bqa\label{lineins}
- \Box \psi_{\mu\nu} \approx 16 \, \pi \, \kappa \, \left[T_{\mu\nu} + t_{\mu\nu}\right],
\eqa
where $t_{\mu\nu}$ follows from (\ref{tmunu}) in the weak field approximation. Thus, in this approximation the energy--momentum pseudo--tensor is responsible for the non--linear part of the perturbations.

Now one can understand the reason for the pseudo--tensorial properties of $t_{\mu\nu}$. In fact, to specify what one means by the energy--momentum in the presence of gravity one has to fix a background and consider energy--momentum carried by perturbations. But what is meant by background and what can be attributed to fluctuations in the total metric tensor {\it does} depend on the choice of coordinates, i.e. this separation of the total metric can change after a generally covariant transformation. At the same time it is not hard to see that all the quantities in (\ref{lineins}) do transform as tensors under the linearized transformations (\ref{transh}).

{\bf 3.} To make further clarifications on the meaning of the pseudo--tensor $t_{\mu\nu}$ let us consider free gravitational wave solutions of the equation (\ref{lineins}). We consider weak fields in the linearized approximation and, hence, for the moment neglect $t_{\mu\nu}$ on the right hand side of this equation.

Let us look for the monochromatic plane--wave solution of this equation with $T_{\mu\nu} = t_{\mu\nu} = 0$:

\bqa
h_{\mu\nu}(x) = Re \left[\epsilon_{\mu\nu}\, e^{- i \, k_\alpha x^\alpha}\right],
\eqa
where $k^\alpha$ is a constant real wave--vector and $\epsilon_{\mu\nu}$ is a constant complex polarization tensor. Such a $h_{\mu\nu}$ solves homogeneous version of (\ref{lineins}) if $k^\alpha$ and $\epsilon_{\mu\nu}$ obey the following relations:

\bqa\label{4nu}
k^\alpha \, k_\alpha = 0, \quad {\rm and} \quad k^\mu \epsilon_{\mu\nu} - \frac12 \, k_\nu \, \epsilon^\mu_\mu = 0.
\eqa
Let us choose such a reference frame where the null vector $k^\mu$ has the form as follows:
$k^\mu = (k,0,0,k)$. This corresponds to a wave traveling along the third direction with the speed of light.
Second (vector) relation in (\ref{4nu}) composes a system of four equations for each value of $\nu = 0,1,2,3$.
The $\nu = 1$ and $\nu = 2$ components of this equation imply the relations as follows:

\bqa
\epsilon_{01} = - \epsilon_{31}, \quad {\rm and} \quad \epsilon_{02} = - \epsilon_{32}.
\eqa
At the same time the sum of the $\nu=0$ and $\nu=3$ components of this equation leads to such a relation as

\bqa
\epsilon_{11} + \epsilon_{22} = 0.
\eqa
After this the $\nu = 0$ component implies

\bqa
\epsilon_{03} = - \frac12 \, \left(\epsilon_{00} + \epsilon_{33}\right).
\eqa
Now let us perform the remaining harmonic gauge transformation $\bar{h}_{\mu\nu}(x) = h_{\mu\nu}(x) + \partial_{(\mu} \epsilon_{\nu)}(x)$ with $\epsilon_\nu(x) = - i \, \epsilon_\nu \, e^{- i \, k_\alpha x^\alpha}$, where $k^\alpha$ is the above null four--vector and $\epsilon_\nu$ is a constant four--vector (hence, $\Box \epsilon_\nu(x) = 0$). Under such a transformation the polarization tensor changes as follows:

\bqa
\bar{\epsilon}_{\mu\nu} = \epsilon_{\mu\nu} - k_\mu \epsilon_\nu - k_\nu \epsilon_\mu.
\eqa
Choosing in this equation $\epsilon_1 = - \epsilon_{13}/k$, $\epsilon_2 = - \epsilon_{23}/k$, $\epsilon_0 = \epsilon_{00}/2k$ and $\epsilon_3 = - \epsilon_{33}/2k$, we achieve that $\bar{\epsilon}_{13}, \bar{\epsilon}_{23}, \bar{\epsilon}_{00}$ and $\bar{\epsilon}_{33}$ are vanishing. As the result the polarization tensor has only two independent non--zero components as follows: $\bar{\epsilon}_{11} = - \bar{\epsilon}_{22}$ and $\bar{\epsilon}_{12} = \bar{\epsilon}_{21}$. In the following we drop off the bar on top of the polarization tensor components.

Thus, the gravitational wave in question propagates in the third direction and is transversally polarized along the first and the second directions. The corresponding metric tensor is:

\bqa\label{234}
ds^2 = dt^2 - \left(1 + h_{11}\right)\, dx^2 - 2\,h_{12} \, dxdy - \left(1 - h_{11}\right)\, dy^2 - dz^2,
\eqa
solves linearized approximation to the Einstein equations. Here

\begin{figure}
\begin{center}
\includegraphics[scale=0.5]{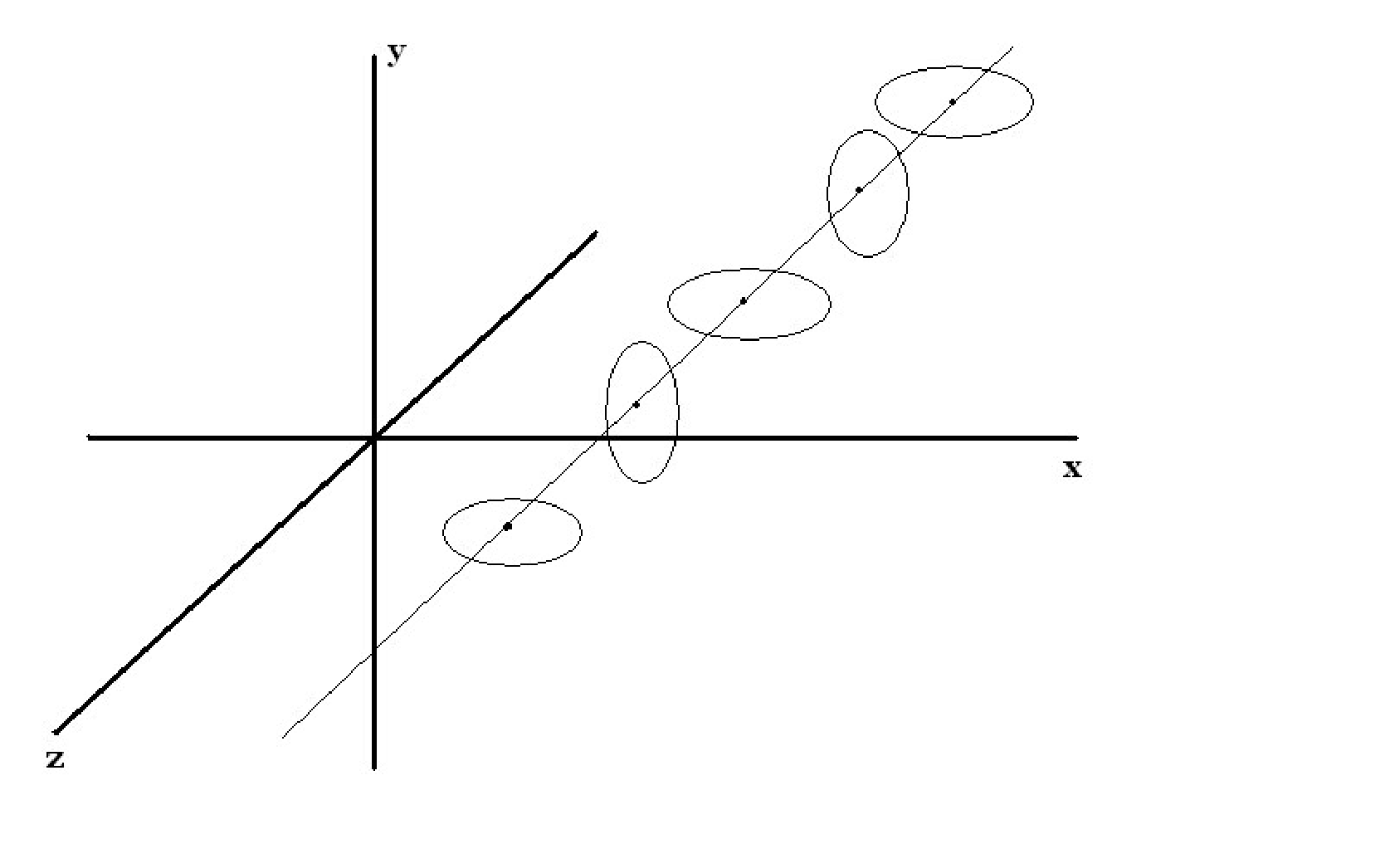}\caption{}\label{fig17}
\end{center}
\end{figure}

\bqa\label{235}
h_{11} = \left|\epsilon_{11}\right| \, \cos\left[k(z-t) + \phi_0\right]\quad {\rm and}
\quad h_{12} = \left|\epsilon_{12}\right|\, \cos\left[k(z-t) + \psi_0\right],
\eqa
where $\phi_0$ and $\psi_0$ are initial phases hidden in the complex components $\epsilon_{11}$ and $\epsilon_{12}$ of the polarization tensor. The corresponding picture of the space--time curving is shown on the fig. (\ref{fig17}). This picture shows for the case of, when $\phi_0 = 0$ and $\psi_0 = \pi/2$, how transversal directions are affinely transformed during the propagation of the wave. Note that the Riemann tensor for the wave (\ref{234}), (\ref{235}) is not zero and, hence, (\ref{234}) cannot be mapped to the Minkowskimetric by a coordinate transformation. So the mentioned above affine transformations correspond to physical expansions and contractions of the transversal directions, which cannot be attributed to a simple coordinate redefinitions.

{\bf 4.} To complete our study of the gravitational energy--momentum pseudo--tensor let us calculate the energy--momentum carried by the above free gravitational wave. As we have seen above, for
the gravitational wave $h^\mu_\mu = 0$. Hence, $|g| \approx |\eta| = 1$. Furthermore, at the linear order

\bqa
G^{\mu\nu}_{,\alpha} \approx g^{\mu\nu}_{,\alpha} \approx - h^{\mu\nu}_{,\alpha}.
\eqa
As the result, for the \underline{free plane--wave} the non--zero contributions to $t_{\mu\nu}$ in (\ref{tmunu}) are contained in the term as follows:

\bqa
\frac12 \, g^{\mu\alpha} \, g^{\nu\beta} \, g_{\gamma\delta} \, g_{\sigma\xi} \, g^{\gamma\xi}_{,\alpha} \, g^{\delta\sigma}_{,\beta} \approx \frac12 \, h^{\alpha,\mu}_\beta \, h^{\beta,\nu}_\alpha.
\eqa
Thus, in this case we obtain that

\bqa %\label{235}
t^{\mu\nu} \approx \frac{1}{32\, \pi \, \kappa} \, h^{\alpha,\mu}_\beta \, h^{\beta,\nu}_\alpha.
\eqa
For the plane--wave under consideration we have that $h_{\mu\nu}(\underline{x}) = h_{\mu\nu}(t-z)$ and the only non--zero components are $h_{11} = - h_{22}$ and $h_{12} = h_{21}$. Then the only non--vanishing non--diagonal component of $t_{\mu\nu}$ is equal to:

\bqa
t^{03} \approx \frac{1}{16 \, \pi \, \kappa} \, \left[\dot{h}_{12}^2 + \dot{h}_{11}^2\right].
\eqa
This means that we have energy flux in the third direction --- in the direction of the wave propagation. Recently (in 2016) gravitational waves have been
observed.

\vspace{10mm}

\centerline{\bf Problems:}

\vspace{5mm}

\begin{itemize}

\item Derive equations (\ref{conslaw}), (\ref{203}) and (\ref{defeta}).

\item Derive equations (\ref{217}) and (\ref{220}).

\item Derive equation (\ref{235})

\end{itemize}

\vspace{10mm}

\centerline{\bf Subjects for further study:}

\vspace{5mm}

\begin{itemize}

\item Linearized perturbations over the Schwarzschild background. (See e.g. ``The mathematical theory of black holes'', S. Chandrasekhar, Oxford University Press, 1992)

\item Experimental observations of gravitational waves.

\end{itemize}

\newpage

\section*{LECTURE X \\{\it  Gravitational radiation by moving massive bodies. Shock gravitational wave or Penrose parallel plane wave.}}

\vspace{10mm}

{\bf 1.} In the previous lecture we have considered free gravitational waves. Here we continue with the creation of such waves by massive bodies. Then, we have to solve the following equation:

\bqa\label{BOXT}
- \Box \psi_{\mu\nu} \approx 16\, \pi \, \kappa \, T_{\mu\nu},
\eqa
in the gauge $\partial^\mu \psi_{\mu\nu} = 0$, where

\bqa
\psi_{\mu\nu} \equiv h_{\mu\nu} - \frac12 \, \eta_{\mu\nu} \, h, \quad {\rm and} \quad h\equiv h^\alpha_\alpha.
\eqa
Here $T_{\mu\nu}$ is the matter stress--energy tensor, which contains terms that are $\sim {\cal O} (h)$ and the gauge under consideration is in the agreement with the stress--energy conservation $\partial^\mu T_{\mu\nu} = 0$ in the linear approximation.

Here we consider the creation of the radiation in the non--relativistic approximation. To solve the eq. (\ref{BOXT}) one has to use the retarded Green function of the d'Alembert operator. From the course of classical electrodynamics it is known that this function is as follows:

\bqa
\Box G_R\left(\underline{x} - \underline{y}\right) = 4\, \pi \, \delta^{(4)}\left(\underline{x} - \underline{y}\right), \nonumber \\
G_R\left(\underline{x} - \underline{y}\right) = \frac{\delta\left(x_0 - y_0 - \left|\vec{x} - \vec{y}\right|\right)}{\left|\vec{x} - \vec{y}\right|},
\eqa
where $\underline{x} = \left(x^0, \, x^1, \, x^2,\, x^3\right)$. Hence, the solution of (\ref{BOXT}) has the form:

\bqa\label{GRrad}
\psi_{\mu\nu}\left(t, \vec{x}\right) \approx - 4\, \kappa \, \int d^3\vec{y} \, \frac{T_{\mu\nu}\left(\vec{y}, t - \left|\vec{x} - \vec{y}\right|\right)}{\left|\vec{x} - \vec{y}\right|}.
\eqa
Suppose now that massive bodies, which create the radiation, perform their motion in a finite region $V$ of the space, i.e. $T_{\mu\nu}\left(t_y, \vec{y}\right) \neq 0$ only if $\vec{y} \in V$. We want to consider the gravitational field in the so called wave zone, i.e. very far away from the region where the radiation was created. Hence, in the above expressions we assume that $\left|\vec{x}\right| \gg \left|\vec{y}\right|$, for $\vec{y} \in V$. Then one can neglect $y$ in comparison with $x$ in the denominator of (\ref{GRrad}). Moreover, if we consider non--relativistic motion of the radiating bodies, then we also can neglect $y$ in comparison with $x$ in the argument of $T_{\mu\nu}$ under the integral in (\ref{GRrad}).

Thus, for $\left|\vec{x}\right| \gg \left|\vec{y}\right|$ the field in (\ref{GRrad}) actually solves the homogeneous form of the equation (\ref{BOXT}), i.e. with $T_{\mu\nu} = 0$, because we consider it outside the region $V$. As the result, the gravitational field, $\psi_{\mu\nu}$ or $h_{\mu\nu}$, in eq. (\ref{GRrad}) has the same properties as free waves of the previous lecture, i.e. $h^\alpha_\alpha = 0$ and $\psi_{\mu\nu} = h_{\mu\nu}$. Moreover, the latter $h_{\mu\nu}$ has only spatially directed non--zero components. All in all the solution that we are looking for is as follows:

\bqa\label{hij}
h_{ij}\left(t, \vec{x}\right) \approx - \frac{4\, \kappa}{\left|\vec{x}\right|} \, \int_V d^3\vec{y} \, T_{ij} \left(t - \left|\vec{x}\right|, \vec{y}\right), \quad i,j = 1,2,3.
\eqa
Here $t$ is the observation time, while $t - \left|\vec{x}\right|$ is the moment of the creation of radiation, which is observed at $t$.

In this expression $T_{ij}$ depends on details of motion of radiating massive bodies rather than just on their mass distribution. To simplify this expression we will transform it into such a form that it will depend only on the mass distribution. For this reason consider the spatial part of the stress--energy conservation: $\partial^\mu T_{\mu i}(y) = 0$. Multiply it by $y_j$ and integrate over a spatial section:

\bqa
0 = \int d^3 \vec{y} \, y_j \, \partial^\mu T_{\mu i} = \int d^3 \vec{y} \, y_j \, \left(\partial_t T_{0 i} - \partial_k T_{ki}\right).
\eqa
After the integration by parts in the second term in the last expression and dropping off the boundary term (as usual), we arrive at the following relation:

\bqa
0 = \partial_t \int d^3 \vec{y} \, y_j \, T_{0 i} + \int d^3 \vec{y} \, T_{ji}.
\eqa
Taking into account that $T_{ij} = T_{ji}$, we obtain the equation:

\bqa
\int d^3 \vec{y} \, T_{ij} = - \frac12 \, \partial_t \int d^3\vec{y} \, \left(y_j \, T_{0i} + y_i \, T_{0j}\right).
\eqa
Similarly multiplying the equation $\partial^\mu T_{\mu 0} = 0$ by $y_k\, y_l$ and integrating over a spatial section we find that:

\bqa
0 = \int d^3 \vec{y} \, y_k \, y_l \, \partial^\mu T_{\mu 0} = \int d^3 \vec{y} \, y_k \, y_l \, \left(\partial_t T_{00} - \partial_m T_{m0}\right) = \nonumber \\ = \partial_t \int d^3\vec{y} \, y_k \, y_l \, T_{00} + \int d^3\vec{y} \, \left(y_k \, T_{0l} + y_l \, T_{0k}\right).
\eqa
Thus, from these relations and from (\ref{hij}) we obtain that:

\bqa\label{hij1}
h_{ij} \approx - \frac{2\kappa}{\left|\vec{x}\right|}  \, \partial_t^2 \, \int d^3\vec{y} \, y_i \, y_j \, T_{00}.
\eqa
Now take into account that for the non--relativistic system under consideration the $T_{00}$ component of the energy--momentum tensor coincides with the mass density: $T_{00} \approx \rho$. Also in the wave zone $h^i_i = 0$
and, hence, we can subtract from $h_{ij}$ in eq. (\ref{hij1}) its trace. As the result, we obtain that

\bqa\label{quadr}
h_{ij} \approx - \frac{2\, \kappa}{3\, \left|\vec{x}\right|} \, \ddot{Q}_{ij}, \quad {\rm where} \nonumber \\
Q_{ij} \equiv \int d^3 \vec{y} \, \rho\left(t - \left|\vec{x}\right|, \vec{y}\right) \, \left(3\, y_i \, y_j - \delta_{ij} \, \vec{y}^2 \right)
\eqa
is the quadruple moment of the mass distribution.

This answer is quite natural and can be understood on general physical grounds. In fact, in the case of electromagnetic radiation the corresponding {\it vector} field is proportional to the first derivative of the dipole moment, $A_i \sim \dot{d}_i$. Hence, in the gravitational case it is natural to expect that the corresponding radiation {\it tensor} field is proportional to the second derivative of the quadruple moment: $h_{ij} \sim \ddot{Q}_{ij}$. Furthermore, in the electromagnetic case the electric and magnetic dipole radiation vanish for such a system in which the gyromagnetic ratios $e/m$ of all charges composing the system are the same. The radiation type that is not zero in such a case is the quadruple one. But for the gravitational radiation the role of the charge is played by the mass. Hence, in such a case ``$e/m$'' is always the same. Then it is natural to expect that the quadruple radiation is the first non--zero contribution in the multiple expansion.

{\bf 2.} Let us find now the intensity of the radiation under consideration. As we have seen in the previous lecture, if the radiated wave is traveling along the third direction, then:

\bqa
t^3_0 \approx \frac{1}{32\, \pi \, \kappa} \, \dot{h}_{ij}\,\dot{h}_{ij},
\eqa
and the indexes here run only over $i,j = 1,2$. However, in generic situation waves are created by randomly moving bodies and, then, radiation is going in all directions. Hence, we have to average the energy flux over all spatial directions. For that reason we need to find $h_{ij}$ corresponding to a wave which is traveling in an arbitrary direction specified by a unit vector $\vec{n}$, $\vec{n}^2 = 1$. Such a $\tilde{h}_{ij}$ tensor should be symmetric, $\tilde{h}_{ij} = \tilde{h}_{ji}$, traceless, $\tilde{h}_{ii} = 0$, and transversal to $\vec{n}$, i.e. $\tilde{h}_{ij} \, n_j = 0$. The transversal tensor has the following form:

\bqa
h_{ij}^\perp = h_{ij} - n_i \, n_k \, h_{kj} - n_j \, n_k \, h_{ik} + n_i \, n_j \, n_k \, n_l \, h_{kl}.
\eqa
where $h_{ij}$ is defined above. The transverse traceless tensor is then equal to:

\bqa
\tilde{h}_{ij} = h^\perp_{ij} - \frac12 \, \left(\delta_{ij} - n_i \, n_j\right)\, h^\perp_{ll}.
\eqa
From this equation and the condition that $n_i^2 = 1$ we find:

\bqa
\tilde{h}_{ij} \, \tilde{h}_{ij} = h_{ij} \, h_{ij} - 2 \, n_i \, n_j \, h_{ki} \, h_{jk} + \frac12 \, \left(n_i \, n_j \, h_{ij}\right)^2.
\eqa
Taking into account the following expressions for the averages over all spatial directions:

\bqa
\left\langle n_i \, n_j \right\rangle = \frac13 \, \delta_{ij}, \nonumber \\
\left\langle n_i \, n_j \, n_k \, n_l \right\rangle = \frac{1}{15} \, \left(\delta_{ij} \, \delta_{kl} + \delta_{ik} \, \delta_{jl} + \delta_{il} \, \delta_{jk}\right),
\eqa
we obtain that:

\bqa
\left\langle \tilde{h}_{ij} \, \tilde{h}_{ij}\right\rangle = \frac{2}{5}\, h_{ij} \, h_{ij}.
\eqa
Here:

\bqa
\left\langle n_{i_1} \dots n_{i_N}\right\rangle \equiv \frac{1}{4\, \pi} \, \int_0^{2\pi} d \varphi \int_0^\pi d\theta \, \sin \theta \, n_{i_1} \dots n_{i_N} \quad {\rm and} \nonumber \\ \vec{n} = \left(\sin(\theta) \, \cos(\varphi), \sin(\theta) \, \sin(\varphi), \cos(\theta)\right).
\eqa
Combining all these formulas together and using (\ref{quadr}), we obtain  the following expression for the radiation intensity:

\bqa
I \equiv \oint d\sigma_i \, t_{0}^i  \approx 4\,\pi \, \left|\vec{x}\right|^2 \, \frac{1}{32\, \pi \, \kappa} \, \left\langle \dot{\tilde{h}}_{ij} \,\dot{\tilde{h}}_{ij}\right\rangle \approx \frac{\kappa}{45} \, \dddot{Q}_{ij} \, \dddot{Q}_{ij},
\eqa
where $d\sigma^i$ is the three--vector orthogonal to the surface at spatial infinity and whose norm is equal to the volume element of that surface. The integral $\oint d\sigma_i \dots$ does the averaging over the angles.
The obtained equation defines the total intensity of the gravitational radiation in the non--relativistic quadruple approximation.

Energy loss due to the radiation of the gravitational waves in binary star systems have been observed at the end of the XX-th century.

{\bf 3.} So far we have considered approximate solutions of the Einstein equations, which describe weak gravitational waves in the linearized approximation. In the remaining part of this lecture we present an exact solution describing so to say gravitational shock wave.

Consider the Schwarzschild metric and make the following change of the radial coordinate

$$
r = \left(1 + \frac{r_g}{4\rho}\right)^2 \, \rho.
$$
Then the metric is transformed into the following form:

\bqa\label{origmetr}
ds^2 = \left(\frac{1 - \frac{r_g}{4\rho}}{1 + \frac{r_g}{4\rho}}\right)^2\, dt^2 - \left(1 + \frac{r_g}{4\rho}\right)^4 \, \left(d\rho^2 + \rho^2 \, d\Omega^2\right).
\eqa
In this metric the spatial part is conformaly flat. Hence, we can make a change from the spherical coordinates, $(\rho,\theta,\varphi)$, to the Cartesian ones $(x,y,z)$: e.g., $\rho^2 = x^2 + y^2 + z^2$. Then, $d\rho^2 + \rho^2 \, d\Omega^2 = dx^2 + dy^2 + dz^2$.

Now we make a coordinate transformation, which describes the Lorentz boost along the $z$ direction:

\bqa
\bar{t} = \gamma \, t + \gamma\, v\, z, \quad {\rm and} \quad \bar{z} = \gamma \, v\, t + \gamma\, z, \quad {\rm where} \quad \gamma = \frac{1}{\sqrt{1 - v^2}}.
\eqa
Because our metric is not Minkowskain such a transformation does change it to:

\bqa\label{1+A}
ds^2 = \left(1 + A\right)^4 \, \left(d\bar{t}^2 - dx^2 - dy^2 - d\bar{z}^2\right) - \left\{\left(1 + A\right)^4 - \left(\frac{1-A}{1+A}\right)^2\right\} \, \left(d\bar{t} - v \, d\bar{z}\right)^2 \, \gamma^2,
\eqa
where

\bqa
A = \frac{r_g}{4\, \gamma \, \sqrt{\left(\bar{z} - v \, \bar{t}\right)^2 + \frac{x^2 + y^2}{\gamma^2}}}.
\eqa
The new metric describes the motion of the black hole with the constant velocity $v$ along the third direction.

Now let us take the limit $\gamma \to \infty$ ($v \to 1$) and $r_g \to 0$ in such a way that $\frac{r_g \, \gamma}{2} \equiv \kappa \, p$ remains finite.
In this limit $A\to 0$ and

\bqa
\lim_{v\to 1}  \frac{r_g \, \gamma}{4\, \sqrt{\left(\bar{z} - v \, \bar{t}\right)^2 + \frac{x^2 + y^2}{\gamma^2}}} = \frac{\kappa \, p}{2\, \left|\bar{z} - \bar{t}\right|}, \quad {\rm for} \quad \bar{z} \neq \bar{t},
\eqa
and this expression is divergent for the case when $\bar{z} = \bar{t}$. (In making these derivations we use the results of ``On the gravitational field of a massless particle'', by P.C.Aichelburg and R.U.Sexl, published in General Relativity and Gravitation, December 1971, Volume 2, Issue 4, pp 303-312).
 As a result, in this limit we obtain the metric as follows:

\bqa
ds^2 = d\bar{t}^2 - dx^2 - dy^2 - d\bar{z}^2 - \frac{4\,\kappa \, p}{\left|\bar{t} - \bar{z}\right|} \, \left(d\bar{t} - d\bar{z}\right)^2.
\eqa
Beyond the surface $\bar{z} = \bar{t}$ the obtained metric is flat. In fact, for $\bar{z} > \bar{t}$, we have:

\bqa
ds_+^2 = du \, dv_+ - dy^2 - dx^2, \quad {\rm where} \quad u = \bar{t} - \bar{z}, \nonumber \\
{\rm and} \quad v_+ = \bar{t} + \bar{z} + 2 \, p \, \kappa \, \log(\bar{z} - \bar{t}),
\eqa
while for $\bar{z} < \bar{t}$ we have that:

\bqa
ds_-^2 = du \, dv_- - dy^2 - dx^2, \quad {\rm where} \quad u = \bar{t} - \bar{z}, \nonumber \\
{\rm and} \quad v_- = \bar{t} + \bar{z} - 2 \, p \, \kappa \, \log(\bar{t} - \bar{z}).
\eqa
To define the metric at $\bar{z} = \bar{t}$ let us make the following coordinate transformation:

\bqa
z' - v \, t' = \bar{z} - v \, \bar{t}, \nonumber \\
z' + v \, t' = \bar{z} + v \, \bar{t} - 4 \,\kappa \,  p \, \log \left[\sqrt{\left(\bar{z} - v \, \bar{t}\right)^2 + \frac{1}{\gamma^2}} - \left(\bar{z} - \bar{t}\right)\right].
\eqa
After such a transformation the leading form of the metric (\ref{1+A}) in the limit in question is as follows:

\bqa
ds^2 = dt^2 - dx^2 - dy^2 - dz^2 - 4 \, \kappa \, p \, \left\{\frac{1}{\sqrt{(\bar{z} - v \, \bar{t})^2 + \frac{x^2 + y^2}{\gamma^2}}} - \frac{1}{\sqrt{(\bar{z} - v \, \bar{t})^2 + \frac{1}{\gamma^2}}}\right\} \, \left(dt - dz\right)^2.
\eqa
Here we have dropped off the primes over $z$ and $t$. Taking into account that

\bqa
\lim_{v \to 1} \left\{\frac{1}{\sqrt{(\bar{z} - v \, \bar{t})^2 + \frac{x^2 + y^2}{\gamma^2}}} - \frac{1}{\sqrt{(\bar{z} - v \, \bar{t})^2 + \frac{1}{\gamma^2}}}\right\} = - \delta\left(\bar{z} - \bar{t}\right) \, \log\left(x^2 + y^2\right),
\eqa
we arrive at the following expression for the metric tensor:

\bqa\label{Hux}
ds^2 = du \, dv + 2 \, H\left(u, \vec{x}_\perp\right)\, du^2 - d\vec{x}_\perp^2,
\eqa
where

\bqa
u = t-z, \quad v = t+z, \quad \vec{x}_\perp = (x, y),
\quad {\rm and} \quad
H\left(u, \vec{x}_\perp\right) = 4 \, \kappa \, p \, \delta(u) \, \log\left|\vec{x}_\perp\right|.
\eqa
Thus, in this limit the Schwarzschild black hole is transformed to a solution describing a shock wave, which is moving with momentum $p$ and with the speed of light along the $z$ direction. It is not hard to check that the only non--zero component of the Ricci tensor for (\ref{Hux}) has the following form --- $R_{uu} = \Delta_{x_\perp} H\left(u,\vec{x}_\perp\right)$, where $\Delta_{x_\perp} = \partial_x^2 + \partial_y^2$.
%(Note that it means that $H\left(u,\vec{x}_\perp\right)$ is a harmonic function $\Box H\left(u,\vec{x}_\perp\right) = 0$ outside the world--line of the wave, $u=0$ and $\vec{x}_\perp = 0$.)

The metric tensor (\ref{Hux}) is the exact solution of the Einstein equations with the matter energy--momentum tensor equal to $T_{uu} = p \, \delta(u) \, \delta^{(2)}\left(\vec{x}_\perp\right)$, because $\Delta_{x_\perp} \log \left|\vec{x}_\perp\right| = 2\, \pi\, \delta^{(2)}\left(\vec{x}_\perp\right)$. This solution is referred to as \underline{Penrose parallel plane wave}. The proof of all the above statements is left as an exercise for the reader. Such an energy--momentum tensor, $T_{uu}$, corresponds to a shock wave traveling with momentum $p$ along the world line $u = 0$ and $\vec{x}_\perp = 0$. The metric in question is defined beyond the world--line of the gravitational source, where there is the singularity. Note that the metric under consideration is not asymptotically flat and has a naked singularity.

\begin{figure}
\begin{center}
\includegraphics[scale=0.5]{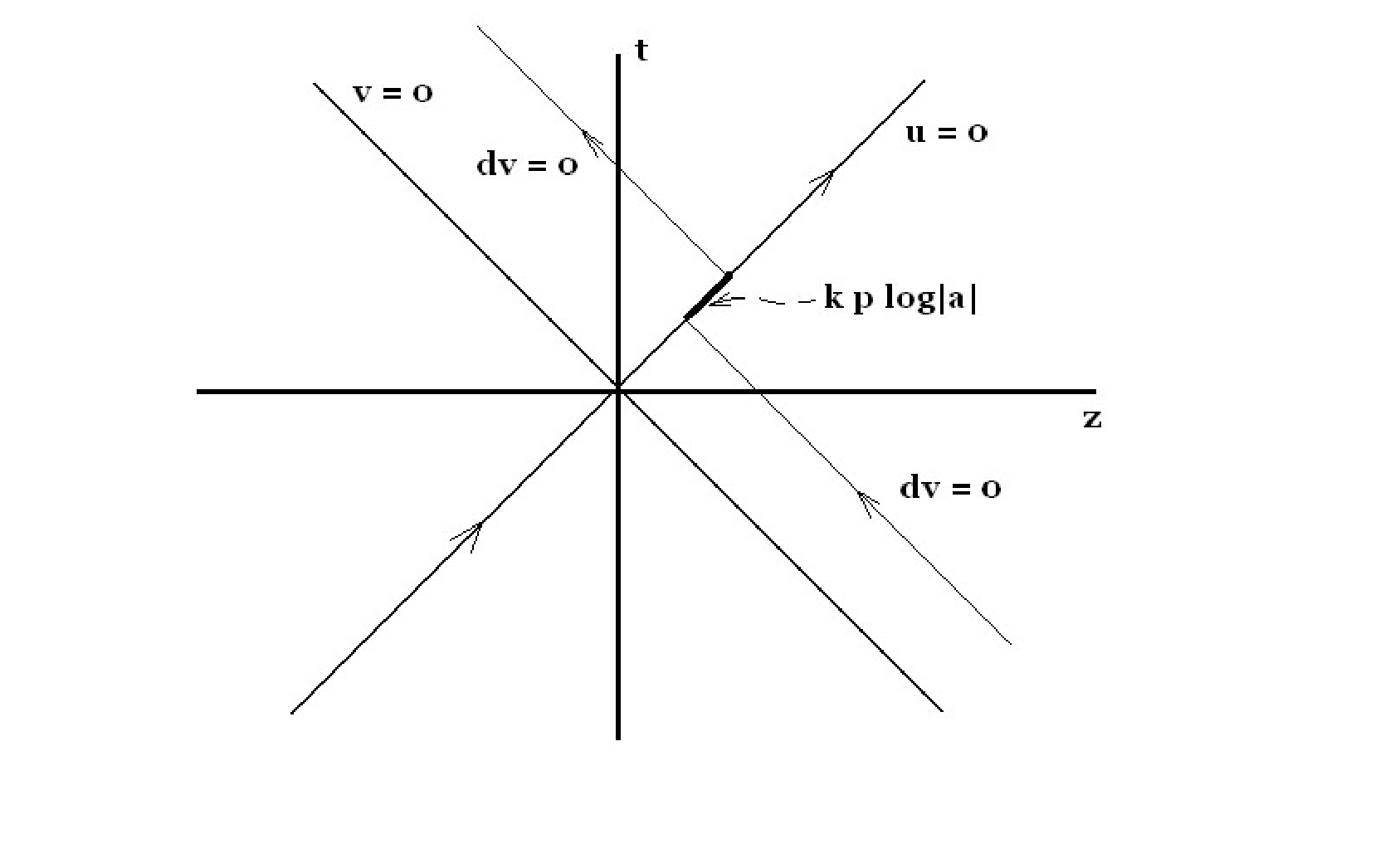}\caption{}\label{fig18}
\end{center}
\end{figure}

{\bf 4.} Let us describe the behavior of light rays in the background of (\ref{Hux}). Light rays which are moving some distance, $\vec{x}_\perp = \vec{a}$ (hence, $d\vec{x}_\perp = 0$), away from the source of the gravity are described by the following equation:

\bqa
0 = du \, \left[dv + 2\, H\left(u, \vec{a}\right)\, du\right].
\eqa
Hence, those light rays which are propagating parallel to the world--line of the shock wave, $u=0$, (i.e. their world--line is $u = const$) never feel its influence and always travel according to the law $du = 0$.
However, those light rays which collide with the shock wave, with the impact parameter $\vec{a}$, are moving along the world--line described by the equation:

\bqa
dv = - 8\, \kappa \, p \, \delta(u) \, \log\left|\vec{a}\right| \, du.
\eqa
Before the collision with the wave (i.e. before $u=0$) they move as if they are in flat space, $dv=0$. (See the fig. (\ref{fig18}).) During the collision the light is carried over by the shock wave for the distance

\bqa
8\, \kappa \, p\, \int_{0-0}^{0+0} du \, \delta(u) \, \log\left|\vec{a}\right| = 8 \, \kappa \, p \, \log\left|\vec{a}\right|
\eqa
along the $v$ direction. And then such a light ray is released to continue its free propagation according to the law $dv = 0$.

\begin{figure}
\begin{center}
\includegraphics[scale=0.5]{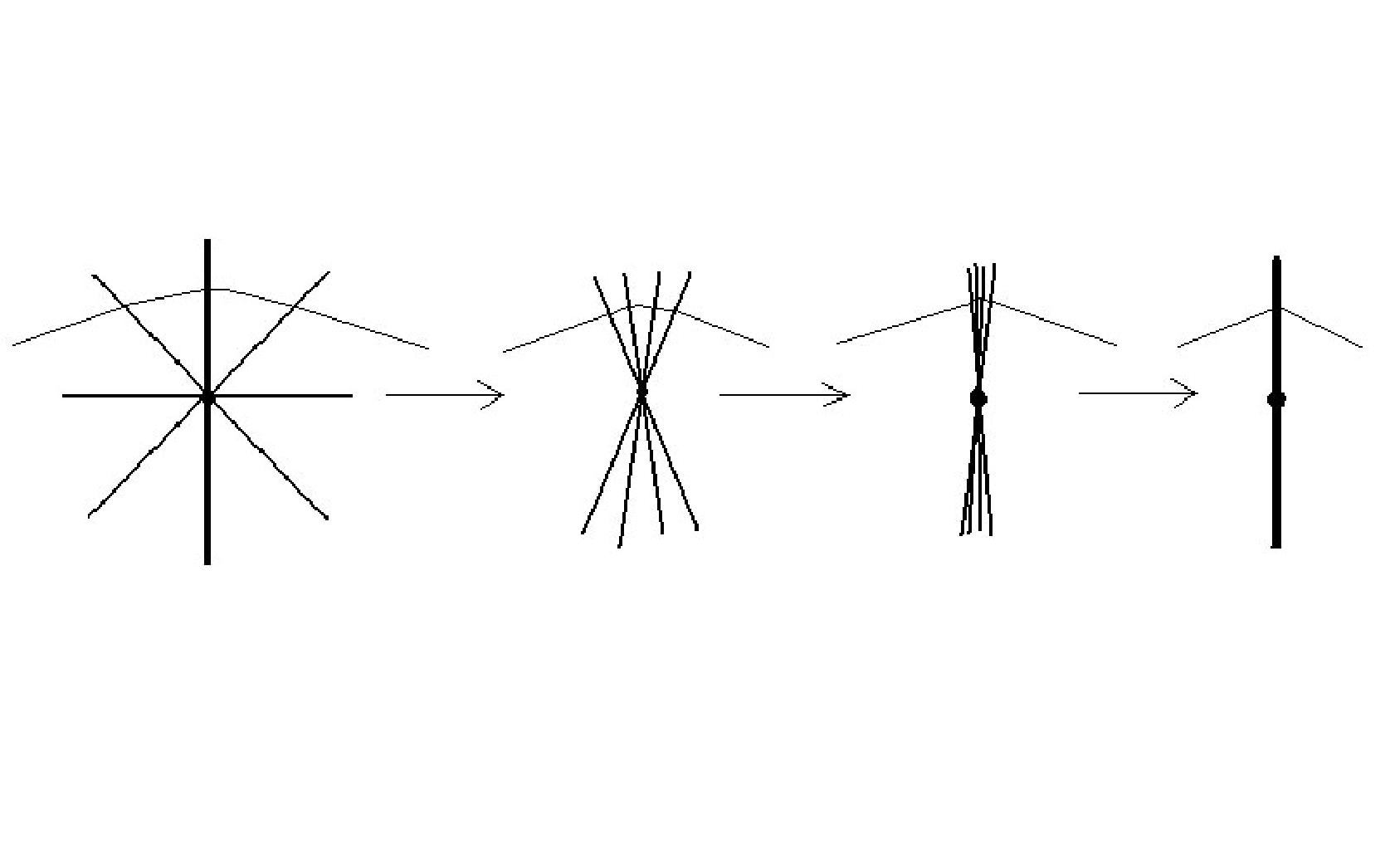}\caption{}\label{fig19}
\end{center}
\end{figure}

The reason for such a behavior of light rays in the background of the shock wave can be understood on general grounds. Field lines of a static massive body are spherically symmetric. However, after a boost, due to the Lorentz contraction, the density of these lines is increased in the directions that are transversal to the boost axis. For a faster boost the density increase is stronger. See the fig. (\ref{fig19}). In the case of the infinite
boost one obtains so to say a pancake spanned in the directions perpendicular to the propagation axis.

A light ray experiences a time delay while passing nearby a gravitating body: The main portion of the delay happens in the dense part of its field lines. As we boost the gravitating body the region where the time delay is happening is shrinking. In the above defined limit the entire time delay is happening when the light ray is crossing the shock wave (pancake), as is shown on the fig. (\ref{fig19}).

\vspace{10mm}

\centerline{\bf Problems:}

\vspace{5mm}

\begin{itemize}

\item Derive eq. (\ref{origmetr}).

\item Show that (\ref{Hux}) solves Einstein equations with the defined in the lecture energy--momentum tensor.

\item Find the change in time of the radius of a double--star system due to the gravitational radiation. (See the corresponding paragraph in Landau and Lifshitz)

\item Find the resonance transformation of the high--energy electromagnetic wave into the gravitational one. (See the corresponding paragraph in Khriplovich book.)

\end{itemize}

\vspace{10mm}

\centerline{\bf Subjects for further study:}

\vspace{5mm}

\begin{itemize}

\item ADM gravitational mass and other definitions of the mass in the General theory of Relativity.

\item Petrov's classification of Einstein space--times.

\item Gravitational memory and Bondi--Metzner--Sachs transformations at asymptotic infinity.

\end{itemize}

\newpage

\section*{LECTURE XI \\{\it  Homogeneous three--dimensional spaces. Friedmann--Robertson--Walker metric. Homogeneous isotropic cosmological solutions. Anisotropic Kasner cosmological solution.}}

\vspace{10mm}

{\bf 1.} At the scales of the Sun system our Universe is highly inhomogeneous. In fact, we see compact gravitating bodies surrounded by large empty spaces. However, on scales of the order of several hundreds of millions of parsecs (bigger than a characteristic size of galactic clusters) the distribution of matter in the Universe is isotropic and homogeneous to a high precision. Moreover, observational data show that distant galaxies in all directions are running away from our own galaxy. In this lecture we would like to find solutions of Einstein equations which describe such a behavior of the homogeneously distributed matter.

The most general four--dimensional metric tensor, which describes isotropic and homogeneous spatial sections as seen by inertial (free floating) observers, is as follows:

\bqa
ds^2 = dt^2 - a^2(t) \, \gamma_{ij}(x) \, dx^i \, dx^j, \quad i,j = 1,2,3.
\eqa
Here $dl^2 = \gamma_{ij}(x) \, dx^i \, dx^j$ describes spatial sections and their size is changing according to the time dependence of $a(t)$. Because of the homogeneity all points of spatial sections and all directions from these points are equivalent to each other. Hence, the three--dimensional metric, $dl^2$, under consideration should correspond to a space of constant curvature, as we will see now.

{\bf 2.} Metric tensor in a space of constant curvature can be written as follows:

\bqa
dl^2 = \gamma_{ij}(x) \, dx^i \, dx^j = \frac{dr^2}{1 - k \, r^2} + r^2\, d\Omega^2, \quad d\Omega^2 = d\theta^2 + \sin^2\theta \, d\varphi^2.
\eqa
Here, $k$ can be either $-1$, $0$ or $1$. In fact,

\begin{itemize}

\item When, $k=0$ we have that $dl^2 = dr^2 + r^2\, d\Omega^2$, which is just the flat three--dimensional Euclidian space represented in the spherical coordinates --- the space of the constant zero curvature.

\item If $k=1$, one deals with the metric:

\bqa\label{269}
dl^2 = \frac{dr^2}{1 - r^2} + r^2 \, d\Omega^2.
\eqa
After the coordinate change $r = \sin\chi$ this metric is transformed into:

\bqa\label{3dsphere}
dl^2 = d\chi^2 + \sin^2\chi \, d\Omega^2,
\eqa
which is just the three--dimensional sphere --- the space of constant positive unit curvature. We have encountered this metric in the lecture on the Oppenheimer--Snyder collapse.

For the future convenience let us point out here that the last metric is induced on the unit sphere

\bqa\label{3dsphere1}
W^2 + X^2 + Y^2 + Z^2 = 1,
\eqa
which is embedded into the four--dimensional Euclidian space

\bqa\label{4dEuclid}
dl_4^2 = dW^2 + dX^2 + dY^2 + dZ^2.
\eqa
In fact, if one solves eq. (\ref{3dsphere1}) as $(W,X,Y,Z) = (\cos\chi, \sin\chi \, \sin\theta \, \cos\varphi, \sin\chi \, \sin\theta \, \sin\varphi, \sin\chi \, \cos\theta)$ and, then, substitutes this into (\ref{4dEuclid}), he obtains the induced metric (\ref{3dsphere}). Note that $SO(4)$ rotations of the four--dimensional Euclidian space (\ref{4dEuclid}) leave unchanged the equation (\ref{3dsphere1}) and, hence, the sphere. At the same time, an arbitrary point on the sphere, say $\left(W, \, X, \, Y, \, Z\right) = (1, \, 0, \, 0, \, 0)$ is not moved by an $SO(3)$ rotation. Hence, the sphere is
the homogeneous $SO(4)/SO(3)$ space.

\item Finally, for the case $k=-1$, we have to work with the metric as follows:

\bqa\label{272}
dl^2 = \frac{dr^2}{1 + r^2} + r^2 \, d\Omega^2.
\eqa
If one will perform a coordinate change $r = \sinh \chi$, he will find that the metric is question is transformed into:

\bqa\label{3dhyperb}
dl^2 = d\chi^2 + \sinh^2\chi \, d\Omega^2.
\eqa
This metric can be obtained as follows. Consider the two--sheeted hyperboloid,

\bqa\label{3dhyperb1}
- W^2 + X^2 + Y^2 + Z^2 = - 1,
\eqa
embedded into the four--dimensional Minkowski space--time:

\bqa\label{4dMinkowski}
ds^2 = - dW^2 + dX^2 + dY^2 + dZ^2.
\eqa
One sheet of the hyperboloid corresponds to $W\geq 1$, while the other --- to $W \leq -1$.
The equation defining the hyperboloid under consideration is invariant under the Lorentz rotations, $SO(3,1)$, of the ambient Minkowski space. Hence, the whole hyperboloid can be generated from a single point if one will act by the $SO(3,1)$ group. Moreover, an arbitrary point of this hyperboloid, say $(W,X,Y,Z) = (1,0,0,0)$, does not move under the action of the $SO(3)$ rotational subgroup of $SO(3,1)$ in the spatial, $(X,Y,Z)$, part. As the result the hyperboloid is a homogeneous $SO(3,1)/SO(3)$ space.
Hence, its every point and every direction are equivalent to each other. Then the space has constant curvature. In fact, the hyperboloid is just the three--dimensional Lobachevsky space --- the space of constant negative unit curvature.

One can solve the condition (\ref{3dhyperb1}) as $(W,X,Y,Z) = (\cosh\chi, \sinh\chi \, \sin\theta \, \cos\varphi, \sinh\chi \, \sin\theta \, \sin\varphi, \sinh\chi \, \cos\theta)$. Then, after the substitution of this solution into eq.(\ref{4dMinkowski}), we obtain the metric tensor (\ref{3dhyperb}).

\end{itemize}

{\bf 3.} Thus, in this lecture we will work with the \underline{Friedmann--Robertson--Walker metric}:

\bqa\label{FRWm}
ds^2 = dt^2 - a^2(t) \, \left[\frac{dr^2}{1 - k\, r^2} + r^2\, d\Omega^2\right].
\eqa
The components of this metric tensor are:

\bqa
g_{00} = 1, \quad g_{11} = - \frac{a^2}{1 - k\, r^2}, \quad g_{22} = - a^2\, r^2, \quad g_{33} = - a^2 \, r^2 \, \sin^2\theta.
\eqa
At the same time, the inverse metric tensor is as follows:

\bqa
g^{00} = 1, \quad g^{11} = - \frac{1 - k\, r^2}{a^2}, \quad g^{22} = - \frac{1}{a^2 \, r^2}, \quad g^{33} = - \frac{1}{a^2 \, r^2 \, \sin^2\theta}.
\eqa
Then the corresponding non--zero components of the Christoffel's symbols (up to their symmetries) are:

\bqa
\Gamma^0_{11} = \frac{a\, \dot{a}}{1 - k\, r^2}, \quad \Gamma^0_{22} = r^2 \, a\, \dot{a}, \quad \Gamma^0_{33} = r^2 \, \sin^2\theta \, a\, \dot{a}, \nonumber \\
\Gamma^1_{01} = \frac{\dot{a}}{a}, \quad \Gamma^1_{11} = \frac{k\, r}{1 - k\, r^2}, \quad \Gamma^1_{22} = - r\, \left(1 - k\, r^2\right), \quad \Gamma^1_{33} = - r\, \left(1 - k\, r^2\right)\, \sin^2\theta, \nonumber \\
\Gamma^2_{02} = \frac{\dot{a}}{a}, \quad \Gamma^2_{12} = \frac{1}{r}, \quad \Gamma^2_{33} = - \sin\theta \, \cos\theta, \nonumber \\
\Gamma^3_{03} = \frac{\dot{a}}{a}, \quad \Gamma^3_{13} = \frac{1}{r}, \quad \Gamma^3_{23} = \cot\theta.
\eqa
As the result the Ricci tensor has the form as follows:

\bqa
R_{00} = - \frac{3\, \ddot{a}}{a}, \quad R_{11} = \frac{a\, \ddot{a} + 2\, \dot{a}^2 + 2\, k}{1 - k\, r^2}, \nonumber \\ R_{22} = r^2 \, \left(a\, \ddot{a} + 2\, \dot{a}^2 + 2\, k\right), \quad R_{33} = r^2 \, \sin^2\theta \, \left(a\, \ddot{a} + 2\, \dot{a}^2 + 2\, k\right),
\eqa
and the Ricci scalar is equal to

\bqa
R = - 6\, \frac{a\, \ddot{a} + \dot{a}^2 + k}{a^2}.
\eqa
Here $\dot{a} \equiv da/dt$ and $\ddot{a} \equiv d^2a/dt^2$.

{\bf 4.} To find the dependence of the scale factor $a$ on time we have to solve the Einstein equations. To do that one has to specify the matter energy--momentum tensor, which respects the symmetries of the problem. The appropriate tensor was defined in one of the previous lectures and has the following form:

\bqa
T_{\mu\nu} = \rho(t) \, u_\mu \, u_\nu + p(t) \left[u_\mu\, u_\nu - g_{\mu\nu}\right].
\eqa
The only peculiarity here, which is relevant for cosmology, is that $\rho$ and $p$ are functions of time only.
For the free falling matter in the metric (\ref{FRWm}) its four--velocity is equal to $u^\mu = (1,0,0,0)$. Then the energy--momentum tensor has the form $T_\mu^\nu = Diag\left[\rho(t), - p(t), - p(t), - p(t)\right]$, which obviously respects the homogeneity of spatial sections in the problem under consideration.
In this lecture we set the cosmological constant $\Lambda$ to zero.

Thus, we have that

\bqa
T_{00} = \rho, \quad T_{11} = \frac{p\, a^2}{1 - k\, r^2}, \quad T_{22} = p \, r^2 \, a^2, \quad T_{33} = p\, r^2 \, a^2 \, \sin^2\theta, \quad {\rm and} \nonumber \\
T^{00} = \rho, \quad T^{11} = \frac{p\, \left(1 - k \, r^2\right)}{a^2}, \quad T^{22} = \frac{p}{a^2 \, r^2}, \quad T^{33} = \frac{p}{a^2 \, r^2 \, \sin^2\theta}.
\eqa
Correspondingly the ``$00$'' and ``$11$'' components of the Einstein equations are as follows:

\bqa\label{Friedmanneq}
3\, \left(\dot{a}^2 + k\right) = 8\, \pi\, \kappa \, \rho\, a^2, \nonumber \\
2\, a\, \ddot{a} + \dot{a}^2 + k = - 8\, \pi \, \kappa \, p \, a^2.
\eqa
These are referred to as \underline{Friedmann equations}. The ``$22$'' and ``$33$'' components lead to the equations that are equivalent to the second relation in (\ref{Friedmanneq}). That should be the case due to the symmetries of the problem. Off diagonal components of the Einstein equations lead in this case to the trivial relations of the form $0=0$.

The temporal  component of the energy--momentum conservation condition, $D_\mu T^{\mu 0} = 0$, acquires the form:

\bqa\label{contFri}
\dot{\rho} + 3\, \left(p + \rho\right)\, \frac{\dot{a}}{a} = 0.
\eqa
Spatial components, $D_\mu T^{\mu j} = 0$, $j=1,2,3$, of the conservation law lead to trivial relations. The obtained equation (\ref{contFri}) is not independent, but follows from (\ref{Friedmanneq}). This should be the case, because, as was explained in the previous lectures, the energy--momentum conservation follows from the Einstein equations.

{\bf 5.} Let us consider various solutions of the above Friedmann's equations for different choices of equations of state $p = p(\rho)$. The standard cosmological equation of state is as follows:

\bqa
p = w \, \rho, \quad {\rm where} \quad w = const.
\eqa
The two peculiar cases are represented by the choices $w = 0$ and $w = 1/3$. The first case $w=0$ is the familiar dust matter. In cosmology it describes galactic matter spread over the universe.
The second case, $w = 1/3$ corresponds, to the traceless, $T_\mu^\mu = 0$, stress--energy tensor. This situation one encounters, e.g., in the case of the Maxwell theory. Hence, this is said to correspond to the radiation.

For $w = 0$ the eq. (\ref{contFri}) is reduced to $\rho(t) \cdot a^3(t) = const$, which just means that if the size of spatial directions is scaling as $a(t)$, i.e. if the spatial volume behaves as $a^3(t)$, then the density of dust should change as $\rho(t) \sim 1/a^3(t)$.

At the same time, when $p = \rho/3$ from (\ref{contFri}) we obtain that $\rho(t) \cdot a^4(t) = const$.
The explanation of this relation is as follows. The {\it number} density of radiation is scaling in the same way as that of the dust $\sim 1/a^3(t)$. But on top of that the wave length of the radiation is scaling as $a(t)$. Hence, the frequency (or energy) is changing as $1/a(t)$. As the result, the radiation {\it energy} density behaves as $\rho(t) \sim 1/a^4(t)$.

To solve the Friedmann's equations it is convenient to change in (\ref{FRWm}) the time coordinate according to $dt = a (\eta) \, d\eta$. Then we obtain the metric as follows:

\bqa
ds^2 = a^2(\eta) \, \left[d\eta^2 - \frac{dr^2}{1 - k\, r^2} - r^2 \, d\Omega^2\right],
\eqa
where $\eta$ is the so called \underline{conformal time}.

From the first equation in (\ref{Friedmanneq}) we find that:

\bqa\label{maint}
dt = \pm \frac{da}{\sqrt{\frac{8\,\pi \, \kappa}{3} \, a^2\, \rho - k}}.
\eqa
Here the ``$+$'' and ``$-$'' signs correspond to the solutions that are related to each other by the time--reversal transformation. From this equation we find the conformal time:

\bqa\label{maineta}
\eta = \pm \int \frac{da}{a\, \sqrt{\frac{8\,\pi \, \kappa}{3} \, a^2\, \rho - k}}.
\eqa
Let us calculate this integral to obtain $a(\eta)$ and then find $t(\eta)$ from $dt = a(\eta) \, d\eta$ for all six cases
$k = -1, 0, 1$ and $w = 0, 1/3$. We restrict ourselves to the ``$+$'' sign in (\ref{maineta}).

\begin{itemize}

\item \underline{If $k=1$, $w=0$}, it is convenient to introduce the notations:

\bqa
\rho a^3 \equiv \frac{M}{2\, \pi^2} = const.
\eqa
Here we have introduced $M$, which is so to say the total mass of the compact Universe.
In this case from (\ref{maineta}) we obtain that

\bqa
a(\eta) = \frac{2\, \kappa \, M}{3\, \pi} \, \left(1 - \cos\eta\right).
\eqa
Furthermore, from $dt = a\, d\eta$ we find:

\bqa
t(\eta) = \frac{2\, \kappa \, M}{3\, \pi}\, \left(\eta - \sin\eta\right).
\eqa
This is the solution that we have encountered in the lecture on the Oppenheimer--Snyder collapse, but written in a bit different terms. The function $a(t)$ starts its growth at $t=0$ ($\eta=0$) as $a(t) \sim t^{2/3}$ for small $t$. Then, it approaches its maximal value, $a = \frac{4\, \kappa \, M}{3\, \pi}$, at $t = \frac{2\, \kappa\, M}{3}$ ($\eta = \pi$).
After that $a(t)$ shrinks back to zero at $t = \frac{4\, \kappa \, M}{3}$ ($\eta = 2\, \pi$). At the initial expanding stage the energy--density behaves as $\rho \sim 1/t^2$.

\item \underline{If $k=1$, $w = 1/3$}, we define:

\bqa
\rho \, a^4 \equiv \frac{3\, a_1^2}{8\, \pi\, \kappa} = const.
\eqa
In this case from (\ref{maineta}) we obtain:

\bqa
a(\eta) = a_1 \, \sin\eta, \quad t(\eta) = a_1\, \left(1 - \cos\eta\right).
\eqa
Then, at the beginning of the expanding stage $a(t) \sim \sqrt{t}$ and $\rho \sim 1/t^2$.
Again in this case the Universe starts its expansion from $a=0$, reaches its maximal size and
then shrinks back.

\item \underline{If $k=-1$, $w=0$}, we introduce notations:

\bqa
\rho \, a^3 \equiv \frac{3\, a_0}{4\,\pi \, \kappa} = const.
\eqa
Then,

\bqa
a(\eta) = a_0 \, \left(\cosh\eta - 1\right), \quad t(\eta) = a_0\, \left(\sinh\eta - \eta\right),
\eqa
and in this case $a(t)$ starts its homogeneous eternal growth at $a=0$. At the initial stage $a\sim t^{2/3}$.

\item \underline{If $k=-1$, $w=1/3$}, we define:

\bqa
\rho \, a^4 \equiv \frac{3\, a_1^2}{8\, \pi\, \kappa} = const.
\eqa
Then,

\bqa
a(\eta) = a_1 \, \sinh\eta, \quad t(\eta) = a_1 \, \left(\cosh\eta - 1\right).
\eqa
Again at the initial stage the spatial conformal factor starts as $a(t) \sim \sqrt{t}$ and then $a(t)$ continues its eternal homogeneous growth.

\item \underline{If $k=0$, $w=0$}, then $\rho\, a^3 = const$ and directly from (\ref{maint}) it follows that for all times we have that $a(t) = const \cdot t^{2/3}$.

\item \underline{If $k=0$, $w=1/3$}, as usual we have that $\rho\, a^4 = const$ and from (\ref{maint}) it follows that $a(t) = const \cdot \sqrt{t}$.

\end{itemize}

With a good precision at present stage of its evolution our Universe corresponds to the case of $k=0$ and $w=0$. At the same time at an early stage of its expansion there was a situation when it was described by the case of $k=0$ and $w=1/3$. Recent observations show, however, that the cosmological constant is not zero, but is very small. This is so called dark energy. Its presence will change the cosmological picture in the future of our Universe. We discuss these issues in the next lecture.

In all the above cases the metric tensor degenerates at $t=0$. This is the physical (curvature rather than metric) singularity because it corresponds to the infinite energy density, $\rho$. Usually the problem of this initial singularity is solved via the assumption that the Universe had an initial exponentially expanding stage, which is described by a space--time of constant curvature, i.e. again with non--zero cosmological constant. There are several observational signs approving this assumption. We will discuss this type of space--times in the next lecture.

{\bf 6.} Let us consider here the simplest vacuum {\it anisotropic} (but spatially {\it homogeneous}) cosmological solution. This is the so called \underline{Kasner solution}.

Consider the metric of the form:

\bqa
ds^2 = dt^2 - X^2(t) \, dx^2 - Y^2(t)\, dy^2 - Z^2(t) \, dz^2.
\eqa
Note that if $X=Y=Z=a$ we encounter the Friedmann--Robertson--Walker solution with $k=0$.

For such a metric the non--zero components of the Ricci tensor are as follows:

\bqa\label{ABC11}
R_0^0 = \dot{\Theta} + A^2 + B^2 + C^2 \quad {\rm and} \nonumber \\ R_1^1 = - \dot{A} - \Theta \, A, \quad
R_2^2 = - \dot{B} - \Theta \, B, \quad R_3^3 = - \dot{C} - \Theta \, C,
\eqa
where

\bqa\label{ABC}
A = \frac{\dot{X}}{X}, \quad B = \frac{\dot{Y}}{Y}, \quad C = \frac{\dot{Z}}{Z}, \quad {\rm and} \quad \Theta \equiv A + B + C.
\eqa
The energy--momentum conservation condition is of the following form:

\bqa
\dot{\rho} + \Theta \, \left(p + \rho\right) = 0.
\eqa
Because we consider the vacuum solution, $T_{\mu\nu} = 0$, the ``$00$'' component of the Einstein equations leads to

\bqa
0 = R_{00} - \frac12 \, R = - AB - BC - CA.
\eqa
As the result, from the definition of $\Theta$ we find that:

\bqa\label{Theta}
\Theta^2 = A^2 + B^2 + C^2.
\eqa
Hence, using (\ref{ABC11}) from the equation

\bqa
0 = R_{00} = \dot{\Theta} + \Theta^2
\eqa
we deduce that

\bqa\label{Theta1}
\Theta = \frac{1}{t - t_0}.
\eqa
By a choice of the origin we can set $t_0 = 0$. Then,

\bqa
0 = R_1^1 = - \dot{A} - A\, \Theta, \quad {\rm and} \quad A = \frac{p}{t},
\eqa
for some constant $p$. Similarly we obtain

\bqa
B = \frac{q}{t}, \quad {\rm and} \quad C = \frac{r}{t},
\eqa
for some constants $q$ and $r$.
Next (\ref{Theta}) and (\ref{Theta1}) with (\ref{ABC}) imply that

\bqa\label{relpqr}
p + q + r = 1, \quad {\rm and} \quad p^2 + q^2 + r^2 = 1.
\eqa
And finally from $A = \dot{X}/X = p/t$ we obtain that $X = X_0 \, t^p$ and similarly for $Y$ and $Z$. Thus, the Kasner's metric can therefore be written as:

\bqa
ds^2 = dt^2 - t^{2p}\, dx^2 - t^{2q} \, dy^2 - t^{2r} \, dz^2,
\eqa
after the appropriate rescalings of $x,y$ and $z$. Here $p,q$ and $r$ are subject to (\ref{relpqr}).

{\bf 7.} Let us consider again the Schwarzschild black hole behind the event horizon. There one can also use the same Schwarzschild metric, but with other $t$ and $r$, which do not have an obvious relation to the standard Schwarzschild coordinates:

\bqa
ds^2 = \frac{dr^2}{\frac{r_g}{r} - 1} - \left(\frac{r_g}{r} - 1\right) \, dt^2 - r^2\, d\Omega^2, \quad {\rm where} \quad r < r_g.
\eqa
Here $r$ plays the role of time, while $t$ is space--like. The metric is time--dependent, it is not spherically symmetric and it is not asymptotically flat.

Consider the limit as $r\to 0$. Then this metric simplifies to

\bqa
ds^2 \approx \frac{r}{r_g}\, dr^2 - \frac{r_g}{r}\, dt^2 - r^2 \, d\Omega^2.
\eqa
Making the change of variables:

\bqa
\sqrt{\frac{r}{r_g}}\, dr = dT, \quad \left(\frac{2\, r_g}{3}\right)^{\frac13} \, dt = dX, \quad \left(\frac32 \, \sqrt{r_g}\right)^{\frac23} \, d\theta = dY, \quad \left(\frac32 \, \sqrt{r_g}\right)^{\frac23} \, d\varphi = dZ,
\eqa
we arrive at the following form of this metric:

\bqa
ds^2 \approx dT^2 - T^{-\frac23} \, dX^2 - T^{\frac43}\, dY^2 - T^{\frac43} \, dZ^2,
\eqa
for the case of small distances in $\varphi$ and $\theta$ directions. The reason why we can take small distances in $\varphi, \theta$ directions is because as $T\to 0$, in the approach towards the singularity, points that have large spatial separations (e.g., in $\varphi$ and $\theta$ directions) are causally disconnected.

Thus, the obtained metric is nothing but the Kasner solution with $T\to 0$. Hence, in the vicinity of the physical singularity of the Schwarzschild black hole we have a Kasner type of the behavior of the metric.
This is actually a particular case of a more general situation: in the vicinity of any time--like singularity one encounters Kasner type solution, which, however, does not necessarily have to be of the vacuum type.

\vspace{10mm}

\centerline{\bf Problems:}

\vspace{5mm}

\begin{itemize}

\item Show that (\ref{269}), (\ref{3dsphere}) and (\ref{272}), (\ref{3dhyperb}) solve 3d Einstein equations $R_{ij}(\gamma) = 2\, k \, \gamma_{ij}$ for $k= \pm 1$.

\item Derive (\ref{3dsphere}) from (\ref{3dsphere1}) and (\ref{4dEuclid}).

\item Derive eq. (\ref{Friedmanneq}) and eq. (\ref{contFri})

\item Draw Penrose--Carter diagrams for all the homogeneous cosmological solutions.

\item Find the metric on the hyperboloids $X^2 + Y^2 - Z^2 = 1$ and $X^2 + Y^2 - Z^2 = - 1$ embedded into the three--dimensional Euclidian space $dl^2 = dX^2 + dY^2 + dZ^2$.

\end{itemize}

\vspace{10mm}

\centerline{\bf Subjects for further study:}

\vspace{5mm}

\begin{itemize}

\item Linearized perturbations in cosmology. (See e.g. S.Weinberg, "Cosmology", Oxford University Press, 2008.)

\item Anisotropic cosmological solutions and cosmological billiards. (See e.g. ``Cosmological billiards'', by
T. Damour, M. Henneaux and H. Nicolai;
Published in Class.Quant.Grav. 20 (2003) R145-R200;
e-Print: hep-th/0212256)

\item Bianchi classification.

\end{itemize}

\newpage

\section*{LECTURE XII \\ {\it  Geometry of the de Sitter space--time. Geometry of the anti--de Sitter space--time. Penrose--Carter diagrams for de Sitter and anti de Sitter. Wick rotation. Global coordinates. Poincare coordinates. Hyperbolic distance.}}

\vspace{10mm}

{\bf 1.} So far we always have been setting the cosmological constant, $\Lambda$, to zero. In this lecture we find the most symmetric and simplest (so to say ground state) solutions of the Einstein equations in the case when $\Lambda \neq 0$, but $T_{\mu\nu} = 0$. Note, however, that the standard energy--momentum tensor,

\bqa
T_{\mu\nu} = \rho \, u_\mu \, u_\nu + p\, \left(u_\mu\, u_\nu - g_{\mu\nu}\right),
\eqa
with $p = - \rho$, i.e. with $w = -1$, describes the same situation, $T_{\mu\nu} = \rho \, g_{\mu\nu}$, as non--zero cosmological constant, if $\rho = const$. Hence, one can find the metrics that are defined below using the methods of the previous lecture, but for many reasons we prefer the pure geometric picture/language adopted here. E.g., it makes symmetries apparent.

The $D$-dimensional space--time (with $D>2$) of constant curvature solves the following equation:

\bqa\label{first}
G_{\alpha\beta} = \pm \frac{(D-1)(D-2)}{2}\, H^2 \, g_{\alpha\beta}, \quad \alpha, \beta = 0, \dots, D-1.
\eqa
Here $G_{\alpha\beta}$ is the Einstein tensor and $H$ is the Hubble constant:

\bqa\label{323}
\Lambda = \frac{(D-1)\, (D-2)}{2} \, H^2.
\eqa
In this equation ``$+$'' sign corresponds to the space--time of constant positive curvature. At the same time ``$-$'' sign corresponds to the space--time of constant negative curvature. The reason for that is explained below. For $D=2$ the relation between $\Lambda$ and $H$ is different from (\ref{first}) and (\ref{323}).

{\bf 2.} The $D$-dimensional \underline{de Sitter space--time}, that of the constant {\it positive} curvature, can be realized as the following hyperboloid,

\bqa\label{hyp}
- \left(X^0\right)^2 + \left(X^1\right)^2 + \dots + \left(X^D\right)^2 \equiv - \eta_{AB} \, X^A \, X^B = H^{-2}, \quad A,B = 0, \dots, D,
\eqa
placed into the ambient $(D+1)$-dimensional Minkowski space--time, with the metric $ds^2 = \eta_{AB} \, dX^A \, dX^B$. One way to see that a metric on such a hyperboloid solves equation (\ref{first}) is to observe that it can be obtained from the sphere via the analytical continuation $X^{D+1} \to i\, X^0$. This is referred to as \underline{Wick rotation}. In fact, after such a change of coordinates the ambient Minkowski space--time is transformed into the Euclidian one. At the same time, the equation (\ref{hyp}) is transformed into the one defining $D$--dimensional sphere. Note that the Wick rotation does not change the sign of the curvature and the sphere does solve such an equation as (\ref{first}) with Euclidian signature.

\begin{figure}
\begin{center}
\includegraphics[scale=0.4]{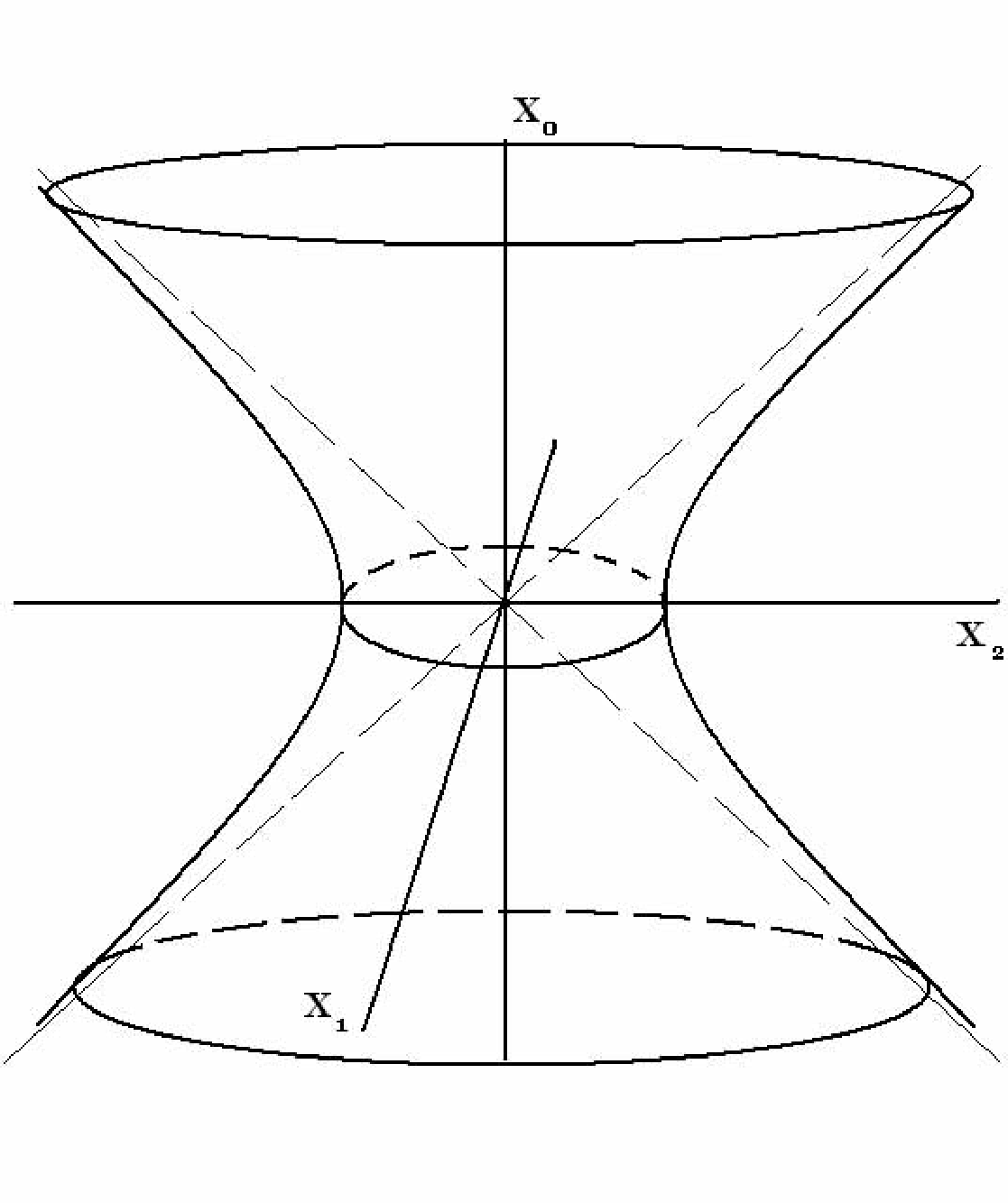}\caption{}\label{fig20}
\end{center}
\end{figure}

For illustrative reasons we depict two-dimensional de Sitter space--time on the fig. (\ref{fig20}). Here each constant $X_0$ slice of this two-dimensional de Sitter space--time is the circle of the radius $(H^{-2} + X_0^2)$.

Another way to see that the hyperboloid in question has a constant curvature is to observe that eq. (\ref{hyp}) is invariant under $SO(D,1)$ Lorentz transformations of the ambient space--time. The stabilizer of an arbitrary point, say $\left(X^0, X^1, X^2, \dots X^D\right) = \left(0,1/H,0, \dots, 0\right)$, obeying (\ref{hyp}) is $SO(D-1,1)$ group. Hence, the de Sitter space--time is \underline{homogeneous}, $SO(D,1)/SO(D-1,1)$, manifold and any its point is equivalent to another one. Furthermore, all directions at every point are equivalent to each other up to the difference between space--like and time--like directions. (Compare this with the sphere, which is $SO(D+1)/SO(D)$.) Thus, $SO(D,1)$ Lorentz group of the ambient Minkowski space is the \underline{de Sitter isometry group}. Isometry group is generated by Killing vectors. E.g. the isometry group of the flat space--time is the Poincare' one. It is generated by the translations in space and time, by spatial rotations and Lorentz boosts in spatial directions.

The geodesic distance, $L_{12}$, between two points, $X_1^A$ and $X_2^A$, on the hyperboloid can be conveniently expressed via the so-called \underline{hyperbolic distance}, $Z_{12}$, as follows:

\bqa
\frac{\cos \left(H \,L_{12}\right)}{H^2} \equiv \frac{Z_{12}}{H^2} \equiv - \eta_{AB}\, X^A_1 \, X^B_2, \quad {\rm where} \quad - \eta_{AB} \, X^A_{1,2} \, X^B_{1,2} = H^{-2}.
\eqa
To better understand the meaning of this expression, it is instructive to compare it to the geodesic distance, $l_{12}$, on a sphere of radius $R$:

$$R^2 \, \cos\left(\frac{l_{12}}{R}\right) \equiv R^2 \, z_{12} \equiv \left(\vec{X}_1, \, \vec{X}_2\right),\quad
{\rm where} \quad \vec{X}^2_{1,2} = R^2.$$
While the spherical distance, $z_{12}$, is always less than unity, the hyperbolic one, $Z_{12}$, can acquire any value because of the Minkowskian signature of the metric. Hyperbolic distance can be found from the spherical one after the above defined Wick rotation.

All geodesics on the hyperboloid of fig. (\ref{fig20}) are curves that are cut out on it by planes going through the origin of the ambient Minkowski space--time. (The situation here is absolutely similar to the case of the sphere, as can be understood after the Wick rotation.) Hence, space-like geodesics are ellipses, time-like ones are hyperbolas and light-like are straight generatrix lines of the hyperboloid, which are also light--like in the ambient Minkowski space--time.

{\bf 3.} To define a metric on the de Sitter space--time, which is induced from the ambient space, one has to find a solution of eq. (\ref{hyp}). One possibility is as follows

\bqa
X^0 = \frac{\sinh (Ht)}{H}, \quad X^i = \frac{n_i \, \cosh(Ht)}{H}, \quad i = 1, \dots, D
\eqa
where $n_i$ is a unit, $n_i^2 = 1$, $D$-dimensional vector. One can choose:

\bqa
n_1 = \cos \theta_1, \quad -\frac{\pi}{2} \leq \theta_1 \leq \frac{\pi}{2} \nonumber \\
n_2 = \sin \theta_1 \, \cos \theta_2, \quad -\frac{\pi}{2} \leq \theta_2 \leq \frac{\pi}{2} \nonumber \\
\dots \\
n_{D-2} = \sin\theta_1 \, \sin \theta_2 \dots \sin \theta_{D-3} \, \cos\theta_{D-2}, \quad -\frac{\pi}{2} \leq \theta_{D-2} \leq \frac{\pi}{2} \nonumber \\
n_{D-1} = \sin\theta_1 \, \sin \theta_2 \dots \sin \theta_{D-2} \, \cos\theta_{D-1}, \quad -\pi \leq \theta_{D-1} \leq \pi \nonumber \\
n_{D} = \sin\theta_1 \, \sin \theta_2 \dots \sin \theta_{D-2} \, \sin\theta_{D-1}. \nonumber
\eqa
Then, the induced metric is:

\bqa\label{global}
ds^2 = dt^2 - \frac{\cosh^2(Ht)}{H^2} \, d\Omega^2_{D-1},
\eqa
where

\bqa
d\Omega^2_{D-1} = \sum_{j=1}^{D-1} \left(\prod_{i=1}^{j-1} \sin^2\theta_i\right) \, d\theta_j^2
\eqa
is the line element on the unit $(D-1)$-dimensional sphere. The metric (\ref{global}) covers the de Sitter space--time totaly and is referred to as \underline{global}. Its constant $t$ slices are compact $(D-1)$-dimensional spheres. Note that one can obtain from (\ref{global}) the metric of the $D$-dimensional sphere after the analytical continuation, $H \, t \to i \left(\theta_D - \frac{\pi}{2}\right)$, which is the same as above Wick rotation.

The hyperbolic distance in these coordinates is given by:

\bqa
Z_{12} = - \sinh(H t_1) \, \sinh(H t_2) + \cosh(H t_1)\, \cosh(H t_2) \, \cos(\omega),
\eqa
where $\cos(\omega) = (\vec{n}_1, \vec{n}_2)$.

\begin{figure}
\begin{center}
\includegraphics[scale=0.4]{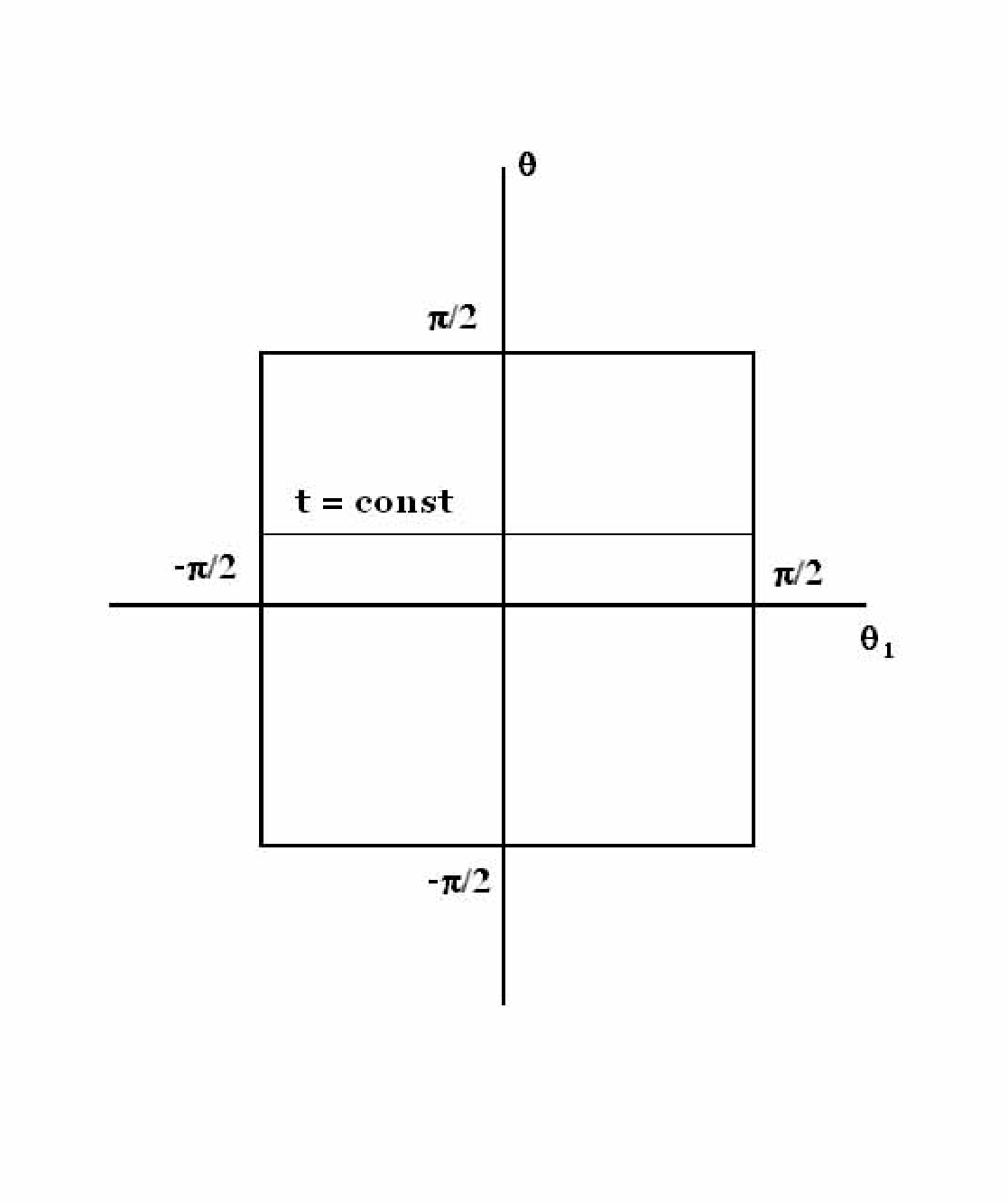}\caption{}\label{fig21}
\end{center}
\end{figure}

{\bf 4.} To understand the causal structure of the de Sitter space--time it is convenient to transform the global coordinates as follows:

$$
\cosh^2(H t) = \frac{1}{\cos^2\theta}, \quad -\frac{\pi}{2} \leq \theta \leq \frac{\pi}{2},
$$
and to obtain the  metric:

\bqa
ds^2 = \frac{1}{H^2 \, \cos^2\theta} \, \left[d\theta^2 - d\Omega^2_{D-1}\right].
\eqa
If $D>2$, then to draw the diagram on the two-dimensional sheet, we should choose, in addition to $\theta$, one of the angles $\theta_j$, $j = 1, \dots, D-1$. The usual choice is $\theta_1$ because the metric in question has the form $d\theta^2 - d\theta_1^2 - \sin^2(\theta_1) d\Omega^2_{D-2}$, i.e., it is flat in the $(\theta-\theta_1)$-plain. The Penrose--Carter diagram for the de Sitter space--time, whose dimension is grater than $2$, is depicted on the fig. (\ref{fig21}). The straight thin line here is the constant $t$ and/or $\theta$ slice.

Note that when $D>2$ angle $\theta_1$ is taking values in the range $\left[-\frac{\pi}{2},\, \frac{\pi}{2}\right]$. At the same time, when $D=2$ we have that $\theta_{D-1} \equiv \theta_1 \in \left[-\pi, \, \pi\right]$. When $D>2$, the problem with the choice of $\theta_1$ in the Penrose--Carter diagram is that then cylindrical topology, $S^{D-1} \times R$, of the de Sitter space--time is not transparent. At the same time, the complication with the choice of $\theta_{D-1} \in \left[-\pi,\, \pi\right]$, instead of $\theta_1$, appears form the fact that the metric in the $(\theta-\theta_{D-1})$--plain is not flat, if $D > 2$. For this reason we prefer to consider just the stereographic projection in the two-dimensional case because it is sufficient to describe the causal structure and also clearly shows the topology of the de Sitter space--time.

\begin{figure}
\begin{center}
\includegraphics[scale=0.4]{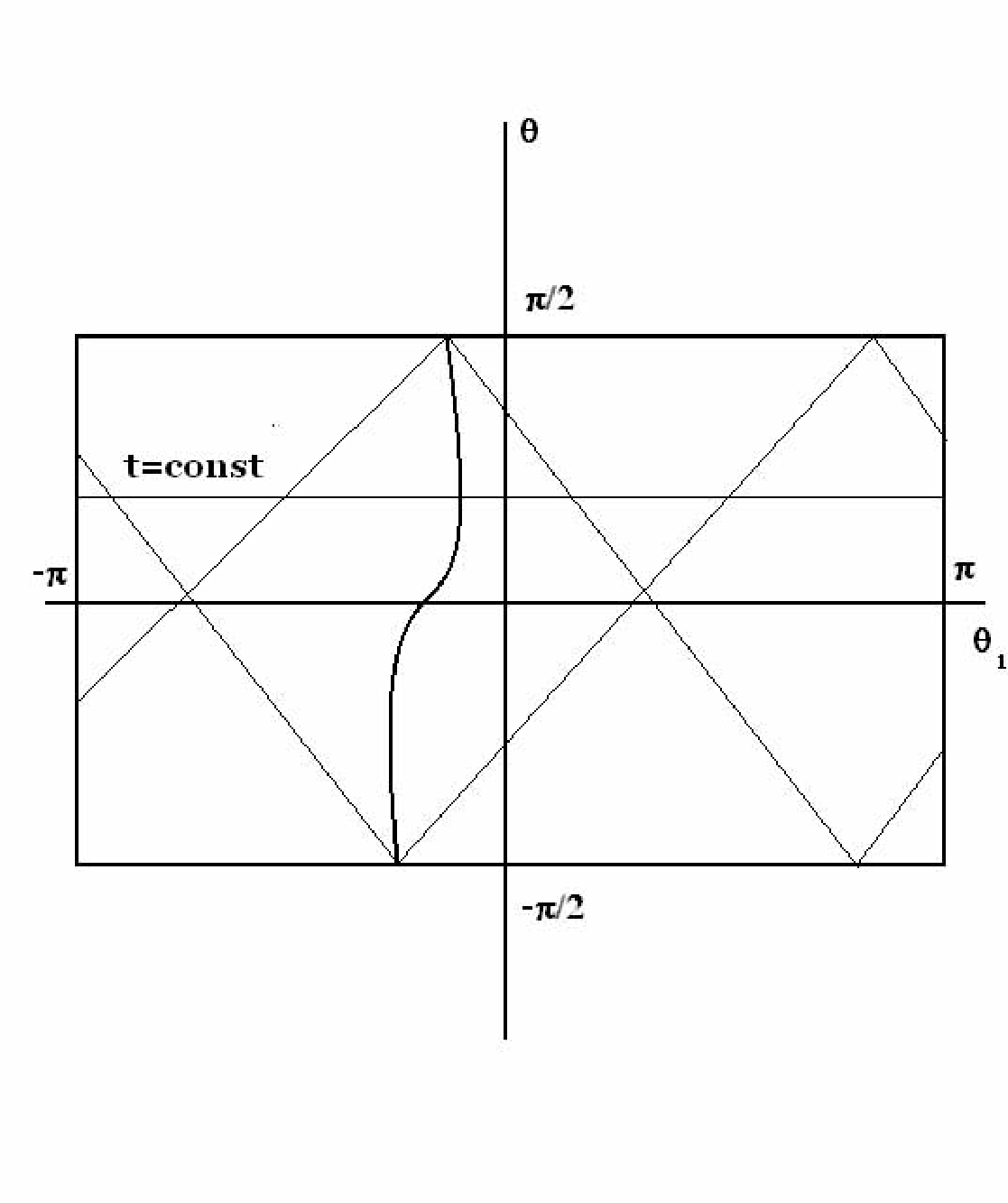}\caption{}\label{fig22}
\end{center}
\end{figure}

The Penrose--Carter diagram of the two-dimensional de Sitter space--time is shown on the fig. (\ref{fig22}). This is the stereographic projection of the hyperboloid from the fig. (\ref{fig20}). What is depicted here is just a cylinder because the left and right sides of the rectangle on this figure are glued to each other. Thus, while on the fig. (\ref{fig21}) the positions $\theta_1 = \pm \frac{\pi}{2}$ sit at the opposite poles of the spherical time slices, on the fig. (\ref{fig22}) the positions $\theta_1 = \pm \pi$ coincide.

The fat solid vertically directed curve on the fig. (\ref{fig22}) is a world line of a massive particle. Thin straight lines, which compose $45^{\rm o}$ angle with both $\theta$ and $\theta_1$ axes, are light rays. From this picture one can see that every observer has a causal diamond within which he can exchange signals. Due to the expansion of the de Sitter space--time there are parts of it that are causally disconnected from the observer.

{\bf 5.} Another possible solution of (\ref{hyp}) is based on the choice:

\bqa
- \left(H\, X^0\right)^2 + \left(H\, X^D\right)^2 = 1 - \left(H\, x^i_+\right)^2 \, e^{2 \, H\, \tau_+}, \nonumber \\
\left(H\, X^1\right)^2 + \dots + \left(H\, X^{D-1}\right)^2 = \left(H\, x^i_+\right)^2 \, e^{2 \, H\, \tau_+}.
\eqa
Then, one can define

\bqa\label{choice}
H \, X^0 = \sinh\left(H \, \tau_+\right) + \frac{\left(H\, x^i_+\right)^2}{2}\, e^{H\,\tau_+}, \nonumber \\
H\, X^i = H x^i_+ \, e^{H \, \tau_+}, \quad i = 1, \dots, D-1, \nonumber \\
H\, X^D = - \cosh\left(H \, \tau_+\right) + \frac{\left(H \, x^i_+\right)^2}{2} \, e^{H\,\tau_+}.
\eqa
With such coordinates the induced from $ds^2 = \eta_{AB} dX^A dX^B$ metric is

\bqa\label{EPP1}
ds^2_+ = d\tau_+^2 - e^{2\, H \, \tau_+} \, d\vec{x}_+^2.
\eqa
Note, however, that in (\ref{choice}) we have the following restriction:
$- X^0 + X^D = -\frac{1}{H} \, e^{H \, \tau_+} \leq 0$, i.e., metric (\ref{EPP1}) covers only half, $X^0 \geq X^D$, of the entire de Sitter space--time. It is referred to as \underline{the expanding Poincar\'{e} patch} (EPP). Another half of the de Sitter space--time, $X^0 \leq X^D$, is referred to as \underline{the contracting Poincar\'{e} patch} (CPP) and is covered by the metric

\begin{figure}
\begin{center}
\includegraphics[scale=0.4]{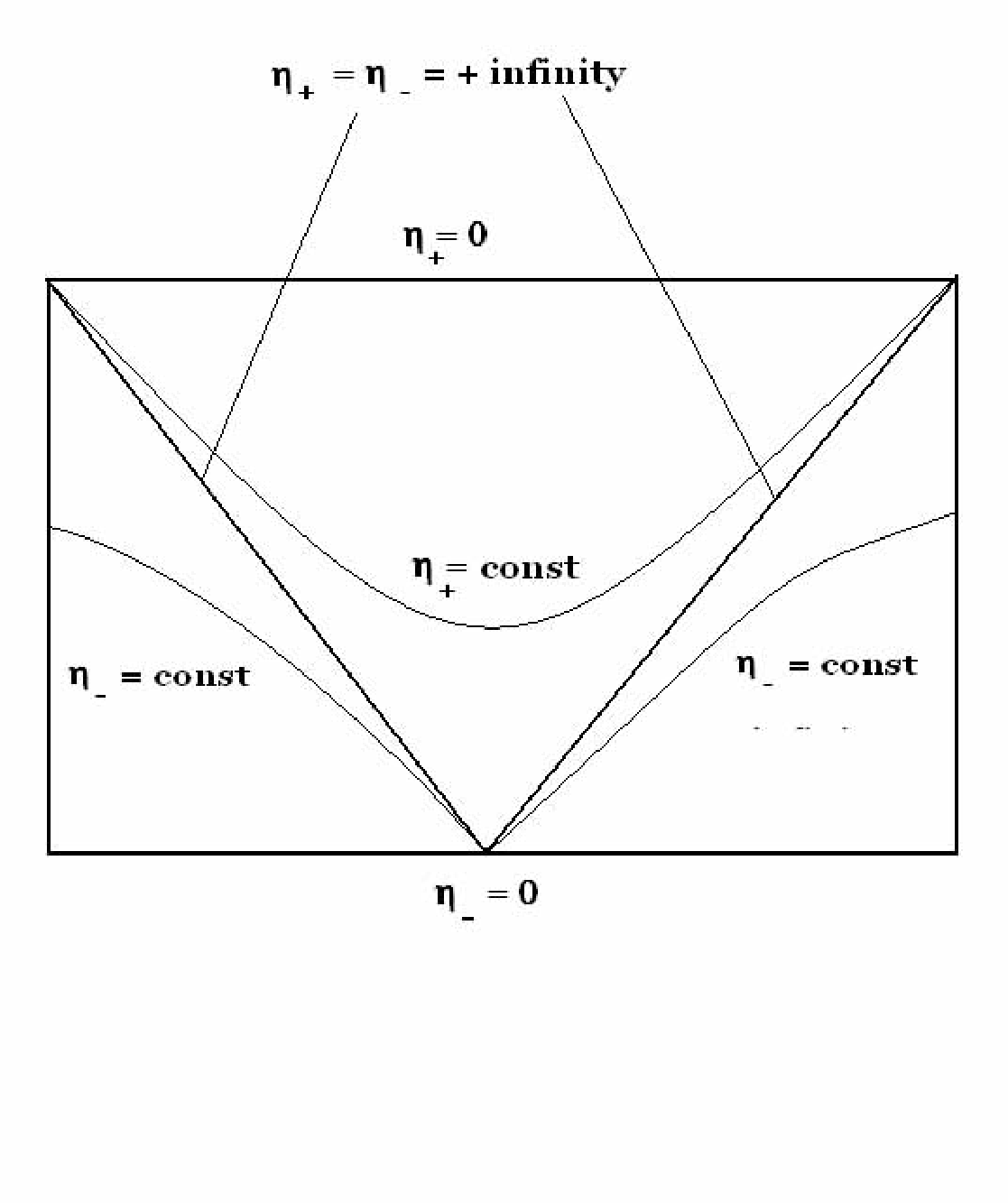}\caption{}\label{fig23}
\end{center}
\end{figure}

\bqa\label{CPP1}
ds^2_- = d\tau_-^2 - e^{- 2\, H \, \tau_-} \, d\vec{x}_-^2.
\eqa
In both patches it is convenient to change the proper time $\tau_\pm$ into the conformal one. Then, the EPP and CPP both possess the same metric:

\bqa\label{PP}
ds^2_\pm = \frac{1}{\left(H \, \eta_{\pm}\right)^2}\,\left[d\eta_\pm^2 - d\vec{x}_\pm^2\right], \quad H\, \eta_\pm = e^{\mp H\, \tau_\pm}.
\eqa
However, while in the EPP the conformal time is changing form $\eta_+ = + \infty$ at past infinity ($\tau_+ = - \infty$) to $0$ at future infinity ($\tau_+ = + \infty$), in the CPP the conformal time is changing from $\eta_- = 0$ at past infinity ($\tau_- = -\infty$) to $+\infty$ at future infinity ($\tau_- = + \infty$).
Both the EPP and CPP are shown on the fig. (\ref{fig23}). The boundary between the EPP and CPP is light-like and is situated at $\eta_\pm = + \infty$. We also show on this picture the constant conformal time slices.

The hyperbolic distance in the EPP and CPP has the form:

\bqa\label{Z12}
Z_{12} = 1 + \frac{\left(\eta_1 - \eta_2\right)^2 - \left|\vec{x}_1 - \vec{x}_2\right|^2}{2 \, \eta_1 \, \eta_2}.
\eqa
It is worth mentioning here that it is possible to cover simultaneously the EPP and CPP with the use of the metric (\ref{PP}), if one makes the changes $\vec{x}_\pm \to \vec{x}$ and $\eta_\pm \to \eta \in (-\infty, +\infty)$. Then, while at the negative values of the conformal time, $\eta = - \eta_+ < 0$, it covers the EPP, at its positive values, $\eta = \eta_- > 0$, this metric covers the CPP. The inconvenience of such a choice of global metric is due to that the boundary between the EPP and CPP simultaneously corresponds to $\eta = \pm \infty$.

We define here the physical and comoving spatial volumes in global de Sitter space--time and in its Poincar\'{e} patches. The spatial sections in all aforementioned metrics contain conformal factors, $\frac{\cosh^2\left(H\, t\right)}{H^2}$ or $\frac{1}{\left(H \, \eta_\pm\right)^2}$. Then, there is the volume form, $d^{D-1}V$, with respect to the spatial metric, which is multiplying the corresponding conformal factor. This form remains constant during the time evolution of the spatial sections and is referred to as \underline{comoving volume}.

It is important to observe that if one considers a dust in the de Sitter space--time, then its density per comoving volume remains constant independently of whether spatial sections are expanding or contracting.

At the same time, if one takes into account the conformal factor, i.e. the expansion (contraction) of the EPP (CPP), then he has to deal with the \underline{physical volume}, $\frac{d^{D-1}V_\pm}{\left(H \, \eta_\pm\right)^{D-1}}$. In the global de Sitter space--time the physical volume is $\frac{\cosh^{D-1}\left(H\, t\right) \, d^{D-1}V_{\rm sphere}}{H^{D-1}}$. Of course the density of the dust with respect to such a volume is changing in time.

{\bf 6.} The case of anti--de Sitter space--time is very similar to the de Sitter one. However, there are certain differences. $D$--dimensional anti--de Sitter is the hyperboloid,

\bqa\label{AdS}
- X_0^2 + \sum_{j=1}^{D-1} X_j^2 - X_D^2 = - \frac{1}{H^2},
\eqa
embedded into the $(D+1)$--dimensional space--time of $(-, +, \dots, +, -)$ signature:

\bqa\label{D22}
ds^2 = - dX_0^2 + \sum_{j=1}^{D-1} dX_j^2 - dX_D^2.
\eqa
Hence, the isometry of the hyperboloid (\ref{AdS}) is $SO(D-1,2)$, while the stabilizer of a generic point is $SO(D-1,1)$. Hence, the anti--de Sitter space--time is homogeneous, $SO(D-1,2)/SO(D-1,1)$, manifold. (For the anti--de Sitter space-time we use different signature of the metric to match with the standard notations in the literature.)

Under the Wick rotation $X_D \to i X_D$ we obtain the two--sheeted hyperboloid,

$$
- X_0^2 + \sum_{j=1}^{D} X_j^2 = - \frac{1}{H^2} \quad {\rm or} \quad \sum_{j=1}^D X_j^2 = X_0^2 - \frac{1}{H^2},
$$
embedded into the $(D+1)$--dimensional Minkowski space--time:

$$
ds^2 = - dX_0^2 + \sum_{j=1}^{D} dX_j^2.
$$
This two--sheeted hyperboloid is the constant negative curvature $D$--dimensional Lobachevski space of Euclidian signature. The three--dimensional version of this space we have encountered in the previous lecture.

Note that after the redefinitions $X \leftrightarrow iX$ and $H \leftrightarrow i H$ we can map all the above defined $D$--dimensional space--times, $D$--dimensional sphere and $D$--dimensional Lobachevski space to each other. The redefinition $H\to i H$ changes the sign of the curvature. Hence, similarly to the sphere and to the de Sitter space--time one can define here the hyperbolic distance both in the Lobachevski space and in the anti--de Sitter space--time. Moreover, this way one also can define the geodesics in Lobachevski space and anti--de Sitter space--time.

\begin{figure}
\begin{center}
\includegraphics[scale=0.4]{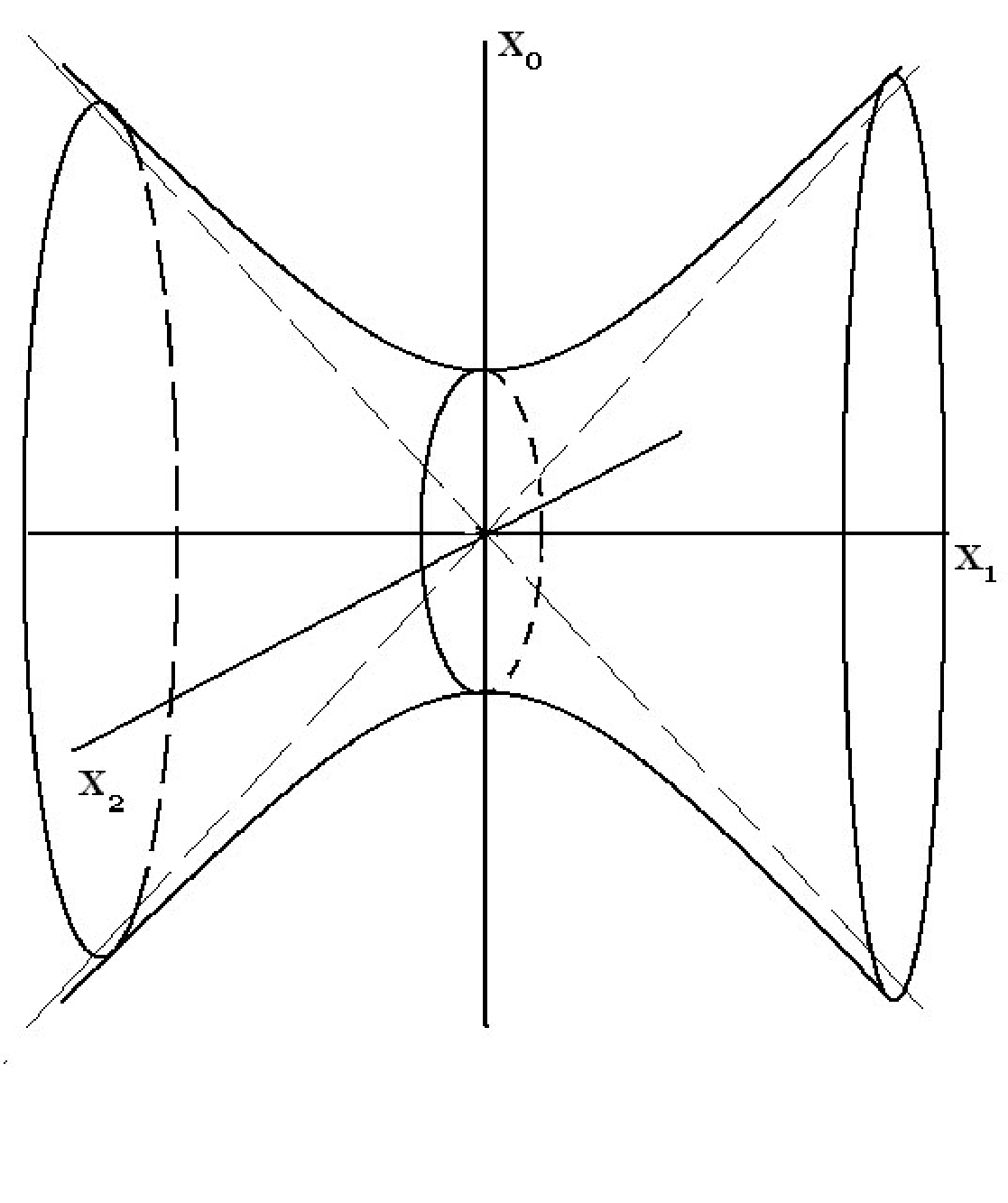}\caption{}\label{fig24}
\end{center}
\end{figure}

{\bf 7.} Solving eq. (\ref{AdS}) as

\bqa
X_0 = \frac{1}{H} \, \cosh\rho \, \cos\tau \quad X_D = \frac{1}{H} \, \cosh\rho \, \sin\tau, \nonumber \\
X_j = \frac{1}{H} \, \sinh\rho \, n_j, \quad n_j^2 = 1, \quad j = 1, \dots, D-1,
\eqa
and plaguing this solution into (\ref{D22}), we obtain the following induced metric on the anti--de Sitter space--time:

\bqa\label{adsmetr}
ds^2 = \frac{1}{H^2} \, \left[- \cosh^2 \rho \, d\tau^2 + d\rho^2 + \sinh^2 \rho \, d\Omega^2_{D-2}\right].
\eqa
Here $\tau \in [0, 2\pi)$ and $d\Omega_{D-2}^2$ is the metric on the $(D-2)$--dimensional sphere, which is parameterized by $n_j$. In the case when $D=2$ the coordinate $\rho$ is ranging from $-\infty$ to $+\infty$.
Then constant time $\tau$ slices are just hyperbolas. The two--dimensional anti--de Sitter space--time looks just as the de Sitter one, but rotated by the angle $\pi/2$ around the $X_2$ axis. See the fig. (\ref{fig24}).

When $D>2$ the range of values of $\rho$ is $[0,+\infty)$. Only in such a case we cover entire anti--de Sitter space
and avoid double--counting. Then constant time $\tau$ slices are Lobachevski hyperboloids
with the metric

\bqa\label{Lobhyp}
dl^2 =  \frac{1}{H^2} \, \left[d\rho^2 + \sinh^2 \rho \, d\Omega^2_{D-2}\right].
\eqa
The three--dimensional variant of this metric (with $H=1$ and $\rho \to \chi$) we have encountered in the previous lecture. Here $\rho = 0$ is just the center of the hyperboloid in question.

Note that above defined time $\tau$ is compact, $\tau \in [0,2\pi)$. Which can be seen on the fig. (\ref{fig24}). This is physically  unfavorable situation. Hence, in physics literature one usually considers universal cover of the space--time under consideration, i.e. the metric (\ref{adsmetr}) with $\tau \in (-\infty, +\infty)$. The spatial boundary, $\rho \to +\infty$, of such a space--time is conformaly equivalent to the cylinder $R\times S^{D-2}$. Here $R$ is for the axis of time $\tau$ and $S^{D-2}$ is for the sphere described by the metric $d\Omega_{D-2}^2$.

\begin{figure}
\begin{center}
\includegraphics[scale=0.4]{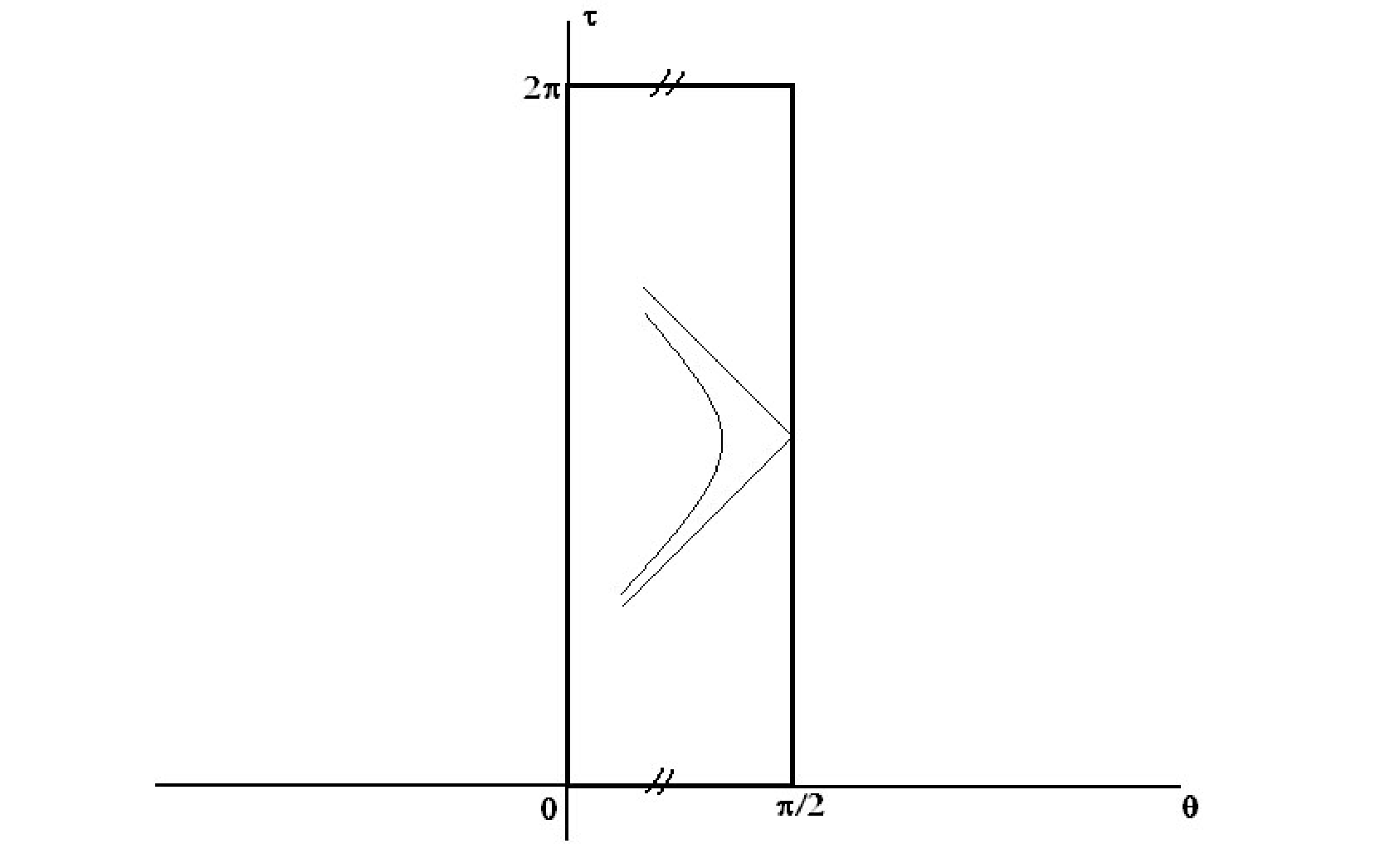}\caption{}\label{fig25}
\end{center}
\end{figure}

{\bf 8.} To draw the Penrose--Carter diagram of the anti--de Sitter space--time, one can make the following change of variables:

$$
\tan \theta = \sinh \rho, \quad {\rm and} \quad ds^2 = \frac{1}{H^2 \, \cos^2 \theta} \, \left[- d\tau^2 + d\theta^2 + \sin^2\theta \, d\Omega_{D-2}^2\right].
$$
Here if $D=2$, then $\theta \in \left[-\pi/2, \pi/2\right]$, while for the case of $D>2$ one has to deal with $\theta \in \left[0,\pi/2\right]$ because of the above mentioned difference of the range of values of $\rho$. Then the Penrose--Carter diagram for the case of $D=2$ is the same as for the two--dimensional de Sitter space--time, which is, however, rotated by the $\pi/2$ angle around the axis perpendicular to the plane of the picture. That is the stereographic projection of the hyperboloid from the fig. (\ref{fig24}). Thus, two--dimensional anti--de Sitter space--time has two boundaries --- at $\theta = - \pi/2$ and $\theta = \pi/2$.

At the same time the Penrose--Carter diagram for the case of $D>2$ is shown on the fig. (\ref{fig25}). On this picture $\theta = \pi/2$ is the spatial boundary of the anti--de Sitter space, but $\theta=0$ is not a boundary, rather it corresponds to $\rho = 0$, which is the center of the $(D-1)$--dimensional Lobachevski space, which is represented by the metric (\ref{Lobhyp}). The Lobachevski space topologically is just a ball.

Also on the fig. (\ref{fig25}) we show light rays and a world--line of a massive particle by the straight and curly thin curves, correspondingly. As can be seen from this picture, starting at any internal point, light rays can reach the spatial boundary of the space--time within finite coordinate and proper times.

\begin{figure}
\begin{center}
\includegraphics[scale=0.4]{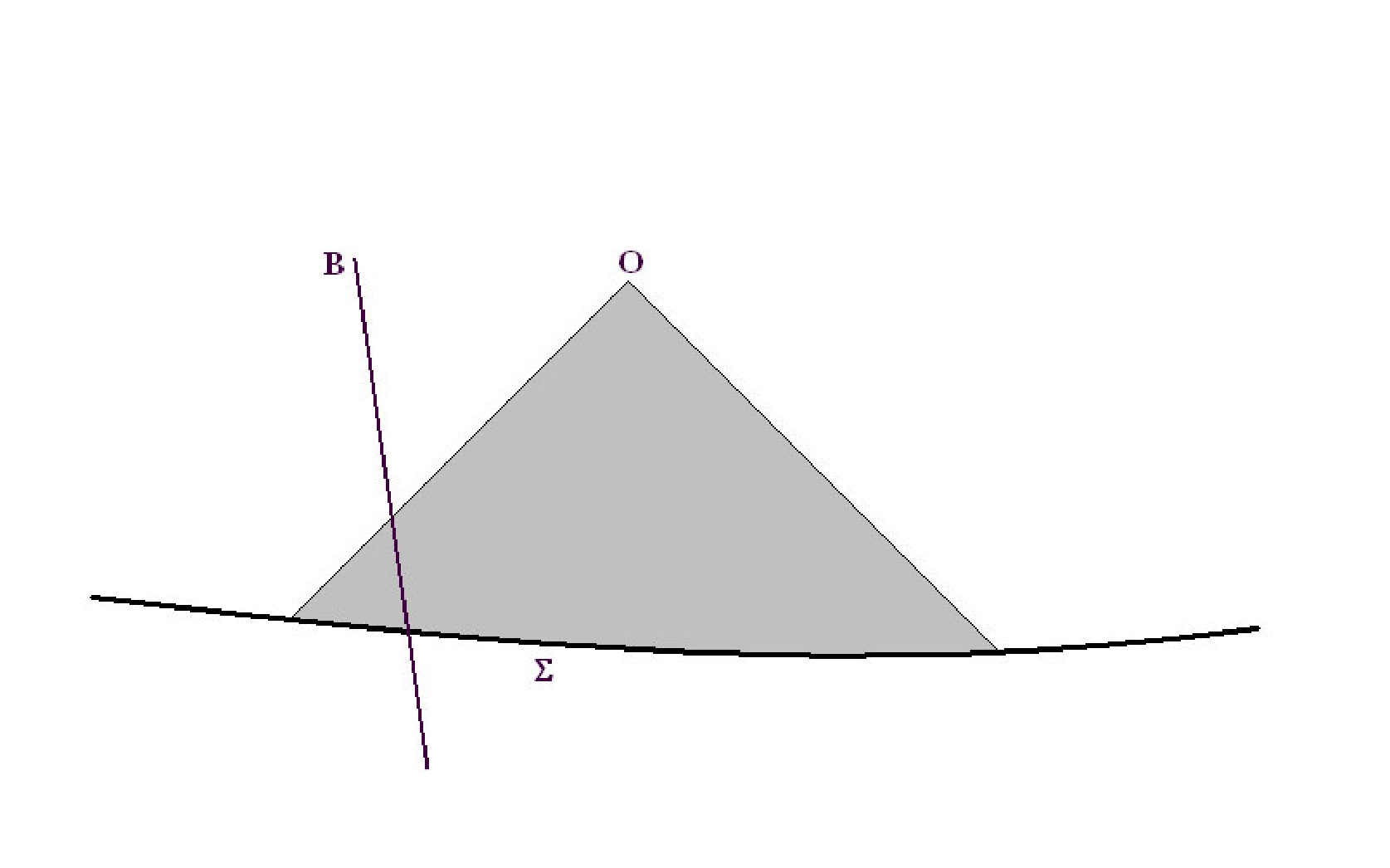}\caption{}\label{fig26}
\end{center}
\end{figure}

As the result to solve a Cauchy problem in the anti--de Sitter space--time one has to specify boundary conditions
on top of the initial ones. Consider the fig. (\ref{fig26}). Indeed, to solve a physical problems one deals with causal differential equations. Then to find a value of the solution at a point O one needs to know its behavior in the grey causal triangle, which is bounded by the past light--cone. In the absence of a boundary this behavior
is totaly defined by initial values given at the Cauchy surface $\Sigma$. In the case under study, however,
we encounter such a situation that a part of the boundary B can lay in the causal past of an internal point. That is the reason one has the above mentioned complication of the Cauchy problem in anti--de Sitter space--time. This property is referred to as the \underline{absence of global hyperbolicity}. Rephrasing this, anti--de Sitter space--time acts as a box: its gravity is such that it attracts back massive particles when they try to escape to the spatial infinity. But for light rays we have to specify boundary conditions at spatial infinity: e.g. that could be of mirror type or say absorbing boundaries.

{\bf 9.} Let us describe the Poincare patch and Poincare coordinates of the anti--de Sitter space--time.
Let us solve the eq. (\ref{AdS}) as follows:

\bqa\label{334}
X_0 = \frac{z}{2} \, \left[1 + \frac{1}{z^2} \, \left(\frac{1}{H^2} + x_\mu \, x^\mu\right)\right], \quad z \geq 0, \nonumber \\
X_1 = \frac{z}{2} \, \left[1 - \frac{1}{z^2} \, \left(\frac{1}{H^2} - x_\mu \, x^\mu\right)\right] \nonumber \\
X_\mu = \frac{x_\mu}{H\,z}, \quad \mu = 2, \dots, D.
\eqa
Here $x_\mu x^\mu = \eta_{\mu\nu} x^\mu x^\nu$ and $||\eta_{\mu\nu}|| = Diag(-,+,\dots,+)$.

This is very similar to the choice of coordinates that we have made in the Poincare patch of the de Sitter space--time. To see that one just has to make the change $z \to \eta = e^{-H\tau}$. Note that here $X_0 - X_1 = \frac{1}{H^2 z} \geq 0$. Hence, these coordinates cover only
half of the entire anti--de Sitter space--time.

As the result from (\ref{D22}) we obtain the induced metric as follows:

\bqa\label{335}
ds^2 = \frac{1}{\left(H\,z\right)^2} \, \left[dz^2 + dx_\mu \, dx^\mu\right].
\eqa
It is very similar to the metric on the Poincare patch of the de Sitter space--time, but with one crucial difference that while $\eta$ is time and, hence, the de Sitter metric is time dependent, $z$ is spatial coordinate. The time coordinate of the anti--de Sitter space--time is hidden in $dx_\mu \, dx^\mu \equiv - dt^2 + d\vec{x}^2$. Thus, both the anti--de Sitter space--time and its Poincare patch are static.

\vspace{5mm}

\centerline{\bf Problems:}

\vspace{5mm}

\begin{itemize}

\item What is the result of Wick rotation of the Rindler matric?

\item Derive the metric (\ref{EPP1}) from eq. (\ref{choice}).

\item Derive the metric (\ref{335}) from eq. (\ref{334}).

\end{itemize}

\vspace{10mm}

\centerline{\bf Subjects for further study:}

\vspace{5mm}

\begin{itemize}

\item  Harmonics and Green's functions in homogeneous spaces of constant curvature. (For the case of de Sitter space--time see e.g. ``Lecture notes on interacting quantum fields in de Sitter space'', by
E.T. Akhmedov; Published in Int.J.Mod.Phys. D23 (2014) 1430001; e-Print: arXiv:1309.2557 )

\item Anti--de Sitter isometry group and the conformal group action on its boundary.

\item Conformal embeddings (C. Fefferman, C.R. Graham, Conformal invariants, in: Elie Cartan et les Mathématiques d'ajourd'hui, Astérisque (hors série) Société Mathématique de France, Paris, 1985, pp. 95–116.)

\item Does a free--falling charge in dS and AdS spaces emit scalar, electromagnetic and gravitational radiation? (On dS space case consider
e.g. Classical radiation by free-falling charges in de Sitter spacetime,
E.T. Akhmedov, Albert Roura, A. Sadofyev,
Published in Phys.Rev. D82 (2010) 044035; e-Print: arXiv:1006.3274; \\
De Sitter space and perpetuum mobile,
Emil T. Akhmedov, P.V. Buividovich, Douglas A. Singleton,
Published in Phys.Atom.Nucl. 75 (2012) 525-529; e-Print: arXiv:0905.2742.)

\end{itemize}

\end{document}